\documentclass[a4paper,12pt]{article}

\usepackage{amsmath}
\usepackage{amssymb}
\usepackage{amsfonts}
\usepackage{amsthm}
\usepackage{braket}
\usepackage{comment}
\usepackage{color}
\usepackage[normalem]{ulem} 
\usepackage{cancel} 
\usepackage{cite}
\usepackage{tikz}
\theoremstyle{definition}
\newtheorem{theorem}{Theorem}[section]

\newtheorem{definition}[theorem]{Definition}

\renewcommand\sout{\bgroup \color{red} \ULdepth=-.5ex \ULset}

\newcommand{\del}{\partial}

 \makeatletter
 \@addtoreset{equation}{section}
 \makeatother

\newtheorem{axiomC}{Axiom} 
\newtheorem{axiomV}{Axiom} 

\usepackage[pdftex]{hyperref}
\newcommand{\dop}{\mathrm{d}} 
\newcommand{\iop}{\iota} 

\newcommand{\dangle}[2]{\langle \! \langle #1, #2 \rangle \! \rangle} 
\newcommand{\pairing}[2]{\dangle{#1}{#2}} 
\newcommand{\rhotilde}{\rho_{\tilde{L}}} 
\newcommand{\dualSchouten}[2]{[#1, #2]^*_{\rm S}} 

\newcommand{\p}{P} 
\newcommand{\Q}{Q} 
\newcommand{\R}{R} 
\newcommand{\e}{\del} 
\newcommand{\te}{\tilde{\del}} 
\newcommand{\boldpar}[1]{\boldsymbol{(} #1 \boldsymbol{)}} 
\newcommand{\etaprod}[1]{\boldpar{#1}_+} 
\newtheorem{propositionLWX}{Proposition}
\newtheorem{lemmaLWX}{Lemma}

\newcommand{\pbra}[1]{\boldsymbol{(}#1\boldsymbol{)}_+}
\newcommand{\mbra}[1]{\boldsymbol{(}#1\boldsymbol{)}_-}
\newcommand{\pmbra}[1]{\boldsymbol{(}#1\boldsymbol{)}_\pm}
\newcommand{\bbra}[1]{\boldsymbol{(}#1\boldsymbol{)}}

\date{empty}
\begin{document}
\begin{titlepage}
\null
\begin{flushright}
January, 2019 
\end{flushright}
\vskip 2cm
\begin{center}
  {\Large \bf 
Doubled Aspects of Vaisman Algebroid and
\\
\vspace{0.5cm}
Gauge Symmetry in Double Field Theory 
}
\vskip 1.5cm
\normalsize
\renewcommand\thefootnote{\alph{footnote}}

{\large
Haruka Mori\footnote{h.mori(at)sci.kitasato-u.ac.jp}, 
Shin Sasaki\footnote{shin-s(at)kitasato-u.ac.jp}
and
Kenta Shiozawa\footnote{k.shiozawa(at)sci.kitasato-u.ac.jp}
}
\vskip 0.7cm
  {\it
  Department of Physics,  Kitasato University \\
  Sagamihara 252-0373, Japan
}
\vskip 2cm
\begin{abstract}
The metric algebroid proposed by Vaisman (the Vaisman algebroid) 
governs the gauge symmetry algebra generated by the {\sf C}-bracket in double
 field theory (DFT).
We show that the Vaisman algebroid is obtained by an analogue of the
 Drinfel'd double of Lie algebroids.
Based on a geometric realization of doubled space-time as a para-Hermitian manifold, 
we examine exterior algebras and a para-Dolbeault cohomology on DFT and
 discuss the structure of the Drinfel'd double behind the DFT gauge
 symmetry. 
Similar to the Courant algebroid in the generalized geometry, 
Lagrangian subbundles $(L,\tilde{L})$ in a para-Hermitian manifold play Dirac-like structures in the
 Vaisman algebroid.
We find that an algebraic origin of the strong constraint in DFT is
 traced back to the compatibility condition needed for $(L,\tilde{L})$ be a Lie bialgebroid.
The analysis provides a foundation toward the ``coquecigrue problem'' for
 the gauge symmetry in DFT.
\end{abstract}
\end{center}

\end{titlepage}

\newpage
\setcounter{footnote}{0}
\renewcommand\thefootnote{\arabic{footnote}}
\pagenumbering{arabic}
\tableofcontents
\section{Introduction} \label{sect:introduction}
String theory is a candidate of consistent quantum gravity. 
A large amount of efforts has been devoted to study string theory and
physicists and mathematicians revealed that the theory exhibits rich
mathematical structures of symmetries.
Among other things, dualities, which relate various different theories
for the same physical phenomena, play important roles to unravel the
overall picture of the theories.
For example, when one considers strings propagating in a compact space-time,
the strings can wrap around non-trivial cycles. The energy spectrum is
determined by the quantized momentum along the compactified direction
(the Kaluza-Klein (KK) modes $n \in \mathbb{Z}$) and the number of windings for the string (the winding modes
$w \in \mathbb{Z}$). The Hamiltonian $H$ is totally
invariant under the exchange of $n$ and $w$ (and the radii of the
compactified directions): $H (n,w) \leftrightarrow H
(w,n)$. This invariance is generalized to that by $O(D,D)$ when the compact
space is $T^D$ and the symmetry is called T-duality. T-duality relates several
consistent string theories and therefore it is important to study its
physical and mathematical properties in order to understand the whole
structures of string theories.

Double field theory (DFT) \cite{Hull:2009mi} is an effective theory of strings
where T-duality is realized manifestly.
DFT is a gravity theory whose dynamical fields are
the generalized dilation and the generalized metric.
They are the $O(D,D)$ tensor fields defined in the {\it doubled} space-time $\mathcal{M}$ of dimension $2D$.
The doubled space-time is characterized by the coordinate $x^M = (x^{\mu},
\tilde{x}_{\mu})$ where $x^{\mu}$ is the Fourier dual of the
KK momentum while $\tilde{x}_{\mu}$ is the dual of the winding momentum.
Since the physical degrees of freedom is doubled by construction, one
needs to impose a constraint (called the strong constraint) to the $O(D,D)$
tensor fields to obtain a physical gravity theory in the ordinary $D$-dimensional space-time.
Along with the $O(D,D)$ T-duality symmetry, the theory involves a gauge
symmetry that originates from the diffeomorphism and the $U(1)$ gauge
symmetry of the NSNS $B$-field. A mathematical structure of the gauge symmetry in DFT has
been studied in \cite{Hull:2009zb, Hohm:2012gk} and it is found that the
algebra associated with the symmetry is governed by the so-called
{\sf C}-bracket~\cite{Siegel:1993th}: 
\begin{align}
[\Xi_1, \Xi_2]_{\sf C} =& \ [X_1, X_2]_{L} + \mathcal{L}_{\xi_1} X_2 -
 \mathcal{L}_{\xi_2} X_1 
- \frac{1}{2} \dop \left( \iop_{X_1} \xi_2 - \iop_{X_2} \xi_1 \right)
\notag \\
& +  [\xi_1, \xi_2]_{\tilde{L}} + \mathcal{L}_{X_1} \xi_2 - \mathcal{L}_{X_2}
 \xi_1 - \frac{1}{2} \tilde{\dop} (\tilde{\iop}_{\xi_1} X_2 - \tilde{\iop}_{\xi_2} X_1),
\label{eq:C-bracket}
\end{align}
where $\Xi_i = \Xi_i^M \del_M \ (i=1,2)$ are vector fields on the doubled
space-time and $X_i = X^{\mu}_i \del_{\mu}$, $\xi_i = \xi_{i,\mu}
\tilde{\del}^{\mu}$ are the components of $\Xi_i = X_i + \xi_i$ corresponding to the
decomposition $x^M = (x^{\mu}, \tilde{x}_{\mu})$. 
The precise definition of each term is found in Appendix \ref{sect:C-bracket}.
Among various properties of the {\sf C}-bracket, the most notable one is that
it does not satisfy the Jacobi identity. Therefore
 the gauge symmetry of DFT must not be prescribed by a Lie group.
The {\sf C}-bracket \eqref{eq:C-bracket} looks quite similar to the Courant bracket
which has been appeared in the 
context of Lie bialgebroids
~\cite{LiWeXu}.
Although, they share various properties, there is a crucial difference
between them. 
The Courant bracket satisfies the strong homotopy Jacobi identity
~\cite{1998math......2118R} 
while the {\sf C}-bracket does not satisfy it in general.
The latter appears only after one solves the strong constraint in DFT.
The strong constraint is a sufficient condition that makes DFT be
a physical theory.
Mathematics of algebra based on the Courant bracket is well-established
(see for example \cite{Gualtieri}).

Mathematical structures of the symmetry algebra based on the {\sf C}-bracket
have been studied in several contexts \cite{Hohm:2012mf,Deser:2016qkw, Deser:2018oyg}. 
Recently, the para-Hermitian structure of the doubled space-time has appeared
to be a basic mathematical framework of the geometry behind DFT
\cite{Freidel:2017yuv, Freidel:2018tkj,Marotta:2018myj}.
Among other things, a mathematical ground of algebroid structures in the para-Hermitian manifold is
discussed \cite{Svoboda:2018rci}. 
Remarkably, based on a tangent bundle on a para-Hermitian manifold, 
an algebra defined by the {\sf C}-bracket is proposed in \cite{Vaisman:2012ke}. 
The author of \cite{Vaisman:2012ke} call this the {\it metric algebroid} (we
call this {\it Vaisman algebroid} after his name).
The Vaisman algebroid is introduced to explain the gauge symmetry in DFT. 
This is a new structure which has never been proposed in the literature.
The bracket of the Vaisman algebroid is defined on a general manifold and it satisfies 
parts of defining axioms of the
Courant algebroid but some of them are not satisfied.
The Vaisman algebroid is re-discovered as the (pre-)DFT algebroid in the study of geometric
structures in DFT \cite{Chatzistavrakidis:2018ztm}.

On the physical side, there are doubled structures as a whole in DFT.
On the other hand, there is an independent notion of ``double'' in
mathematics. For example, the idea of the {\it Drinfel'd double} of Lie
bialgebras has been established in \cite{Drinfeld:1983ky}.
Even more, an analogue of the Drinfel'd double for Lie bialgebroids 
is proposed \cite{LiWeXu}.
Notably, some aspects of doubled structures for the Vaisman algebroid 
are studied \cite{Vaisman:2012px}.
We are therefore guided to find out a correspondence between the
``doubled'' structures in DFT from physics and mathematical sides.
Although, the doubled structure of Courant algebroids based on the
Courant bracket, both from the
viewpoints of physics \cite{Deser:2014mxa} and mathematics
\cite{LiWeXu} has been studied, the relation between the {\sf C}-bracket and
the Vaisman algebroid needs more investigation. 

In this paper, we study doubled aspects of the Vaisman algebroid
and discuss the mathematical structure of the gauge symmetry in DFT.
We push forward the work in \cite{Vaisman:2012px} and make the
 algebroid structures of the DFT gauge symmetry be more explicit.
We also study another aspect of the prescription from the pre-DFT algebroid
to a Courant algebroid in \cite{Chatzistavrakidis:2018ztm} where a
geometric origin of the strong constraint is discussed.
We will show that an algebraic origin of the strong constraint, needed
for the closure of the DFT gauge algebra, is traced back to a
compatibility condition 
for a pair of dual Lie algebroids 
in the Vaisman algebroid.
The condition makes the Vaisman algebroid be a Courant algebroid through
the Drinfel'd double of a Lie bialgebroid.

The organization of this paper is as follows.
In the next section, we introduce the notions of the Drinfel'd double
and the Manin triple for Lie bialgebras.
We also provide definitions of Lie algebroids, Lie bialgebroids, Courant
algebroids and their doubled structures.
Dirac structures as subbundles of Courant algebroids are also discussed.
In Section \ref{sect:Vaisman}, we then introduce the metric algebroid proposed by Vaisman
(the Vaisman algebroid) which is nothing but the algebraic structure
defined by the {\sf C}-bracket in DFT.
We examine an analogue of the Drinfel'd double for the
Vaisman algebroid. We show that there is a natural pair of Lie algebroids
$(E, E^*)$ behind Vaisman algebroids.
They are much like the Lie bialgebroid but not by itself. 
We show that even though the derivation
condition of Lie bialgebroids is not satisfied, a remnant of the
Drinfel'd double survives in the Vaisman algebroid. 
We also study an analogue of the Manin triple for the Vaisman algebroid.
In Section \ref{sect:DFT}, in the DFT language, we demonstrate that there is a
doubled Lie algebroid structure of the gauge symmetry.
After introducing a geometric framework based on the para-Hermitian manifold and the
Dirac structures $(L, \tilde{L})$, we construct an exterior algebra in DFT.
We show that a para-Dolbeault cohomology is naturally
defined on the doubled space-time and they make the structure of the
double of Lie algebroids be more apparent. 
We will explicitly show that the double $L \oplus \tilde{L}$ defines a
Vaisman algebroid governed by the {\sf C}-bracket.
We also discuss physical aspects of the gauge symmetry based on these doubled structures.
Section \ref{sect:conclusion} is devoted to conclusion and discussions where the
``integration'' of the infinitesimal gauge symmetry is briefly discussed.
A quick introduction to DFT is found in Appendix \ref{sect:DFT_review}.
Relevant calculus on the {\sf C}-bracket are found in Appendix
\ref{sect:C-bracket}. Detailed calculations of useful formulae are found
in Appendix \ref{sect:calculations}.

The notations used in this paper 
are summarized as follows.

\begin{center}
\begin{tabular}{ll}
$M$ 
	& manifold \\
${\mathcal M}$
	& doubled space-time (or para-Hermitian manifold) \\
$\mathfrak{g}$
	& Lie algebra \\
$E$ ($L$, $\tilde{L}$)
	& Lie algebroid (Lagrangian subspace; Dirac structure) \\
$E^*$
	& dual space of $E$ \\
$\rho_\bullet: \bullet \to TM$
	& anchor map \\
${\mathcal C}$ 
	& Courant algebroid \\
${\mathcal V}$
	& Vaisman algebroid \\
$[\cdot, \cdot]$
	& Lie bracket on $TM$ \\
$[\cdot, \cdot]_{E,L,\tilde{L}}$
	& Lie bracket on $E$, $L$, $\tilde{L}$ \\
$[\cdot, \cdot]_{\rm S}$
	& Schouten(-Nijenhuis) bracket \\
$[\cdot, \cdot]_{\rm c}$
	& Courant bracket on ${\mathcal C}$ \\
$[\cdot, \cdot]_{\rm V}$
	& Vaisman bracket on ${\mathcal V}$ \\
$[\cdot, \cdot]_{\sf C}$
	& {\sf C}-bracket \\
$\langle \cdot, \cdot \rangle$
	& inner product \\
$\bbra{ \cdot, \cdot }$
	& bilinear form \\
\end{tabular}
\end{center}

\section{Drinfel'd double, Manin triple and Courant algebroid} \label{sect:DD}
In this section, we give a brief review on the Drinfel'd double of Lie
bialgebras, the Manin triple and related topics.
We then generalize these notions to the cases for Lie algebroids, 
bialgebroids and their double. We then discuss the relation of these
notions and Courant algebroids. 
The material presented in this section is nicely summarized, for
example, in \cite{Kosmann-Schwarzbach3, Marle} and references therein.

\subsection{Drinfel'd double of Lie bialgebra and Manin triple}
We first introduce the notion of Lie bialgebras.
Let $(V,[\cdot, \cdot])$ be a Lie algebra over a field $K$ defined by a vector space
$V$ together with a skew-symmetric bilinear bracket (the Lie bracket) $[\cdot,\cdot] :
V \times V \to V$ satisfying the Jacobi identity. 
We denote a Lie algebra by $\mathfrak{g}$.
Since $V$ is a vector space, we can define the dual Lie algebra
$\mathfrak{g}^*$ based on the dual vector space $V^*$ equipped with the dual Lie bracket
$[\cdot,\cdot]_{*}$. Since they are dual with each other as a vector space, a natural bilinear inner product $\langle \cdot , \cdot \rangle$
between $\mathfrak{g}$ and $\mathfrak{g}^*$ taking value in $K$ is defined.
Once a Lie algebra $\mathfrak{g}$ is given, we can consider a vector space $M$ of a representation $\varrho$ of
$\mathfrak{g}$.
For $x \in \mathfrak{g}$ and $m \in M$, we say that $x$ acts on $M$ as
$\varrho (x) \cdot m$.
A Lie algebra element $x \in \mathfrak{g}$ acts on itself by the adjoint
representation $\mathrm{ad}: x \in \mathfrak{g} \mapsto \mathrm{ad}_x \in
\mathop{\mathrm{End}} \mathfrak{g}$ by $\mathrm{ad}_x (y) = [x,y]$ for $x,y \in
\mathfrak{g}$.
More generally, $x \in \mathfrak{g}$ acts on any tensor products
$\otimes^p \mathfrak{g} = \mathfrak{g} \otimes \mathfrak{g} \otimes
\cdots \otimes \mathfrak{g}$ as
\begin{align}
\varrho (x) \cdot (y_1 \otimes \cdots \otimes y_p) =& \
 \mathrm{ad}_x^{(p)} (y_1 \otimes \cdots \otimes y_p)
\notag \\
=& \ \mathrm{ad}_x (y_1) \otimes y_2 \cdots \otimes y_p
+ y_1 \otimes \mathrm{ad}_x (y_2) \otimes \cdots \otimes y_p 
+ \cdots
\notag \\
& \cdots  + y_1 \otimes \cdots \otimes \mathrm{ad}_x (y_p).
\end{align}
Therefore the adjoint action satisfies the Leibniz rule.
By the same way, $x \in \mathfrak{g}$ acts on 
a $p$-th exterior product of $\mathfrak{g}$, $\wedge^p \mathfrak{g} = \mathfrak{g} \wedge \mathfrak{g} \wedge \cdots \wedge
\mathfrak{g}$, which is naturally defined by the totally anti-symmetrized tensor products $\otimes^p \mathfrak{g}$
for any positive integers $p$. The action on it is defined, for example, as 
\begin{align}
\varrho (x) \cdot y_1 \wedge y_2 = [x,y_1] \wedge y_2 + y_1 \wedge [x,y_2].
\end{align}
As an analogue of an exterior derivative $\dop$ in the cotangent bundle
over a manifold, we can define an exterior derivative $\dop : \wedge^p
\mathfrak{g}^* \to \wedge^{p+1} \mathfrak{g}^*$ which satisfies $\dop^2 = 0$.
Similarly, an exterior derivative $\dop_*: \wedge^p \mathfrak{g} \to \wedge^{p+1}
\mathfrak{g}$, $\dop_*^{2} = 0$ is also defined in the dual side.
Using these, we can further define the Lie algebra cohomology on $\mathfrak{g}$ \cite{Chevalley:1948zz}.

It is worthwhile to discuss a generalization of the Lie bracket
to the one in $\wedge^p \mathfrak{g}$.
The skew-symmetric Schouten-Nijenhuis bracket $[\cdot,\cdot]_{\mathrm{S}} :
\wedge^p \mathfrak{g}\times\wedge^q \mathfrak{g} \to \wedge^{p+q-1}
\mathfrak{g}$ is defined by the following properties \cite{Schouten-Nijenhuis}:
\begin{enumerate}
\renewcommand{\labelenumi}{(\roman{enumi})}
\item $[a,b]_{\mathrm{S}} = - (-)^{(p-1)(q-1)} [b,a]_{\mathrm{S}}$.
\item $[a, b \wedge c]_{\mathrm{S}} = [a,b]_{\mathrm{S}} \wedge c +
      (-)^{(p-1)q} b \wedge [a,c]_{\mathrm{S}}$.
\item $(-)^{(p-1)(r-1)} [a,[b,c]_{\mathrm{S}}]_{\mathrm{S}} +
      (-)^{(q-1)(r-1)} [b,[c,a]_{\mathrm{S}}]_{\mathrm{S}} +
      (-)^{(r-1)(q-1)} [c,[a,b]_{\mathrm{S}}]_{\mathrm{S}} = 0$.
\item The bracket of an element $\wedge^p \mathfrak{g}$ and an element
      in $\wedge^0 \mathfrak{g} = K$ is 0.
\end{enumerate}
Here $a \in \wedge^p \mathfrak{g}$, $b \in \wedge^q \mathfrak{g}$ and $c
\in \wedge^r \mathfrak{g}$.
Indeed, the Schouten-Nijenhuis bracket is a unique generalization of the
Lie bracket that makes $\wedge^p \mathfrak{g}$ be a Gerstenhaber algebra.

We next examine an algebraic structure between the Lie algebra
$\mathfrak{g}$ and its dual $\mathfrak{g}^*$.
Let the Lie bracket $[\cdot,\cdot]$ be a bilinear map $\mu : \wedge^2 \mathfrak{g}
\to \mathfrak{g}$. We can then define a co-bracket $\delta :
\mathfrak{g} \to \wedge^2 \mathfrak{g}$ as an adjoint of the dual Lie
bracket $\mu_*: \wedge^2 \mathfrak{g}^* \to \mathfrak{g}^*$.
The adjoint of a map $\mu_*$, denoted as $\mu^*_*$, is defined through the inner product
$\langle \cdot, \cdot \rangle$ between $\wedge^{\bullet} \mathfrak{g}$ and
$\wedge^{\bullet} \mathfrak{g}^*$ 
by $\langle x, \mu_* (\xi) \rangle = \langle \mu^*_* (x), \xi
\rangle$ where $x \in \mathfrak{g}$ and $\xi \in \wedge^2 \mathfrak{g}^*$.
Here, $\wedge^{\bullet}$ stands for any powers of
the wedge products. 
The co-bracket satisfies the co-Jacobi identity. This is equivalent to
say that $\delta_*$ defines the Lie bracket $[\cdot,\cdot]_{*}$ on
$\mathfrak{g}^*$.
Even more, if $\delta$ satisfies the 1-cocycle condition:
\begin{align}
\delta ([x,y]) = \mathrm{ad}_x^{(2)} \delta (y) - \mathrm{ad}_y^{(2)}
 \delta (x), \quad x,y \in \mathfrak{g},
\label{eq:1-cocycle}
\end{align}
then, the structure $(\mathfrak{g}, \mu, \delta)$ is called the 
{\it Lie bialgebra}. Since if $(\mathfrak{g}, \mu,
\delta)$ is a Lie bialgebra,
then $(\mathfrak{g}^*, \delta_*, \mu_*)$ is so too, we use the
notation $(\mathfrak{g}, \mathfrak{g}^*)$ for the Lie bialgebra defined by $\mathfrak{g}$.
Given a Lie bialgebra $(\mathfrak{g}, \mathfrak{g}^{*})$, we can define a
non-degenerate symmetric bilinear form $\bbra{\cdot,\cdot}$ on $\mathfrak{d} = \mathfrak{g} \oplus
\mathfrak{g}^*$ by 
\begin{align}
 \bbra{x,y} = \bbra{\xi, \eta} = 0, \qquad \bbra{x,\xi} = \langle \xi, x \rangle, 
\qquad x,y \in \mathfrak{g}, \ \xi, \eta \in \mathfrak{g}^*.
\end{align}
We then require that there is a skew-symmetric bracket
$[\cdot,\cdot]_{\mathfrak{d}}$ on $\mathfrak{d} = \mathfrak{g} \oplus
\mathfrak{g}^*$ under which the bilinear form is invariant and 
$\mathfrak{g}, \mathfrak{g}^*$ are subalgebras of $\mathfrak{d}$.
A natural definition of such kind of bracket is 
\begin{align}
[x,y]_{\mathfrak{d}} = [x,y], \quad 
[\xi, \eta]_{\mathfrak{d}} = [\xi,\eta]_*, 
\qquad 
x,y \in \mathfrak{g}, \ \xi,\eta \in \mathfrak{g}^*.
\label{eq:LB1}
\end{align}
For the mixing term $[x,\xi]_{\mathfrak{d}}$, the invariant condition implies that 
\begin{align}
\bbra{y, [x, \xi]_{\mathfrak{d}}} = \bbra{[y,x]_{\mathfrak{d}}, \xi} 
= \bbra{[y,x], \xi} = \langle \xi, [y,x] \rangle = \langle \xi, -
 \mathrm{ad}_x (y) \rangle 
= \langle \mathrm{ad}^{*}_x \xi, y \rangle
= \bbra{y, \mathrm{ad}_x^{*} \xi}.
\end{align}
The first equality follows from the definition of the invariance.
The second comes from the fact that $\mathfrak{g}$ is a subalgebra of
$\mathfrak{d}$.
Here we have defined the co-adjoint $\mathrm{ad}_x^* = - (\mathrm{ad}_x)^*$ for $ x \in \mathfrak{g}$.
Similarly, we have $\bbra{\eta, [x, \xi]_{\mathfrak{d}}} = - \bbra{\eta,
\mathrm{ad}^{*}_{\xi} x}$. These facts result in the definition:
\begin{align}
[x, \xi]_{\mathfrak{d}} = - \mathrm{ad}^{*}_{\xi} x + \mathrm{ad}^*_x
 \xi.
\label{eq:LB2}
\end{align}
One can confirm that the bracket $[\cdot,\cdot]_{\mathfrak{d}}$ defined by \eqref{eq:LB1} and
\eqref{eq:LB2} satisfies the Jacobi identity due to the properties of
the Lie bialgebra $(\mathfrak{g}, \mathfrak{g}^*)$.
Therefore $(\mathfrak{d}, [\cdot,\cdot]_{\mathfrak{d}})$ defines a new Lie algebra.
This is called the {\it Drinfel'd double} of a Lie bialgebra $(\mathfrak{g},
\mathfrak{g}^*)$ \cite{Drinfeld:1983ky}.
Since $\mathfrak{g}, \mathfrak{g}^*$ are subalgebras of $\mathfrak{d} =
\mathfrak{g} \oplus \mathfrak{g}^*$ and the scalar product $\bbra{\cdot,\cdot}$
vanishes on $\mathfrak{g}, \mathfrak{g}^*$, they are complementally
isotropic with respect to $\bbra{\cdot,\cdot}$.
Let $\mathfrak{p}$ be a Lie algebra with an invariant, non-degenerate,
symmetric bilinear form $\bbra{\cdot,\cdot}$, and $\mathfrak{a}, \mathfrak{b}$ are
complementally isotropic Lie subalgebras of $\mathfrak{p}$, then the
structure $(\mathfrak{p}, \mathfrak{a}, \mathfrak{b})$ is called {\it the Manin triple} \cite{Drinfeld2}.
It is shown that if $(\mathfrak{p}, \mathfrak{a}, \mathfrak{b})$ is a
finite dimensional Manin triple, then one can make $(\mathfrak{a},
\mathfrak{b})$ be a Lie bialgebra by defining the co-bracket on
$\mathfrak{a}$ by the dual of the Lie bracket on $\mathfrak{b}$.

\subsection{Double of Lie bialgebroid and Courant algebroid}
We then generalize the above discussion to those for Lie algebroids.
A Lie algebroid is a generalization of a Lie algebra defined over a base
manifold $M$ \cite{Pradines}.
Let $E$ be a vector bundle over $M$, $E
\xrightarrow{\pi} M$. A Lie bracket $[\cdot,\cdot]_E : \Gamma (E) \times \Gamma
(E) \to \Gamma (E)$ as a skew-symmetric bilinear form on the section
$\Gamma (E)$ is defined. We demand that the Lie bracket satisfies the Jacobi identity.
A bundle map called the anchor $\rho : E \to TM$ is defined such that it satisfies 
$\rho ([X,Y]_E) = [\rho (X), \rho(Y)]$ where 
$[\cdot,\cdot]$ is a Lie
bracket on $TM$. Here $X,Y \in \Gamma (E)$. 
For a function $f \in C^{\infty} (M)$, 
the Lie bracket $[\cdot,\cdot]_E$ satisfies the condition $[X,f Y]_E = (\rho (X)
\cdot f) Y + f [X,Y]_E$. Here $\rho (X) \cdot f$ represents that $\rho
(X)$ acts on $f$ as a differential operator.
Then $(E, [\cdot,\cdot]_E, \rho)$ defines a Lie algebroid over $M$.
Given a Lie algebroid, we can define the dual Lie algebroid $(E^*,
[\cdot,\cdot]_{E^*}, \rho_*)$ on the same base manifold. 
Again, there is a natural inner product $\langle \cdot,
\cdot \rangle$ between $E$ and $E^*$.
As a generalization of ordinary calculus for (multi)vectors and
forms in $\Gamma(TM)$ and $\Gamma (T^*M)$, we define exterior algebras in $\Gamma
(\wedge^\bullet E)$ and $\Gamma (\wedge^\bullet E^*)$.
A natural inner product $\langle \xi, X \rangle$ between $\wedge^p E$
and $\wedge^p E^*$ is defined.
We then define a Lie algebroid differential as a map $\dop : \Omega_p (E)
= \Gamma (\wedge^p E^*) \to \Omega_{p+1} (E)$ where $\Omega_p (E)$ is a
generalization of $p$-form on $T^*M$. More explicitly, the exterior
derivative $\dop$ is defined through the action of $\xi \in \Gamma
(\wedge^p E^*)$ on vectors $X_i \in \Gamma (E)$ \cite{Mackenzie}:
\begin{align}
\dop \xi (X_1, \ldots, X_{p+1}) =& \ \sum^{p+1}_{i=1} (-)^{i+1} \rho (X_i) \cdot
\left(
\xi (X_1, \ldots, \check{X_i}, \ldots, X_{p+1})
\right)
\notag \\
&+ \ \sum_{i<j} (-)^{i+j} \xi ([X_i,X_j]_E, X_1, \ldots, \check{X_i},
 \ldots, \check{X_j}, \ldots, X_{p+1}),
\end{align}
where the notation $\check{X_i}$ stands for that the term is omitted in
the expression.
We sometimes use the notation such as $\xi (X) = \langle \xi, X \rangle$ for the natural scalar
product between $\xi \in \Omega_p (E)$ and $X \in \Omega_p (E^*)$.

The exterior derivative, in particular, satisfies the following properties:
\begin{align}
& \dop (\xi \wedge \eta) = \dop \xi \wedge \eta + (-)^{|\xi|}
 \xi \wedge \dop \eta,
\notag 
\\
& \dop f (X) = \rho (X) \cdot f, 
\notag 
\\
& \dop \xi (X,Y) = \rho (X) \cdot (\xi (Y)) - \rho (Y) \cdot (\xi (X)) -
 \xi ([X,Y]_E),
\end{align}
where $X,Y \in \Gamma (E)$, $\xi, \eta \in \Gamma (E^*)$.
Similarly, a Lie derivative $\mathcal{L}_X : \Gamma (\wedge^p E^*) \to
\Gamma (\wedge^p E^*)$ by $X \in \Gamma (E)$ is defined by
\begin{align}
\mathcal{L}_X (\xi) (Y_1, \ldots, Y_p) = \rho (X) \cdot ( \xi (Y_1,
 \ldots, Y_p) ) - \sum_{i=1}^p \xi \left( Y_1, \ldots, [X,Y_i]_E,
 \ldots, Y_p \right),
\label{eq:algebroid_Lie_derivative}
\end{align}
where $Y_1, \ldots, Y_p \in \Gamma (E)$, $\xi \in \Gamma (\wedge^p E^*)$.
The interior product $\iop_X: \Gamma (\wedge^{p} E^*) \to \Gamma (\wedge^{p-1}
E^*)$ by $X \in \Gamma (E)$ is defined by
\begin{align}
(\iop_X \xi) (Y_1, \ldots, Y_{p-1}) = \xi (X, Y_1, \ldots, Y_{p-1}),
\end{align}
where $Y_1, \ldots, Y_{p-1} \in \Gamma (E), \xi \in \Gamma (\wedge^p E^*)$.
They satisfy the following relations:
\begin{align}
& \mathcal{L}_{[X,Y]_E} = \mathcal{L}_X \cdot \mathcal{L}_Y -
 \mathcal{L}_Y \cdot \mathcal{L}_{X},
\notag 
\\
& \iop_{[X,Y]_E} = \mathcal{L}_X \cdot \iop_Y - \iop_Y \cdot \mathcal{L}_X,
\notag 
\\
& \mathcal{L}_X = \dop \cdot \iop_X + \iop_X \cdot \dop,
\notag 
\\
& \mathcal{L}_{fX} (\xi) = f \mathcal{L}_X (\xi) + \dop f \wedge \iop_X (\xi),
\notag 
\\
& \mathcal{L}_X \langle \xi, Y \rangle = \langle \mathcal{L}_X \xi, Y
 \rangle + \langle \xi, \mathcal{L}_X Y \rangle,
\label{eq:Lie_derivative_relations}
\end{align}
where $X,Y \in \Gamma (E)$, $f \in C^{\infty} (M)$, 
$\xi \in \Gamma (\wedge^{\bullet} E^*)$.

As we have discussed in the previous subsection, the Lie bracket $[\cdot,\cdot]_E$
can be generalized to those for 
multi-vectors $\Gamma (\wedge^p
E)$. For $X \in \Gamma (\wedge^{p+1} E), Y \in \Gamma (\wedge^{q+1} E)$
and $f \in C^{\infty} (M)$, 
the Schouten-Nijenhuis bracket satisfies the following properties:
\begin{enumerate}
\renewcommand{\labelenumi}{(\roman{enumi})}
\item $[X,Y]_{\mathrm{S}} = - (-)^{pq} [Y,X]_{\mathrm{S}}$.
\item $[X,f]_{\mathrm{S}} = \rho (X) \cdot f$ \ for $X \in \Gamma (E)$.
\item For $X \in \Gamma (\wedge^{p+1} E)$, the bracket $[X, \cdot
      ]_{\mathrm{S}}$ acts on $\Gamma (\wedge^{q} E)$ as a degree-$p$
      derivation.
\end{enumerate}
We also define an exterior derivative $\dop_*$, the interior product
and the Lie derivative on $\Gamma (\wedge^{\bullet} E)$.
We note that when the base manifold $M$ consists of a point, then $\Gamma (E)$
represents a globally defined vector. In this case, $(E, [\cdot,\cdot]_E, \rho = 0)$ becomes a Lie
algebra. We also note that $E = TM$, $\rho = \mathrm{id}$, $[X,Y]_E =
\mathcal{L}_X Y$ defines a Lie algebroid with a trivial structure.

Once a Lie algebroid is defined, we can define a {\it Lie bialgebroid}.
This is a generalization of the Lie bialgebra discussed in the previous
subsection.
Let $(E, [\cdot,\cdot]_E, \rho)$ be a Lie algebroid $E \xrightarrow{\pi} M$
and $(E^*, [\cdot,\cdot]_{E^*}, \rho_*)$ be its dual.
For $X,Y \in \Gamma (\wedge^{\bullet} E)$ and $\dop_*: \Gamma
(\wedge^{\bullet} E) \to \Gamma (\wedge^{\bullet + 1}
E)$, if the following compatibility condition
\begin{align}
\dop_{*} [X,Y]_{\rm S} = [\dop_* X, Y]_{\rm S} + [X,\dop_* Y]_{\rm S}
\label{eq:derivation}
\end{align}
is satisfied, then $(E,E^*)$ is called a Lie bialgebroid over $M$.
This implies that $\dop_*$ acts on the Schouten-Nijenhuis bracket of
$\Gamma (\wedge^\bullet E)$ as a derivation.
Therefore we call \eqref{eq:derivation} the derivation condition.
The notion of a Lie bialgebroid was first introduced in \cite{Mackenzie}.
If $M$ is a point and $\rho$ is trivial, then $(E,E^*)$ becomes a Lie
bialgebra and the condition \eqref{eq:derivation} becomes the 1-cocycle
condition \eqref{eq:1-cocycle}.

Now we consider the Drinfel'd double of a Lie bialgebroid $(E,E^*)$.
We may expect, from the discussion on the Lie bialgebra,
that a double $E \oplus E^*$ possesses a Lie algebroid structure.
However, the result is not the case.
Before discussing this issue, we introduce the notion of {\it Courant
algebroids} \cite{Courant, LiWeXu}.
Let $\mathcal{C} \xrightarrow{\pi} M$ be a vector bundle over $M$.
We introduce a non-degenerate, symmetric bilinear form $\bbra{\cdot,\cdot}$ on the bundle.
We also introduce a skew-symmetric bracket $[\cdot,\cdot]_{\rm c}$ on $\Gamma (\mathcal{C})$ and an anchor map $\rho_{\rm c} : \mathcal{C} \to TM$.
An isomorphism $\beta : E \to E^*$ and a map
$\mathcal{D}: C^{\infty} (M) \to \Gamma (\mathcal{C})$ 
defined by $\mathcal{D} =
\frac{1}{2} \beta^{-1} {} \rho_{\rm c}^* \dop_0$, in which $\dop_0$ is a natural
exterior derivative on $T^*M$, such that $\bbra{\mathcal{D}f, e} =
\frac{1}{2} \rho_{\rm c} (e) \cdot f$ for $f \in C^{\infty} (M)$, $e \in \Gamma (\mathcal{C})$,
are also introduced.
A bracket $[\cdot,\cdot]_{\rm c}$ that satisfies the following axioms \ref{axiom:C1}-\ref{axiom:C5}
is called the Courant bracket~\cite{LiWeXu}:
\begin{axiomC}
\label{axiom:C1}
For any $e_1, e_2, e_3 \in \Gamma (\mathcal{C})$, the Jacobiator of
      $[\cdot,\cdot]_{\rm c}$ is given by 
\begin{align}
[[e_1, e_2]_{\rm c}, e_3]_{\rm c} + \text{c.p.} = \mathcal{D} T (e_1, e_2, e_3),
\label{eq:C1}
\end{align}
 where $T (e_1, e_2, e_3) = \frac{1}{3} \bbra{ [e_1, e_2]_{\rm c}, e_3 } +
       \text{c.p.}$ and $\text{c.p.}$ is terms obtained by the
 cyclic permutations.
\end{axiomC}
\begin{axiomC}
\label{axiom:C2}
For any $e_1, e_2 \in \Gamma (\mathcal{C})$, 
\begin{align}
\rho_{\rm c} ([e_1, e_2]_{\rm c}) =  [\rho_{\rm c} (e_1), \rho_{\rm c} (e_2)],
\label{eq:C2}
\end{align}
where $[\cdot,\cdot]$ is the Lie bracket on $TM$.
\end{axiomC}
\begin{axiomC}
\label{axiom:C3}
For any $e_1, e_2 \in \Gamma (\mathcal{C})$, $f \in C^{\infty} (M)$,
\begin{align}
[e_1, f e_2]_{\rm c} = f [e_1, e_2]_{\rm c} + (\rho_{\rm c} (e_1) \cdot f) e_2 - \bbra{e_1, e_2}
 \mathcal{D}f.
\label{eq:C3}
\end{align}
\end{axiomC}
\begin{axiomC}
\label{axiom:C4}
$\rho_{\rm c} \cdot \mathcal{D} = 0$, namely, for any $ f,g \in
      C^{\infty} (M)$, we have 
\begin{align}
\bbra{\mathcal{D}f, \mathcal{D}g} = 0.
\label{eq:C4}
\end{align}
\end{axiomC}
\begin{axiomC}
\label{axiom:C5}
For any $e_1, e_2, e_3 \in \Gamma (\mathcal{C})$, we have the compatibility
 between the bilinear form $\bbra{\cdot,\cdot}$ and the anchor $\rho_{\rm c}$:
\begin{align}
\rho_{\rm c} (e_1) \cdot \bbra{e_2,e_3} = \bbra{[e_1, e_2]_{\rm c} + \mathcal{D} \bbra{e_1,e_2}, e_3}
+\bbra{e_2, [e_1, e_3]_{\rm c} + \mathcal{D} \bbra{e_1,e_3}}.
\label{eq:C5}
\end{align}
\end{axiomC}
Then $(\mathcal{C}, [\cdot,\cdot]_{\rm c}, \rho_{\rm c}, \bbra{\cdot,\cdot})$ defines a Courant algebroid.
We note that axioms \ref{axiom:C1}-\ref{axiom:C5} are not independent.
Indeed, it is shown that the axioms \ref{axiom:C3} and \ref{axiom:C4} are followed 
from \ref{axiom:C5} and \ref{axiom:C2} respectively
~\cite{2002math......4010U}.
There is a definition of Courant algebroids based on a non-skew-symmetric bracket for which the Jacobi identity holds. 
In the following, we employ the definition based on a skew-symmetric bracket.

Now we discuss the double of a Lie bialgebroid $(E,E^*)$.
This notion was first introduced by Liu, Weinstein and Xu in
\cite{LiWeXu}. Given a Lie bialgebroid $(E,E^*)$, they considered the following doubled structure on $\mathcal{C} = E
\oplus E^*$:
\begin{enumerate}
\renewcommand{\labelenumi}{(\Roman{enumi})}
 \item For $X_1, X_2 \in \Gamma (E), \xi_1, \xi_2 \in \Gamma (E^*)$, a non-degenerate, bilinear forms $\pmbra{\cdot,\cdot}$ are defined by
\begin{align}
\pmbra{X_1 + \xi_1, X_2 + \xi_2} = \frac{1}{2} \left\{ \frac{}{} \langle \xi_1,
       X_2 \rangle \pm \langle \xi_2, X_1 \rangle \right\},
\end{align}
where $\langle \cdot, \cdot \rangle$ is a natural inner product between $E$ and $E^*$
\item A skew-symmetric bracket $[\cdot,\cdot]_{\rm c}$ on $\Gamma (\mathcal{C})$ is defined by 
\begin{align}
[e_1, e_2]_{\rm c} =& \  [X_1, X_2]_E + \mathcal{L}_{\xi_1} X_2 -
 \mathcal{L}_{\xi_2} X_1 - \dop_* \mbra{e_1,e_2}
\notag \\
&+  [\xi_1, \xi_2]_{E^*} + \mathcal{L}_{X_1} \xi_2 -
 \mathcal{L}_{X_2} \xi_1 + \dop \mbra{e_1, e_2},
\label{eq:Courant_bracket}
\end{align}
where $e_i = X_i + \xi_i \in \Gamma (\mathcal{C}), \ (i=1,2)$. 
\item An anchor $\rho_{\rm c} : \mathcal{C} \to TM$ is defined by $\rho_{\rm c} = \rho +
      \rho_*$. Namely, $\rho_{\rm c} (X+\xi) = \rho (X) + \rho_*
      (\xi)$ for ${}^\forall X \in \Gamma (E)$, ${}^\forall \xi \in
      \Gamma (E^*)$. 
\item An exterior derivative $\mathcal{D} = \dop + \dop_*$ on
      $\mathcal{C} = E \oplus E^*$ is defined.
\end{enumerate}
Given these structures, the authors in \cite{LiWeXu} showed that 
$(\mathcal{C} = E \oplus E^*, [\cdot,\cdot]_{\rm c}, \rho_{\rm c}, \pbra{\cdot,\cdot})$ becomes a Courant
algebroid satisfying the axioms \ref{axiom:C1}-\ref{axiom:C5}.
We 
stress that the Jacobiator of the Courant bracket $[\cdot,\cdot]_{\rm c}$ does not
vanish in general. Therefore a Courant algebroid, obtained by the double
of a Lie bialgebroid, is not a Lie algebroid. This is in contrast to the
double of a Lie bialgebra.

The authors in \cite{LiWeXu} also showed that if there are
complementally isotropic subbundles $E,E^*$ with respect to the bilinear
product $\bbra{\cdot,\cdot}$ in a Courant algebroid $\mathcal{C}$, and if they are closed under the
Courant bracket $[\cdot,\cdot]_{\rm c}$, then there is a natural Lie bialgebroid
structure on $(E,E^*)$. When $E,E^*$ are maximally isotropic, 
{\it i.e.}\ $\dim E = \dim E^* = \frac{1}{2} \dim \mathcal{C}$, then 
$E,E^*$ are called Dirac structures and they provide therefore a natural
generalization, a Lie algebroid analogue, of the Manin triple
$(\mathcal{C},E,E^*)$.
Indeed, when $M$ consists of a point, a Courant algebroid becomes a quadratic Lie
algebra, namely, a Lie algebra with the non-degenerate bilinear form
$\bbra{\cdot,\cdot}$. 
This is just the double of a Lie bialgebra.

The Courant bracket naturally appears in the context of generalized
geometry \cite{Hitchin:2004ut} where the generalized tangent bundle
$\mathbb{T}M = TM \oplus T^*M$ is prepared in order to realize manifest T-duality.
We note that the original Courant bracket on $TM \oplus T^*M$ introduced
by T.~Courant is defined by 
\begin{align}
[X_1 + \xi_1, X_2 + \xi_2]_{\rm c} = [X_1, X_2] + (\mathcal{L}_{X_1} \xi_2 -
\mathcal{L}_{X_2} \xi_1) + \frac{1}{2} \dop_0 (\xi_1 (X_2) - \xi_2 (X_1)),
\label{eq:c-bracket}
\end{align}
for $X_i \in \Gamma (TM)$, $\xi_i \in \Gamma (T^*M)$.
We call \eqref{eq:c-bracket} the c-bracket.
It of course satisfies the axioms 
\ref{axiom:C1}-\ref{axiom:C5}.
We will comment on the relations of DFT and generalized geometry in Section \ref{sect:DFT}.

A few comments are in order.
First, it is not always true that a Courant algebroid is defined by Lie bialgebroids~\cite{1999math.....10078R}.
Second, we consider a class of Courant algebroids called exact~\cite{exactCourant} in the following.

\section{Doubled aspects of Vaisman algebroid} \label{sect:Vaisman}
In this section, we study doubled aspects of Vaisman algebroids.
It has been discussed that the gauge symmetry algebra of DFT, which is
governed by the {\sf C}-bracket, is characterized by an algebroid proposed by
Vaisman \cite{Vaisman:2012ke,Vaisman:2012px}.
We call this the Vaisman algebroid. 
In the following, we introduce the notion of the Vaisman algebroid and
discuss its doubled structures.

\subsection{Vaisman and Courant algebroids}
The definition of the Vaisman algebroid is the following. 
Let $\mathcal{V} \xrightarrow{\pi} M$ be a vector bundle over a manifold $M$. 
We introduce a non-degenerate symmetric bilinear form $\bbra{\cdot,\cdot} :
\Gamma (\mathcal{V}) \times \Gamma (\mathcal{V}) \to M \times \mathbb{C}$ and an anchor
$\rho_{\rm V} : \mathcal{V} \to TM$. 
A map 
$\mathcal{D}: C^{\infty} (M) \to \Gamma (\mathcal{V})$ is defined
through $\bbra{\mathcal{D} f, e} = \frac{1}{2} \rho_{\rm V} (e) \cdot f$ for $e \in \Gamma
(\mathcal{V})$ and $f \in C^{\infty} (M)$.
When a skew-symmetric bracket $[\cdot,\cdot]_{\mathrm{V}}$ on $\Gamma (\mathcal{V})$ 
satisfies the following axioms \ref{axiom:V1}-\ref{axiom:V2}, then a Vaisman algebroid is defined by
$(\mathcal{V}, [\cdot,\cdot]_{\mathrm{V}}, \rho_{\rm V}, \bbra{\cdot,\cdot})$:
\begin{axiomV}
\label{axiom:V1}
$[e_1, f e_2]_{\mathrm{V}} = f [e_1, e_2]_{\mathrm{V}} + (\rho_{\rm V} (e_1) \cdot f) e_2 - \bbra{e_1, e_2}
 \mathcal{D}f$.
\end{axiomV}
\begin{axiomV}
\label{axiom:V2}
$\rho_{\rm V} (e_1) \cdot \bbra{e_2,e_3} = \bbra{[e_1, e_2]_{\mathrm{V}} + \mathcal{D}
      \bbra{e_1,e_2},e_3} + \bbra{e_2, [e_1, e_3]_{\mathrm{V}} + \mathcal{D} \bbra{e_1,e_3}}$.
\end{axiomV}
Here $e_1, e_2 \in \Gamma (\mathcal{V})$, $f \in C^{\infty} (M)$. 
We employ the definition in \cite{Vaisman:2013} where the bracket is
skew-symmetric. An alternative (but equivalent) definition based on a
non-skew-symmetric bracket is found in \cite{Vaisman:2012ke, Vaisman:2012px}. 
The above conditions are nothing but the axioms \ref{axiom:C3} and \ref{axiom:C5} for the Courant
algebroid.
If a Vaisman algebroid further satisfies the axioms \ref{axiom:C1},
\ref{axiom:C2}, and \ref{axiom:C4} in the previous section, it becomes a Courant algebroid.
Properties of Vaisman algebroids are discussed in detail in \cite{Vaisman:2012ke, Vaisman:2012px}. 
As its definition stands for, the Vaisman algebroid is a weakened cousin
of the Courant algebroid. It follows therefore that any Courant
algebroids are Vaisman algebroids. 
The author of \cite{Vaisman:2012ke, Vaisman:2012px} showed that on a
flat para-Hermitian manifold $M$, when there are two Lagrangian foliations 
$L = T \mathcal{F},\tilde{L} = T \tilde{\mathcal{F}}$, where $TM = L \oplus \tilde{L}$, then there is a natural
structure of the Vaisman algebroid on $TM$ whose anchor is trivial $\rho
= \mathrm{id}$.
It was shown that the bracket $[\cdot,\cdot]_{\mathrm{V}}$ defined on $TM$ is nothing but the
{\sf C}-bracket discussed in the physics literature. 
Indeed, as discussed in \cite{Vaisman:2012ke, Vaisman:2012px, Freidel:2017yuv, Freidel:2018tkj},
a natural geometry underlying the doubled space-time in DFT is a para-Hermitian manifold.
One notices that the {\sf C}-bracket \eqref{eq:C-bracket} appearing in the DFT language is quite similar to
the Courant bracket \eqref{eq:Courant_bracket} introduced in Section \ref{sect:DD}. 
However, it was discussed that the {\sf C}-bracket lacks appropriate properties of
Courant brackets \cite{Vaisman:2012ke, Vaisman:2012px, Chatzistavrakidis:2018ztm, Svoboda:2018rci}
and it does not define a Courant algebroid in general.
In the following, we make it be more apparent with showing that this nature follows from
the failure of the derivation condition of Lie bialgebroids. 

\subsection{Drinfel'd double behind Vaisman algebroid}
Following the prescription given by Liu-Weinstein-Xu \cite{LiWeXu} on 
Courant algebroids, we examine doubled aspects of Vaisman algebroids.
Given a Lie algebroid $(E,[\cdot,\cdot]_E, \rho)$ and its dual $(E^*,
[\cdot,\cdot]_{E^*}, \rho_*)$ over a manifold $M$, we consider a vector bundle
$\mathcal{V} = E \oplus E^*$. 
Note that we never assume the Lie bialgebroid structures on $(E,E^*)$.
We then define non-degenerate bilinear forms
$\pmbra{\cdot,\cdot}$ on $\mathcal{V}$ as 
\begin{equation}
\pmbra{e_1,e_2} = \frac{1}{2} 
\Bigl( 
\langle \xi_1,X_2 \rangle \pm \langle  \xi_2,X_1 \rangle 
\Bigr),
\end{equation}
where $e_i = X_i + \xi_i \in \Gamma (\mathcal{V}) \ (i=1,2)$, $X_i \in \Gamma
(E), \xi_i \in \Gamma (E^*)$ and $\langle \cdot, \cdot \rangle$ is the inner product between $E$ and $E^*$.
We next define a skew-symmetric bracket $[\cdot, \cdot]_{\rm V}$ in $\Gamma (\mathcal{V})$ as
\begin{align}
[e_1, e_2]_{\mathrm{V}} =& \ [X_1, X_2]_E + \mathcal{L}_{\xi_1} X_2 -
 \mathcal{L}_{\xi_2} X_1 - \dop_{*} \mbra{e_1,e_2}
\notag \\
&+   [\xi_1, \xi_2]_{E^*} + \mathcal{L}_{X_1} \xi_2 - \mathcal{L}_{X_2}
 \xi_1 + \dop \mbra{e_1,e_2},
\label{eq:Vaisman_bracket}
\end{align}
where $\mathcal{L}_X, \mathcal{L}_{\xi}$, $\dop_*, \dop$ are natural Lie
derivatives and de Rham differentials defined on $\Gamma(E),\Gamma (E^*)$.
We employ the morphism $\rho_{\rm V} = \rho + \rho_*$ as the anchor in
$\mathcal{V}$. The map $\mathcal{D}: C^\infty(M) \rightarrow \Gamma(\mathcal{V})$ is defined by
$\bbra{\mathcal{D} f,e} = \frac{1}{2}\rho_{\rm V} (e) \cdot f$ which is expressed as $\mathcal{D} = \dop + \dop_*$.
The expression \eqref{eq:Vaisman_bracket} is nothing but the one
defined in \eqref{eq:Courant_bracket} but we here again stress that we never assume that
$(E,E^*)$ is a Lie bialgebroid, {\it i.e.}\ the derivation condition
\eqref{eq:derivation} is not satisfied in general.
In the following, we show that $(E \oplus E^*, \rho + \rho_*, [\cdot,\cdot]_{\mathrm{V}},
\pbra{\cdot,\cdot})$ introduced above indeed defines a Vaisman algebroid, but not a Courant algebroid.

We now examine to what extent the axioms \ref{axiom:C1}-\ref{axiom:C5} for the Courant algebroid
are lacked due to the failure of the derivation condition.
Before the discussion, let us define the following quantity:
\begin{equation}
  T(e_1,e_2,e_3) = \frac{1}{3}
\Bigl(
\pbra{[e_1,e_2]_{\rm V},e_3} + \text{c.p.}
\Bigr).
\end{equation}
Here $\text{c.p.}$ represents cyclic permutations. 
For later convenience, we first rewrite the above quantity.
By the definition of the bracket $[\cdot,\cdot]_{\mathrm{V}}$ in
\eqref{eq:Vaisman_bracket} and the bilinear form $\pbra{\cdot,\cdot}$, we have 
\begin{align}
\pbra{[e_1,e_2]_{\rm V},e_3}
  &= \frac{1}{2} 
\left\{ \frac{}{}
\langle \xi_3, [X_1,X_2]_{\rm V} \rangle 
+ 
\langle \xi_3,\mathcal{L}_{\xi_1}X_2 \rangle 
- 
\langle \xi_3,\mathcal{L}_{\xi_2}X_1 \rangle 
- \rho_* (\xi_3) \cdot \mbra{e_1,e_2}
\right.
\notag \\
 &\qquad 
\left. \frac{}{}
+
\langle [\xi_1,\xi_2]_{\rm V},X_3 \rangle 
+ 
\langle 
\mathcal{L}_{X_1}\xi_2,X_3
\rangle
 - 
\langle \mathcal{L}_{X_2}\xi_1,X_3
\rangle
 + \rho (X_3) \cdot \mbra{e_1,e_2}
\right\}
.
 \label{T_1}
\end{align}
By defining properties of Lie derivatives in \eqref{eq:Lie_derivative_relations}
we have
\begin{align}
 \langle \xi_3,\mathcal{L}_{\xi_1}X_2 \rangle &= \mathcal{L}_{\xi_1}
 \langle \xi_3,X_2 \rangle - \langle [\xi_1,\xi_3]_{\rm V},X_2 \rangle
,  \notag\\
 \langle \mathcal{L}_{X_1}\xi_2,X_3 \rangle &= \mathcal{L}_{X_1} \langle
 \xi_2,X_3 \rangle - \langle \xi_2,[X_1,X_3]_{\rm V} \rangle.
\end{align}
The first terms in each equation can be rewritten as 
\begin{align}
\mathcal{L}_{X_1} \langle \xi_2,X_3 \rangle =& \ 
\iop_{X_1}{\dop} \langle \xi_2,X_3 \rangle 
= \rho (X_1) \cdot \langle \xi_2,X_3 \rangle,
  \label{Liederiv_f1}
 \\
  \mathcal{L}_{\xi_1} \langle \xi_3, X_2 \rangle =& \ 
\iop_{\xi_1} {\dop}_* \langle \xi_3,X_2 \rangle 
= \rho_* (\xi_1) \cdot \langle \xi_3,X_2 \rangle.
  \label{Liederiv_f2}
\end{align}
Note that the definition of the exterior derivative in a Lie algebroid
has been used.
Therefore, the equation \eqref{T_1} is calculated as 
\begin{align}
  \pbra{[e_1,e_2]_{\rm V},e_3}
  &= \frac{1}{2} 
\Bigl\{ 
\langle \xi_3,[X_1,X_2]_{\rm V} \rangle  + \langle [\xi_1,\xi_2]_{\rm
 V},X_3 \rangle  + \rho (X_3) \cdot \mbra{e_1,e_2} \! - \rho_* (\xi_3)
 \cdot \mbra{e_1,e_2} \! + \text{c.p.} 
\Bigr\} \notag\\
  &\quad + \frac{1}{2} \rho_{\rm V} (e_1) \cdot \pbra{e_2,e_3} - \frac{1}{2} \rho_{\rm V}
 (e_2) \cdot \pbra{e_3,e_1}.
  \label{T_2}
\end{align}
By summing up all the contributions from the cyclic permutation parts,
we then have the following expression for $T (e_1, e_2, e_3)$:
\begin{equation}
  T(e_1,e_2,e_3) = \frac{1}{2}
\left\{ 
\frac{}{}
\langle \xi_3, [X_1,X_2]_{\mathrm{V}} \rangle + \langle [\xi_1,\xi_2]_{\mathrm{V}}, X_3
\rangle + \rho (X_3) \cdot \mbra{e_1,e_2} \! - \rho_* (\xi_3) \cdot \mbra{e_1,e_2} 
\right\} 
+ \text{c.p.}
\label{Lemma3.2}
\end{equation}
We also derive other useful relations.
By expanding the bracket $[\cdot, \cdot]_{\mathrm{V}}$ and the bilinear form
$\pmbra{\cdot,\cdot}$, and using the relations \eqref{Liederiv_f1},\eqref{Liederiv_f2}, we find
\begin{align}
  \mbra{[e_1,e_2]_{\rm V},e_3} + \pbra{[e_1,e_2]_{\rm V},e_3}
  &= \langle [\xi_1,\xi_2]_{\rm V},X_3 \rangle + \rho (X_1) \cdot \langle
 \xi_2,X_3 \rangle - \langle \xi_2,[X_1,X_3]_{\rm V} \rangle \notag\\
  &\quad - \rho (X_2) \cdot \langle \xi_1,X_3 \rangle + \langle
 \xi_1,[X_2,X_3]_{\rm V} \rangle + \rho (X_3) \cdot \mbra{e_1,e_2}.
   \label{T_4}
\end{align}
The summation over all the permutations of \eqref{T_4} results in 
\begin{align}
  \mbra{[e_1,e_2]_{\rm V},e_3} + \text{c.p.} =& \ 
T(e_1,e_2,e_3) 
\notag \\
& +  
\left\{ 
\left(
\frac{}{}
\rho (X_3) \cdot \mbra{e_1,e_2} + 2 \rho_* (\xi_3) \cdot \mbra{e_1,e_2}
- \langle [\xi_1,\xi_2]_{\rm V},X_3 \rangle 
\right) 
+ \text{c.p.} 
\right\},
 \label{Lemma3.4}
\end{align}
where we have used the relation \eqref{Lemma3.2}.
The equations \eqref{Lemma3.2} and \eqref{Lemma3.4} are just the
formulas of Lemma 3.2 and Lemma 3.4 in \cite{LiWeXu}. 
These relations will be useful in the following discussions.

\paragraph{Analysis on \ref{axiom:C1}}
We first examine axiom \ref{axiom:C1} for Courant algebroids.
Given the bracket structure \eqref{eq:Vaisman_bracket}, where $(E,E^*)$ is not
necessarily assumed to be a Lie bialgebroid, then the left hand side of \eqref{eq:C1} is calculated as 
\begin{align}
&[[e_1,e_2]_{\rm V},e_3]_{\rm V} + \text{c.p.}
= I_1 + I_2, 
\\
& I_1 =  
[[\xi_1,\xi_2]_{\rm V}, \xi_3]_{\rm V} 
+ [\mathcal{L}_{X_1} \xi_2 
- \mathcal{L}_{X_2} \xi_1, \xi_3]_{\rm V} 
+ [{\dop}\mbra{e_1,e_2},\xi_3]_{\rm V}
\notag\\
& \qquad 
+ \mathcal{L}_{[X_1,X_2]_E + \mathcal{L}_{\xi_1}X_2 
	- \mathcal{L}_{\xi_2}X_1 - {\dop}_*\mbra{e_1,e_2}}\xi_3 
\notag \\
& \qquad 
- \mathcal{L}_{X_3} [\xi_1,\xi_2]_{\rm V} 
- \mathcal{L}_{X_3} \mathcal{L}_{X_1} \xi_2 
+ \mathcal{L}_{X_3} \mathcal{L}_{X_2} \xi_1 
- \mathcal{L}_{X_3}{\dop}\mbra{e_1,e_2}
+ {\dop} \mbra{[e_1,e_2]_{\rm V},e_3}
+ \text{c.p.},
\label{C1_2}
\\
&I_2 = 
[[X_1,X_2]_{\rm V},X_3]_{\rm V} 
+ [\mathcal{L}_{\xi_1} X_2
- \mathcal{L}_{\xi_2} X_1, X_3]_{\rm V} 
- [{\dop}_* \mbra{e_1,e_2},X_3]_{\rm V}
\notag \\
& \qquad 
+ \mathcal{L}_{[\xi_1,\xi_2]_{E^*} + \mathcal{L}_{\xi_1}X_2 
	- \mathcal{L}_{\xi_2}X_1 + {\dop} \mbra{e_1,e_2}}X_3 
\notag \\
& \qquad 
- \mathcal{L}_{\xi_3} [X_1,X_2]_{\rm V} 
- \mathcal{L}_{\xi_3} \mathcal{L}_{\xi_1} X_2 
+ \mathcal{L}_{\xi_3} \mathcal{L}_{\xi_2} X_1 
+ \mathcal{L}_{\xi_3}{\dop}_*\mbra{e_1,e_2} \!
- {\dop}_*\mbra{[e_1,e_2]_{\rm V},e_3} \!
+ \text{c.p.}
\label{C1_1}
\end{align}
Now we define the $\Gamma(E^*)$ and $\Gamma(E)$ parts as
$I_1$ and $I_2$ respectively.
Since the calculations are the same for $I_1$ and $I_2$, we consider
$I_1$ only in the following.

First, because of the Jacobi identity of the Lie algebroid $E^*$, we have 
$[[\xi_1,\xi_2]_{\rm V},\xi_3]_{\rm V} + \text{c.p.} = [[\xi_1,\xi_2]_{E^*},\xi_3]_{E^*} + \text{c.p.} = 0$.
We then decompose the Lie bracket into each part
\begin{align}
\mathcal{L}_{
[X_1,X_2]_{E} + 
\mathcal{L}_{\xi_1}X_2 -
\mathcal{L}_{\xi_2}X_1 - 
{\dop}_*\mbra{e_1,e_2}
 } = 
\mathcal{L}_{[X_1,X_2]_E} 
+ 
\mathcal{L}_{\mathcal{L}_{\xi_1}X_2 -
\mathcal{L}_{\xi_2}X_1} - 
\mathcal{L}_{\dop_*\mbra{e_1,e_2}}.
\end{align}
Since it is the ordinary Lie derivative, the first term is 
$ \mathcal{L}_{[X_1,X_2]_{E}} =
\mathcal{L}_{X_1} \mathcal{L}_{X_2} - \mathcal{L}_{X_2}
\mathcal{L}_{X_1}$.
Therefore this with the cyclic permutations give a vanishing contribution in \eqref{C1_2}.
Then the remaining parts in $I_1$ are 
\begin{align}
I_1
&= 
[\mathcal{L}_{X_1} \xi_2 - \mathcal{L}_{X_2} \xi_1, \xi_3]_{\rm V} 
+ [{\dop}\mbra{e_1,e_2}, \xi_3]_{\rm V} 
+ \mathcal{L}_{\mathcal{L}_{\xi_1} X_2 - \mathcal{L}_{\xi_2}X_1} \xi_3 
- \mathcal{L}_{ {\dop}_*\mbra{e_1,e_2}} \xi_3 
\notag \\
& \quad 
- \mathcal{L}_{X_3} [\xi_1, \xi_2]_{\rm V} 
- \mathcal{L}_{X_3}{{\dop} \mbra{e_1,e_2}}
+ {\dop} \mbra{[e_1,e_2]_{\rm V},e_3}  
+ \text{c.p.}
\label{C1_3}
\end{align}
Now we focus on the term $\mathcal{L}_{X_3}[\xi_1,\xi_2]_{\rm V}$ in the
second line. One finds that this with the cyclic permutations give (see
Appendix \ref{sect:calculations} for detail)
\begin{align}
\mathcal{L}_{X_3}[\xi_1,\xi_2]_{\rm V} 
+ \text{c.p.}
& = [\mathcal{L}_{X_1}\xi_2 
- \mathcal{L}_{X_2} \xi_1, \xi_3]_{\rm V} 
+ \mathcal{L}_{\mathcal{L}_{\xi_1} X_2 - \mathcal{L}_{\xi_2}X_1} \xi_3 
\notag \\
& \quad 
+ 2 [{\dop} \mbra{e_1,e_2}, \xi_3]_{\rm V} 
+ 2 {\dop} (\rho_* (\xi_3) \cdot \mbra{e_1,e_2}) 
- {\dop} \langle [\xi_1,\xi_2]_{\rm V}, X_3 \rangle 
\notag \\
& \quad 
+ \iop_{X_3} 
( {\dop}[\xi_1,\xi_2]_{\rm V} 
- \mathcal{L}_{\xi_1} {\dop} \xi_2 
+ \mathcal{L}_{\xi_2} {\dop} \xi_1 ) 
+ \text{c.p.}
\label{C1_4}
\end{align}
Substituting the expression \eqref{C1_4} into \eqref{C1_3} and using
\eqref{Lemma3.4}, we find that there are several cancellations among terms.
The result is 
\begin{align}
  I_1 &= {\dop} T(e_1,e_2,e_3)  - \{K_1 + K_2\} + \text{c.p.},
\end{align}
where we have defined the following quantities:
\begin{align}
  &K_1 = \iop_{X_3}({\dop}[\xi_1,\xi_2]_{\rm V} - \mathcal{L}_{\xi_1}{\dop}\xi_2 + \mathcal{L}_{\xi_2}{\dop}\xi_1), \notag\\
  &K_2 = 
  \mathcal{L}_{{\dop}_* \mbra{e_1,e_2}}\xi_3 + [{\dop}\mbra{e_1,e_2},\xi_3]_{\mathrm{V}}.
\end{align}
By the same way, we have 
\begin{align}
  &I_2 = {\dop}_* T(e_1,e_2,e_3)  - \{K_3 + K_4\} + \text{c.p.}, \notag\\
  &K_3 = \iop_{\xi_3}({\dop}_*[X_1,X_2]_{\mathrm{V}} - \mathcal{L}_{X_1}{\dop}_*X_2 + \mathcal{L}_{X_2}{\dop}_*X_1), \notag\\
  &K_4 = - \Bigl(
  \mathcal{L}_{{\dop} \mbra{e_1,e_2}}X_3 + [{\dop}_*\mbra{e_1,e_2},X_3]_{\mathrm{V}}
  \Bigr).
\end{align}
With the above results at hand, the Jacobiator of the bracket $[\cdot,\cdot]_{\mathrm{V}}$ is evaluated as 
\begin{align}
  [[e_1,e_2]_{\mathrm{V}},e_3]_{\mathrm{V}} + \text{c.p.} = I_1 + I_2 =
 \mathcal{D}T(e_1,e_2,e_3) - ( J_1 + J_2 + \text{c.p.}).
\label{eq:Jac_V}
\end{align}
Here we have defined the following quantities:
\begin{align}
  J_1
  &= K_1 + K_3 \notag\\
  &= \iop_{X_3}
\Bigl(
{\dop}[\xi_1,\xi_2]_{\rm V} - \mathcal{L}_{\xi_1}{\dop}\xi_2 +
 \mathcal{L}_{\xi_2}{\dop}\xi_1
\Bigr)
+ 
\iop_{\xi_3}
\Bigl(
{\dop}_*[X_1,X_2]_{\rm V} - \mathcal{L}_{X_1}{\dop}_*X_2 +
 \mathcal{L}_{X_2}{\dop}_*X_1
\Bigr), 
\notag\\
  J_2
  &= K_2 + K_4 \notag\\
  &= 
\Bigl(
\mathcal{L}_{{\dop}_* \mbra{e_1,e_2}}\xi_3 + [{\dop}\mbra{e_1,e_2},\xi_3]_{\mathrm{V}}
\Bigr)
 - 
\Bigl(
\mathcal{L}_{{\dop} \mbra{e_1,e_2}}X_3 + [{\dop}_*\mbra{e_1,e_2},X_3]_{\mathrm{V}}
\Bigr).
\label{eq:J1J2}
\end{align}
It is obvious that the last term $J_1 + J_2 + \text{c.p.}$ in
\eqref{eq:Jac_V} does not vanish in general.
Therefore, we find that $(E \oplus E^*,[\cdot,\cdot]_{\rm
V},\rho_{\rm V} ,\pbra{\cdot,\cdot})$ fails to satisfy the relation \eqref{eq:C1}
in axiom \ref{axiom:C1}.

\paragraph{Analysis on \ref{axiom:C2}}
We next examine the relation 
\begin{equation}
 \rho_{\rm V} ([e_1,e_2]_{\rm V}) \cdot f = [\rho_{\rm V} (e_1),\rho_{\rm V} (e_2)] f,
\label{eq:C2_eq}
\end{equation}
in axiom \ref{axiom:C2}. 
We evaluate $\text{LHS}-\text{RHS}$ for $(E \oplus E^*,[\cdot,\cdot]_{\rm
V},\rho_{\rm V} ,\pbra{\cdot,\cdot})$ and examine whether it vanishes or not. 
Given 
the definition of $\rho_{\rm V}$ and $[\cdot,\cdot]_{\mathrm{V}}$ in \eqref{eq:Vaisman_bracket},
the left hand side of the above equation is
evaluated as 
\begin{align}
& \rho_{\rm V} ([e_1,e_2]_{\rm V}) \cdot f \notag \\
 &= [\rho (X_1),\rho (X_2)] \cdot f + \rho (\mathcal{L}_{\xi_1}X_2)
 \cdot f - \rho (\mathcal{L}_{\xi_2}X_1) \cdot f - \frac{1}{2} \rho
 \rho_*^* {\dop}_0(\braket{\xi_1,X_2} - \braket{\xi_2,X_1}) \cdot f \notag\\
 &+ [\rho_* (\xi_1),\rho_* (\xi_2)] \cdot f +
 \rho_*(\mathcal{L}_{X_1}\xi_2) \cdot f - 
\rho_*(\mathcal{L}_{X_2}\xi_1) \cdot f 
+ \frac{1}{2} \rho_* \rho^*{\dop}_0(\braket{\xi_1,X_2} -
 \braket{\xi_2,X_1}) \cdot f,
 \label{C2_1}
\end{align}
where we have used the fact ${\dop}_* = \rho_*^*{\dop}_0, {\dop} =
\rho^*{\dop}_0$ and $E, E^*$ are Lie algebroids. Here $\dop_0$ is the
ordinary exterior derivative defined on $\Gamma (T^*M)$.

On the other hand, the right hand side of \eqref{eq:C2_eq} is 
\begin{align}
[\rho (X_1), \rho (X_2)] f + [\rho_* (\xi_1), \rho (X_2)]f 
+ [\rho (X_1), \rho_* (\xi_2)]f
+ [\rho_* (\xi_1), \rho_* (\xi_2)] f.
\end{align}
We examine the second and the third cross terms in the above. 
In order to evaluate these terms, we use the following expression:
\begin{align}
\rho_* (\xi) \rho (X) \cdot f = 
-  \rho \rho_*^* {\dop_0}\braket{\xi,X} \cdot f
+ \braket{\xi,\mathcal{L}_{\dop f}X}
+ \langle \dop f, \mathcal{L}_{\xi} X \rangle.
\label{eq:C2_eq2}
\end{align}
This follows from the relation
\begin{align}
 \rho \rho_*^* {\dop_0}\braket{\xi,X} \cdot f
 &= \braket{\dop_*\braket{\xi,X},\dop f} \notag\\
 &= \mathcal{L}_{\dop f}\braket{\xi,X} \notag\\
 &= 
- \langle \mathcal{L}_\xi \dop f, X \rangle + \braket{\xi,\mathcal{L}_{\dop f}X},
\end{align}
together with 
\begin{align}
  \braket{ \mathcal{L}_\xi \dop f,X}
 &= \mathcal{L}_\xi \braket{\dop f,X} - \braket{\dop f, \mathcal{L}_\xi X} \notag\\
 &= \iop_\xi \dop_* \braket{\dop f,X} - \braket{\dop f, \mathcal{L}_\xi X}  \notag\\
 &= \rho_*(\xi) \rho (X) \cdot f -  \braket{\dop f, \mathcal{L}_\xi X}.
 \label{C2_3}
\end{align}
Here we have used the relation $\mathcal{L}_\xi f = \braket{\xi,\dop_*
f}$ for the Lie derivative and the definition of the anchors $\rho,
\rho_*$.
Then we obtain
\begin{align}
 [\rho (X), \rho_* (\xi)] f
 &= (\rho (X) \rho_* (\xi) - \rho_* (\xi) \rho (X)) \cdot f
\notag \\
 &= 
 \rho (X) \rho_* (\xi) \cdot f
 + (\rho \rho_*^* {\dop}_0\braket{\xi,X}) \cdot f 
 - \braket{\xi,\mathcal{L}_{{\dop}f}X}
 - \rho (\mathcal{L}_{\xi}X) \cdot f. 
 \label{C2_4}
\end{align}
Using this expression, we find that $\text{LHS} - \text{RHS}$ in
\eqref{eq:C2_eq} is evaluated as 
\begin{align}
& 
\rho_{\rm V} ([e_1,e_2]_{\rm V}) \cdot f - 
[\rho_{\rm V} (e_1),\rho_{\rm V} (e_2)] f
\notag \\
& =
 - \braket{\xi_1 , \bigl(\mathcal{L}_{{\dop}f}X_2 - [X_2,{\dop}_*f]_E \bigr)} 
+ \braket{\xi_2 , \bigl(\mathcal{L}_{{\dop} f}X_1 - [X_1,{\dop}_*f]_E
 \bigr) }\notag\\
& \quad \,
+ \frac{1}{2} 
\Bigl( 
\rho \rho_*^*
+
\rho_* \rho^* 
\Bigr)
{\dop}_0(\braket{\xi_1,X_2} 
- \braket{\xi_2,X_1}) \cdot f.
 \label{C2_7}
\end{align}
We find that the right hand side in \eqref{C2_7} does not vanish in general.
Therefore we conclude that $(E \oplus E^*,[\cdot,\cdot]_{\rm V},\rho_{\rm V}. \pbra{\cdot,\cdot})$ does not
satisfy axiom \ref{axiom:C2}.

\paragraph{Analysis on \ref{axiom:C3}}
For axiom \ref{axiom:C3}, the left hand side of \eqref{eq:C3} is expanded as 
\begin{align}
  [e_1,fe_2]_{\rm V}
  &= [X_1, fX_2]_{\rm V} + [X_1, f\xi_2]_{\rm V} + [\xi_1, fX_2]_{\rm V} + [\xi_1,f\xi_2]_{\rm V}.
  \label{C3_0}
\end{align}
By the definition of the bracket $[\cdot,\cdot]_{\rm V}$ \eqref{eq:Vaisman_bracket}, we find 
\begin{equation}
  [X_1, f\xi_2]_{\rm V}
  = -\mathcal{L}_{f\xi_2}X_1 + \frac{1}{2}{\dop}_*(f \langle \xi_2,X_1
 \rangle) + \mathcal{L}_{X_1}(f\xi_2) - \frac{1}{2}{\dop}(f \langle \xi_2,X_1 \rangle).
 \end{equation}
By using \eqref{eq:Lie_derivative_relations} and the definition
 $\mathcal{D}$, this expression is rewritten as 
 \begin{align}
  [X_1, f\xi_2]_{\rm V}
  &= -f \mathcal{L}_{\xi_2}X_1 - {\dop}_*f \langle X_1,\xi_2 \rangle
 + \frac{1}{2} {\dop}_*f \langle X_1,\xi_2 \rangle + \frac{1}{2}f{\rm
 d}_* \langle X_1,\xi_2 \rangle \notag\\
  &\quad + f{\dop} \langle X_1, \xi_2 \rangle + \iop_X {\dop} f \xi_2 +
 f \iop_X {\dop} \xi_2 - \frac{1}{2} {\dop}f
\langle \xi_2,X_1 \rangle - \frac{1}{2}f{\dop} \langle \xi_2,X_1
 \rangle \notag \\
  &= f[X_1,\xi_2]_{\rm V} + (\rho (X_1) \cdot f) \xi_2 - \frac{1}{2} \mathcal{D} f
 \langle \xi_2,X_1 \rangle.
 \label{C3_1}
\end{align}
Similarly, $[\xi_1, fX_2]_{\rm V}$ is evaluated as
\begin{equation}
  [\xi_1, fX_2]_{\rm V} = f[\xi_1,X_2]_{\rm V} + (\rho_* (\xi_1)
  \cdot f) X_2 - \frac{1}{2}\mathcal{D} f \langle \xi_1,X_2 \rangle.
  \label{C3_2}
\end{equation}
On the other hand, since $E$ and $E^*$ are both Lie algebroids, their
Lie brackets satisfy the following relations,
\begin{align}
  &[X_1, fX_2]_{\rm V} = [X_1, fX_2]_{E} = f[X_1,X_2]_{\rm V} + (\rho
 (X_1) \cdot f ) X_2,
\notag  \\
  &[\xi_1,f\xi_2]_{\rm V} = [\xi_1, f\xi_2]_{E^*} = f [\xi_1,\xi_2]_{\rm V}
 + (\rho_* (\xi_1) \cdot f) \xi_2.
  \label{C3_4}
\end{align}
Therefore, by summing up all the contributions \eqref{C3_1}, \eqref{C3_2}
and \eqref{C3_4}, we find
\begin{align}
  [e_1,fe_2]_{\rm V}
  &= f[e_1,e_2]_{\rm V} + (\rho(e_1)f)e_2 - \mathcal{D} f \pbra{e_1,e_2}.
\end{align}
This is nothing but the relation in axiom \ref{axiom:C3}.

\paragraph{Analysis on \ref{axiom:C4}}
We now confirm axiom \ref{axiom:C4}.
To this end, we evaluate the left hand side of \eqref{eq:C4}.
The result is 
\begin{align}
  \pbra{\mathcal{D}f,\mathcal{D}g}
  &= \pbra{{\dop}f + {\dop}_*f,{\dop}g + {\dop}_*g} \notag\\
  &= \frac{1}{2}( \langle {\dop}f,{\dop}_*g \rangle  + \langle {\rm
 d}_*f,{\dop}g \rangle) \notag\\
  &= \frac{1}{2}(\rho_*({\dop}f) \cdot g + \rho({\dop}_*f) \cdot g) \notag\\
  &= \frac{1}{2} 
\Bigl(
\rho_* \rho^* + \rho \rho_*^* 
\Bigr)(\dop_0 f) \cdot g.
 \label{C4_1}
\end{align}
Now, we remember that 
an operator $\mathcal{O} : E^* \to E$ is skew-symmetric 
when $\langle \mathcal{O} X, X \rangle = 0$ for ${}^{\forall}X \in \Gamma (E)$. 
For $\mathcal{O} = \rho \rho_*^*$, this implies 
$0 = \langle \rho \rho_*^* X,X \rangle = \frac{1}{2} \langle
\rho \rho_*^* X, X \rangle + \frac{1}{2} \langle X, (\rho
\rho_*^*)^* X \rangle$ leading to the expression $\rho \rho_*^* = - \rho_* \rho^*$.
Then, if the operator $\rho \rho_*^*$ is skew-symmetric, 
the axiom \ref{axiom:C4} holds $\pbra{\mathcal{D}f,\mathcal{D}g} = 0$. 
In order to examine the skew-symmetric nature of the operator $\rho
\rho_*^*$, we first derive a variant of the proposition 3.4 in
\cite{Mackenzie}. 

Using the properties of Lie algebroids, one finds
the following relation for any $X,Y \in \Gamma (E)$ (see Appendix
\ref{sect:calculations} for detail):
\begin{align}
&
\left(
\mathcal{L}_{\dop f} X + [\dop_* f,X]_E 
\right) \wedge Y
\notag \\
&= - f
\Bigl(
\dop_* [X,Y]_E + \mathcal{L}_{Y} \dop_* X - \mathcal{L}_{X} \dop_* Y
\Bigr) + 
\Bigl(
\dop_* [X,f Y]_E - \mathcal{L}_X \dop_* (f Y) + \mathcal{L}_{fY} \dop_* X
\Bigr).
\label{eq:C4_V}
\end{align}
One finds that the right hand side of the above expression vanish when
the derivation condition \eqref{eq:derivation} is satisfied.
This means that the relation
\begin{align}
\mathcal{L}_{\dop f} X + [\dop_* f,X]_E = 0
\label{eq:MXprop.3.4}
\end{align}
holds {\it if the derivation condition is satisfied}.
Equivalently, the right hand side in \eqref{eq:MXprop.3.4} is generically
non-zero without imposing the derivation condition.
The equation \eqref{eq:MXprop.3.4} is just the one in proposition
3.4 in \cite{Mackenzie}. 

In the following, we show that $\rho \rho_*^*$ is skew-symmetric if the
equation \eqref{eq:MXprop.3.4} is satisfied, {\it i.e.}\ if the derivation
condition is satisfied.
By substituting $X = \dop_* f$ in \eqref{eq:MXprop.3.4}, we find
\begin{align}
\dop_* 
\Bigl( 
\rho \rho_*^* (\dop_0 f) \cdot f
\Bigr)  = 0.
\label{eq:PP34-1}
\end{align}
Here we have used the relations in \eqref{eq:Lie_derivative_relations}.
By replacing $f$ with $f^2$, this becomes
\begin{align}
\dop_* 
\Bigl( 
\rho \rho_*^* (\dop_0 f^2) \cdot f^2
\Bigr) = 0.
\label{eq:PP34-2}
\end{align}
On the other hand, since we have
\begin{align}
\rho \rho_*^* (\dop_0 f^2) \cdot f^2 = \langle \dop_0 f^2, \rho \rho_*^*
 \dop_0 f^2 \rangle = \langle \dop f^2, \dop_* f^2 \rangle = 4 f^2
 \langle \dop f, \dop_* f \rangle,
\end{align}
we find 
\begin{align}
\Bigl(
\rho \rho_*^* (\dop_0 f) \cdot f 
\Bigr)
\dop_* f^2
=& \  \dop_* 
\Bigl\{
\bigl(
\rho \rho_*^* (\dop_0 f) \cdot f 
\bigr)
f^2
\Bigr\}
-
\dop_* \Bigl( \rho \rho_*^* (\dop_0 f) \cdot f \Bigr) f^2 
\notag \\
=& \ \frac{1}{4} \dop_* 
\Bigl(
\rho \rho_*^* (\dop_0 f^2) \cdot f^2
\Bigr)
-
\dop_* \Bigl( \rho \rho_*^* (\dop_0 f) \cdot f \Bigr) f^2.
\end{align}
The right hand side vanishes due to the relations \eqref{eq:PP34-1}, \eqref{eq:PP34-2}.
Then we find
\begin{align}
\Bigl(
\rho \rho_*^* (\dop_0 f) \cdot f 
\Bigr)
\rho^*_* \dop_0 f^2 = 0.
\end{align}
Note that $\dop_* = \rho^*_* \dop_0$.
Applying $\rho$ and taking the inner product with $\dop_0 f$, we finally obtain
\begin{align}
2 f 
\Bigl(
\rho \rho^*_* (\dop_0 f) \cdot f
\Bigr)^2 = 0.
\label{eq:PP34-3}
\end{align}
This implies that $0 = \rho \rho^*_* (\dop_0 f) \cdot f = \langle \rho
\rho^*_* (\dop_0 f), \dop_0 f \rangle$, namely, $\rho \rho^*_*$ is a
skew-symmetric operator.

Now we go back to the discussion on $E \oplus E^*$. Since we have not
assumed any Lie bialgebroid structures on $E \oplus E^*$, the derivation
condition is not satisfied anymore. Then, 
\eqref{eq:MXprop.3.4} does
not follow and the relation \eqref{eq:PP34-3} never holds.
Therefore the right hand side in \eqref{eq:C4_V} does not vanish in
general and $(E \oplus E^*, [\cdot,\cdot]_{\mathrm{V}}, \rho_{\rm V}, \pbra{\cdot,\cdot})$
fails to satisfy axiom \ref{axiom:C4}.

\paragraph{Analysis on \ref{axiom:C5}}
Finally, we examine the relation \eqref{eq:C5} in axiom \ref{axiom:C5}.
It is useful to start from the relation \eqref{T_2}.
This implies 
\begin{align}
  \pbra{[e,e_1]_{\rm V},e_2} &= T(e,e_1,e_2) + \frac{1}{2}\rho_{\rm V} (e) \cdot
 \pbra{e_1,e_2} - \frac{1}{2} \rho_{\rm V} (e_1) \cdot \pbra{e,e_2}, \notag\\
  \pbra{e_1,[e,e_2]_{\rm V}} &= T(e,e_2,e_1) + \frac{1}{2} \rho_{\rm V} (e) \cdot
 \pbra{e_2,e_1} - \frac{1}{2} \rho_{\rm V} (e_2) \cdot \pbra{e,e_1}. 
\end{align}
By summing up those, we find 
\begin{equation}
  \rho_{\rm V} (e) \cdot \pbra{e_1,e_2} = \pbra{[e,e_1]_{\rm V},e_2} +
  \pbra{e_1,[e,e_2]_{\rm V}} + \frac{1}{2} \rho_{\rm V} (e_1) \cdot \pbra{e,e_2} +
  \frac{1}{2} \rho_{\rm V} (e_2) \cdot \pbra{e,e_1}.
  \label{C5_2}
\end{equation}
Note that the contributions coming from $T$ are canceled due to their skew-symmetric nature.
By their defining properties of $\rho_{\rm V}$ and $\mathcal{D}$, we evaluate
the third term as 
\begin{align}
  \frac{1}{2}\rho_{\rm V} (e_1) \cdot \pbra{e,e_2}
  &= \frac{1}{2} \rho (X_1) \cdot \pbra{e,e_2} + \frac{1}{2} \rho_* (\xi_1)
 \cdot \pbra{e,e_2}\notag\\
  &= \frac{1}{2} \Bigl( \langle {\dop}\pbra{e,e_2}, X_1 \rangle + \langle
 \xi_1,{\dop}_*\pbra{e,e_2} \rangle \Bigr) \notag\\
  &= \pbra{ {\dop}\pbra{e,e_2} + {\rm d_*}\pbra{e,e_2} , X_1 + \xi_1 } \notag\\
  &= \pbra{\mathcal{D}\pbra{e,e_2} , e_1}.
\end{align}
Similarly, we have 
\begin{equation}
  \frac{1}{2} \rho_{\rm V} (e_2) \cdot \pbra{e,e_1} = \pbra{\mathcal{D}\pbra{e,e_1} , e_2}.
\end{equation}
Substituting these into \eqref{C5_2}, we find
\begin{align}
  \rho_{\rm V} (e) \cdot \pbra{e_1,e_2}
  &= \pbra{[e,e_1]_{\rm V} + \mathcal{D}\pbra{e,e_1} , e_2} + \pbra{e_1 ,
 [e,e_2]_{\rm V} + \mathcal{D}\pbra{e,e_2}}.
  \label{C5_3}
\end{align}
The result \eqref{C5_3} shows that the relation \eqref{eq:C5} in axiom \ref{axiom:C5}
is satisfied for ($E \oplus
E^*$, $[\cdot,\cdot]_{\rm V}$, $\rho_{\rm V}$, $\pbra{\cdot,\cdot}$).

Collecting all together, we have demonstrated that ($E \oplus
E^*$, $[\cdot,\cdot]_{\rm V}$, $\rho_{\rm V}$, $\pbra{\cdot,\cdot}$) 
indeed satisfies the axioms \ref{axiom:C3} and \ref{axiom:C5} of Courant
algebroids. 
As clarified in~\cite{2002math......4010U}, the axiom \ref{axiom:C5} implies \ref{axiom:C3}. The result presented here is consistent with this. 
This is nothing but the axioms \ref{axiom:V1} and
\ref{axiom:V2} of Vaisman algebroids.
This means that given a Lie algebroid $E$ and its dual $E^*$, one can
always define a Vaisman algebroid $\mathcal{V} = E \oplus E^*$ with the appropriate
bracket \eqref{eq:Vaisman_bracket}, the anchor $\rho_{\rm V} = \rho + \rho_*$
and the bilinear form $\pbra{\cdot,\cdot}$. 
This is an analogue of the Drinfel'd double for a
Courant algebroid \cite{LiWeXu}. 
As we have discussed in Section \ref{sect:DD}, when one imposes the derivation condition
\eqref{eq:derivation} as the compatibility condition between $E$ and
$E^*$, then $(E,E^*)$ becomes a Lie bialgebroid.
As we have explicitly shown, the equations \eqref{eq:J1J2}, \eqref{C2_7},
\eqref{C4_1}, \eqref{eq:MXprop.3.4} all vanish when the derivation
condition is satisfied. This means axioms \ref{axiom:C1}, \ref{axiom:C2}
and \ref{axiom:C4} are satisfied for a Lie bialgebroid $(E,E^*)$.
In this case, the discussion here reduces to the ones in
\cite{LiWeXu} and the double $\mathcal{V} = E \oplus E^*$ becomes a Courant algebroid.


We will see that the doubled structure in the Vaisman algebroid naturally appears in DFT. 
The strong constraint implies the derivation condition which enable one to
find the doubled structure behind the DFT gauge symmetry.

\subsection{Dirac structures in Vaisman algebroid}
In this subsection, we study Dirac structures in Vaisman algebroids.
It is shown that when there are Dirac subbundles $L, \tilde{L}$ of a Courant algebroid
$(\mathcal{C},[\cdot,\cdot]_{\rm c}, \rho_{\rm c}, \bbra{\cdot,\cdot})$, namely, 
$L, \tilde{L}$ are maximally isotropic with respect to
$\bbra{\cdot,\cdot}$, satisfing $\mathcal{C} = L \oplus \tilde{L}$ and
involutive (integrable), then 
the vector bundle $\tilde{L}$ is regarded as the dual bundle of $L$ under
the natural paring $2 \bbra{\cdot,\cdot}$.
Given these structures, it is shown that $(L, \tilde{L})$ becomes a Lie
bialgebroid.
We briefly demonstrate this fact following the discussion in \cite{LiWeXu}.
Before showing the above statement, we first refer the Proposition 2.3
in \cite{LiWeXu}:
\begin{propositionLWX} 
If $L$ is an integrable isotropic subbundle of a Courant algebroid
 $(\mathcal{C},[\cdot,\cdot]_{\rm c}, \rho_{\rm c}, \bbra{\cdot,\cdot})$, then $(L,
 [\cdot,\cdot]_{\rm c}, \rho_{\rm c}|_L)$ becomes a Lie algebroid.
\end{propositionLWX}
Here the isotropy is defined with respect to the bilinear form
$\bbra{\cdot,\cdot}$. Namely, for any $X,Y \in \Gamma (L)$, they satisfy
$\bbra{X,Y} = 0$.
This proposition is confirmed by showing that the bracket
$[\cdot,\cdot]_{\rm c}$ on $L$
satisfies the Jacobi identity.
This immediately follows from the relation \eqref{eq:C1} in axiom \ref{axiom:C1} of Courant algebroids and the
isotropic nature of $L$.
Due to the proposition 2.3, any Dirac structures $L, \tilde{L}$ in a Courant
algebroid become Lie algebroids. Their anchors are defined by $\rho =
\rho_{\rm c}|_{L}$, $\rho_* = \rho_{\rm c}|_{\tilde{L}}$.

By its defining axiom \ref{axiom:C5}, for $X \in \Gamma (L)$,
$\xi \in \Gamma (\tilde{L})$, one can show that 
\begin{align}
[X, \xi]_{\rm c} = - \mathcal{L}_{\xi} X + \frac{1}{2} \dop_{*} \langle \xi, X
 \rangle + \mathcal{L}_X \xi - \frac{1}{2} \dop \langle \xi, X \rangle.
\label{eq:th26-1}
\end{align}
Here in deriving \eqref{eq:th26-1}, we have used the fact that $L,
\tilde{L}$ are Lie algebroids and isotropic.
With this relation, the following Lemma 5.2 in \cite{LiWeXu} follows:
\begin{lemmaLWX} 
Given Dirac structures $L, \tilde{L}$ such that $\mathcal{C} = L \oplus
 \tilde{L}$ for a Courant algebroid $\mathcal{C}$, then the following relations hold:
\begin{align}
\mathcal{L}_{\dop_* f} \xi = - [\dop f, \xi]_{\tilde{L}}, \quad 
\mathcal{L}_{\dop f} X = - [\dop_* f, X]_L.
\label{eq:lemma5.2}
\end{align}
Here $\dop, \dop_*$ are induced de Rham differentials on $L$ and $\tilde{L}$.
\end{lemmaLWX}
This is shown as follows.
By the axiom \ref{axiom:C4}, one first find the relation
\begin{align}
\rho_* \cdot \dop = - \rho \cdot \dop_*.
\label{eq:th26-2}
\end{align}
Then using this relation, we find 
\begin{align}
[\rho_* (\xi), \rho (X)] = \rho (\mathcal{L}_{\xi} X) - \rho_{*} (\mathcal{L}_X \xi)
 + \rho_* (\dop \langle \xi, X \rangle),
\end{align}
where we have assumed the axiom \ref{axiom:C2} and used the relation
\eqref{eq:th26-2}.
On the other hand, by using the properties of Lie algebroids,
we calculate 
\begin{align}
\rho_* (\dop \langle \xi, X \rangle) \cdot f = 
[\rho_* (\xi), \rho (X)] f - \rho (\mathcal{L}_{\xi} X) \cdot f + \rho_* (\mathcal{L}_X
 \xi) \cdot f + \langle \mathcal{L}_{\dop_* f} \xi + [\dop f, \xi]_{\tilde{L}}, X \rangle.
\end{align}
comparing the above relations, one proofs the first part in \eqref{eq:lemma5.2}.
Performing the same calculus by exchanging $\xi \leftrightarrow X$ the
latter also follows.

Given the Lemma 5.2, now we focus on the Jacobiator of the Courant bracket.
As we have shown before, if $L, \tilde{L}$ are Lie algebroids, we have 
\begin{align}
[[e_1,e_2]_{\rm c}, e_2]_{\rm c} + \text{c.p.} = \mathcal{D} T (e_1,e_2,e_3) - (J_1 + J_2 + \text{c.p.}),
\end{align}
where $J_1, J_2$ are given in \eqref{eq:J1J2}.
Since $\mathcal{C} = L \oplus \tilde{L}$ satisfies the axiom \ref{axiom:C1}, we have $J_1 + J_2 + \text{c.p.} = 0$.
Due to the Lemma 5.2, one can show that $J_2 = 0$ and the above condition implies $J_1 + \text{c.p.}=0$.
If we take $e_1 = X_1$, $e_2 = X_2$, $e_3 = \xi_3$, then this condition yields
\begin{align}
\dop_* [X_1, X_2]_L - \mathcal{L}_{X_1} \dop_* X_2 + \mathcal{L}_{X_2} \dop_* X_1 = 0.
\end{align}
This is nothing but the derivation condition \eqref{eq:derivation} for
Lie bialgebroids.
As we have mentioned before, Dirac structures $L,\tilde{L}$ in a Courant algebroid defines a
Manin triple $(\mathcal{C},L,\tilde{L})$.

We then in turn switch to the discussion on Vaisman algebroids.
A Dirac structure on a Vaisman algebroid $\mathcal{V}$ is defined by a maximally
isotropic subbundle in $\mathcal{V}$ with respect to a bilinear form $\bbra{\cdot,
\cdot}$ defined on $\Gamma (\mathcal{V})$. 
Now we assume that there are Dirac structures $L, \tilde{L}$ such that
$\mathcal{V} = L \oplus \tilde{L}$ in a Vaisman algebroid $\mathcal{V}$.
Indeed, there is a Dirac structure in a Vaisman algebroid defined
in a para-K\"ahler manifold \cite{Vaisman:2012ke,Vaisman:2012px}.
For Vaisman algebroids, however, only the axioms \ref{axiom:C3} and \ref{axiom:C5} of Courant algebroids are satisfied.
Obviously, the proposition 2.3 in \cite{LiWeXu} does not follow since it requires the axiom \ref{axiom:C1}.
Therefore, the bracket does not satisfy the Jacobi identity and $L,
\tilde{L}$ are not Lie algebroid in general.
Even though they have Lie algebroid structures, since $\mathcal{V} = L \oplus \tilde{L}$ does not satisfy the axioms \ref{axiom:C2} and \ref{axiom:C4}, 
Lemma 5.2 in \cite{LiWeXu} does not hold.
Therefore we conclude that the Dirac structures $L, \tilde{L}$ in
Vaisman algebroids do not satisfy the derivation condition and  
they never define a Lie bialgebroid in general.
It is known that a Lie algebroid $L$ and its dual $L^*$ form a Lie bialgebroid $(L,L^*)$
if and only if the pair $(L,L^*)$ defines differential Gerstenhaber
algebras \cite{Kosmann-Schwarzbach1}.
This means that a differential operator $\dop^*$ ($\dop$) is compatible with the
Schouten-Nijenhuis bracket $[\cdot,\cdot]_{\rm S}$ ($[\cdot,\cdot]_{\rm S}^*$) in $L$ ($L^*$).
This will be explicitly seen in the DFT viewpoint in the next section.
In particular, we will explicitly show that the exterior algebras of
DFT defined on the Kaluza-Klein and winding spaces are incompatible with
the derivation condition 
that is required for the Lie bialgebroid.
\section{Gauge symmetry algebra in DFT} \label{sect:DFT}
In this section, we study doubled aspects of the gauge symmetry in
DFT. We first introduce a para-Hermitian manifold and its foliation
structures as a geometric realization of doubled space-time
\cite{Vaisman:2012ke,Vaisman:2012px, Freidel:2017yuv, Freidel:2018tkj,
Svoboda:2018rci}.
Subbundles $L,\tilde{L}$ on the doubled space-time are naturally
introduced due to the para-complex structure.
We then study a para-Dolbeault cohomology in the DFT framework.
Based on this result, we examine the Lie algebroid structures on
$L,\tilde{L}$ and discuss the relation between the strong constraint and
the derivation condition for Lie bialgebroids.
We also address the relations among Lie bialgebroids, Vaisman and
Courant algebroids realized in DFT.

\subsection{Para-Hermitian manifold for doubled space-time geometry}
The doubled space-time was introduced such that its local coordinate is
characterized by a pair of KK and winding coordinates $x^M = (x^{\mu},
\tilde{x}_{\mu})$.
It was proposed that this structure is naturally incorporated in a
para-Hermitian (K\"{a}hler) manifold~\cite{Vaisman:2012ke,Vaisman:2012px}.
The para-Hermitian structure is a basic ingredient 
to understand the doubled nature of space-time behind DFT.
In the following, we exhibit basic materials related to para-Hermitian
geometries \cite{Freidel:2017yuv, Freidel:2018tkj} and then discuss
algebroid structures realized in DFT.

Before discussing the para-Hermitian structure, we first define an
almost para-complex manifold.
\begin{definition}
An almost para-complex manifold $({\mathcal M}, K)$ is 
a differential manifold ${\mathcal M}$ with a vector bundle
 endomorphism $K: T{\mathcal M} \to T{\mathcal M}$ where $K^2 = +1$.
This $K$ is called the almost para-complex structure 
that satisfies the condition $\text{dim ker}(K + 1) = \text{dim ker}(K - 1)$.
\end{definition}
Obviously, the almost para-complex structure is a real analogue of the
almost complex structure $J^2 = -1$.
Given an almost para-complex structure $K$, the tangent bundle
$T{\mathcal M}$ is decomposed into the eigenbundles $L, \tilde{L}$
associated with the eigenvalues $K = \pm 1$.
This decomposition is performed via the projection operators $P, \tilde{P}$ that
map elements in $T{\mathcal M}$ to those in $L$ or $\tilde{L}$:
\begin{align}
P &= {1 \over 2} (1 + K), \qquad
\tilde{P} = {1 \over 2} (1 - K).
\end{align}
The subbundles $L,\tilde{L}$ are 
distributions
of $T\mathcal{M}$. We stress that the para-complex structure $K$
provides a natural decomposition of vectors in doubled space-time.

We now discuss the notion of integrability.
The integrability of a distribution is properly represented by the
Frobenius theorem. The Frobenius theorem is understood as a property of
vector fields. 
For any vector fields $X,Y \in \Gamma (L)$ where $L$ is a distribution, if their Lie bracket
$[X,Y]_L$ belongs to $L$, then the distribution $L$ is called involutive.
The Frobenius theorem states that a distribution $L$ (resp.\ $\tilde{L}$)
is completely integrable if and only if $L$ (resp.\ $\tilde{L}$) is
involutive.
When the eigenbundle $L$ (resp.\ $\tilde{L}$) defined on an almost
para-Hermitian manifold is involutive, then the tensors $N_P,
N_{\tilde{P}}$ defined in the following vanish:
\begin{align}
N_P (X,Y) &= \tilde{P} [P(X), P(Y)], \qquad
N_{\tilde{P}} (X,Y) = P [\tilde{P}(X), \tilde{P}(Y)],
\label{eq:Nijenhuis_onL}
\end{align}
where $X,Y \in \Gamma (T\mathcal{M})$.
We can define the Nijenhuis tensor associated with $K$ by adding the two tensors in \eqref{eq:Nijenhuis_onL}:
$N_K (X,Y) = N_P (X,Y) + N_{\tilde{P}} (X,Y)$.
This is again a real analogue of the Nijenhuis tensor defined on an ordinary
complex manifold:
\begin{align}
N_K(X,Y)
	= {1 \over 4} \big\{ [K(X), K(Y)] + [X,Y]
		- K \big( [K(X),Y] + [X,K(Y)] \big) \big\}.
\label{eq:Nijenhuis_onP}
\end{align}
The Nijenhuis tensor is a torsion on a (para-)complex manifold.
When $N_K$ vanishes, $K$ is integrable.
Then the definition of a para-complex manifold is given as follows:
\begin{definition}
When $K$ is integrable, namely, the Nijenhuis tensor $N_K$ vanishes
 identically, then an almost para-complex manifold $({\mathcal M}, K)$
 is a para-complex manifold.
\end{definition}
Contrast to the ordinary complex manifolds, the notion of integrability
for the two distributions $L$ and $\tilde{L}$ are totally independent
with each other.
Namely, the integrability of $L$ is defined through the condition $N_P
(X,Y) = 0$ for any $X,Y \in \Gamma (T\mathcal{M})$. This does not imply
$N_{\tilde{P}} = 0$ in general.
Since the integrability condition is independent for $L$ and
$\tilde{L}$, we can define a half-integrability in a para-complex
manifold \cite{Freidel:2017yuv,Freidel:2018tkj}:
\begin{definition}
An $L$-para-complex manifold is an almost para-complex manifold
 $({\mathcal M}, K)$ where only $L$ is integrable.
The same is true for $\tilde{L}$.
When the $L$-para-complex and the $\tilde{L}$-para-complex conditions are
 satisfied simultaneously, then $({\mathcal M}, K)$ is a para-complex manifold.
\end{definition}

We next define an almost para-Hermitian manifold by introducing a metric $\eta$:
\begin{definition}
An almost para-Hermitian manifold $({\mathcal M}, \eta, K)$ is an
almost para-complex manifold ${\mathcal M}$ equipped with a neutral
 metric $\eta : T{\mathcal M} \times T{\mathcal M} \to \mathbb{R}$ which satisfies the compatibility condition 
$\eta(K \cdot, K \cdot) = - \eta(\cdot,\cdot)$. $\eta$ is called the
 para-Hermitian metric.
\end{definition}
By its definition, the distribution $L$ is maximally isotropic with respect to $\eta$.
Namely, for any $X,Y \in \Gamma (L)$, since they are elements of the eigenbundle
with $K=1$, we have $\eta(X,Y) = 0$ for a para-Hermitian metric $\eta$.
The same is true even for $\tilde{L}$.
Since $\eta$ is neutral, it follows that $L$ and $\tilde{L}$ have the
same rank $D = {1 \over 2} {\rm dim\,} {\mathcal M}$.
Given an almost para-complex structure $K$ and a compatible metric
$\eta$, then we can define a non-degenerate 2-form $\omega = \eta K$.
This can be seen as an almost symplectic structure on $\mathcal{M}$ and it is not closed
in general $\dop \omega \neq 0$.
This means that an almost para-Hermitian manifold $({\mathcal M}, K, \eta)$ is
an almost symplectic manifold $({\mathcal M}, \omega)$ and vice-versa.
When $\omega$ is closed, $(\mathcal{M}, K, \eta)$ and $(\mathcal{M},
\omega)$ are said to be almost para-K\"{a}hler and symplectic,
respectively (see Table \ref{tb:integrable_vs_closed}.)
We note that a symplectic manifold is a Poisson manifold.
\begin{table}[tb]
\begin{center}
\begin{tabular}{c|c|c}
 & $\dop \omega \not= 0$ & $\dop \omega = 0$ \\
\hline
$N_K \not= 0$ & 
\begin{tabular}{c}
almost para-Hermitian \\
(almost symplectic) 
\end{tabular}
& 
\begin{tabular}{c}
almost para-K\"{a}hler \\ 
(symplectic) 
\end{tabular}
\\
\hline
$N_K = 0$ & 
\begin{tabular}{c}
para-Hermitian \\
(almost symplectic) 
\end{tabular}
& 
\begin{tabular}{c}
para-K\"{a}hler \\
(symplectic)
\end{tabular}
\end{tabular}
\caption{The integrability and closeness of $\omega$.}
\label{tb:integrable_vs_closed}
\end{center}
\end{table}
The compatibility between $\eta$ and $\omega$ results in that 
$L$ and $\tilde{L}$ are Lagrangian subbundles with respect to $\omega$.
Namely, for any $X,Y \in \Gamma (L)$ (resp.\ $\Gamma (\tilde{L})$), we have $\omega (X,Y) =
0$. 
We note that even for the case where $\omega$ is not closed, we can define a Lagrangian
subspace of $\omega$.
Given the almost structures, an analogue of a Hermitian manifold is defined:
\begin{definition}
When $({\mathcal M}, K)$ is an $L$-para-complex manifold, then 
an almost para-Hermitian manifold $({\mathcal M}, \eta, K)$ is an $L$-para-Hermitian manifold.
This is also the same for $\tilde{L}$.
An almost para-Hermitian manifold that satisfies both the $L$- and
 $\tilde{L}$-integrability conditions is a para-Hermitian manifold.
\end{definition}
The subbundles $L,\tilde{L}$ on a para-Hermitian manifold is therefore
Dirac structures. Namely, they are maximally isotropic with respect to
$\eta$ and involutive.

An alternative representation of the Frobenius theorem states that 
a subbundle $E \subset T \mathcal{M}$ is integrable if and only if
it is defined by a regular foliation of $\mathcal{M}$.
Namely, an integrable subbundle $E \subset T \mathcal{M}$ defines the
tangent bundle of a foliation $\mathcal{F}$ in $\mathcal{M}$.
Therefore when $L$ and $\tilde{L}$ are integrable, then
they have foliation structures:
\begin{align}
	L = T{\mathcal F}
	\qquad \mbox{and} \qquad
	\tilde{L} = T\tilde{\mathcal F}.
\end{align}
Here the foliation ${\mathcal F}$ (resp.\ $\tilde{\mathcal{F}}$) is given by the union of leaves $\coprod_{[p]} M_{[p]}$.
A leaf $M_p$ is a subspace of ${\mathcal F}$
(resp.\ $\tilde{\mathcal{F}}$) that pass through a point $p \in {\mathcal
M}$ and its tangent vectors are specified by $L$ (resp.\ $\tilde{L}$).
The index space in the union is the leaf space ${\mathcal M}/\mathcal{F}$ or $\mathcal{M}/\tilde{\mathcal{F}}$.
For $\mathcal{F}$, the local coordinate $x^{\mu}$ is given along a leaf $M_p$ while the one
for the transverse directions to leaves is $\tilde{x}_{\mu}$.
This means that $\tilde{x}_\mu$ is a constant on a leaf $M_p$ in $\mathcal{F}$.

The metric $\eta$ over ${\mathcal M}$ can be seen as a map $\eta:
T{\mathcal M} = L \oplus \tilde{L} \to T^*{\mathcal M} = L^* \oplus \tilde{L}^*$.
Then the metric $\eta$ defines the following two isomorphisms:
\begin{align}
	\phi^+ : \tilde{L} \to L^*
	\qquad \mbox{and} \qquad
	\phi^- : L \to \tilde{L}^*.
\end{align}
They map vectors in $\tilde{L}$ (resp.\ $L$) to forms in $L^*$ (resp.\ $\tilde{L}^*$).
The converse is also true. 
Given these isomorphisms, the following new isomorphisms are naturally defined:
\begin{align}
	\Phi^+ : T{\mathcal M} \to L \oplus L^*
	\qquad \mbox{and} \qquad
	\Phi^- : T{\mathcal M} \to \tilde{L} \oplus \tilde{L}^*.
\end{align}
In particular, the map $\Phi^+$ is utilized to relate DFT and
generalized geometry and it is called the natural isomorphism.

\subsection{Para-Dolbeault cohomology}
In this subsection, we define a para-Dolbeault cohomology
in $L,\tilde{L}$.
It is always true that there is a natural exterior algebra on the tangent bundle over an almost
para-complex manifold $\mathcal{M}$. 
We introduce the section of $\wedge^k T{\mathcal
M}$ (the totally anti-symmetric $k$-th tensor products of
$T\mathcal{M}$) and denote it as $\hat{\mathcal A}^k({\mathcal M})$.
Since $L,\tilde{L}$ are subbundles in $T\mathcal{M}$, we can
define exterior algebras in $\Gamma (L)$ and $\Gamma (\tilde{L})$.
If we define ${\mathcal A}^{r,s}({\mathcal M})$ as the section of
$(\wedge^r L) \wedge (\wedge^s \tilde{L})$, then, we obtain the
following decomposition:
\begin{align}
	\hat{\mathcal A}^k({\mathcal M}) = \bigoplus_{k=r+s} {\mathcal A}^{r,s}({\mathcal M}).
\end{align}
Here we have defined the canonical projection operator
$\pi^{r,s}: \hat{\mathcal A}^{r+s}({\mathcal M}) \to {\mathcal
A}^{r,s}({\mathcal M})$ that is induced by $P$ and $\tilde{P}$
 (see the explicit example in the next subsection).
We then define the exterior derivatives acting on $L$ and $\tilde{L}$:
\begin{align}
	\tilde{\dop} &:
		{\mathcal A}^{r,s}({\mathcal M}) \to {\mathcal A}^{r+1,s}({\mathcal M}), \\
	\dop &:
		{\mathcal A}^{r,s}({\mathcal M}) \to {\mathcal A}^{r,s+1}({\mathcal M}).
\end{align}
They are called the para-Dolbeault operators and have the following properties:
\begin{align}
	\dop^2 = 0, \qquad
	\tilde{\dop}^2 = 0, \qquad
	\dop \tilde{\dop} + \tilde{\dop} \dop = 0.
\end{align}
Due to the nilpotency of the para-Dolbeault operators, we can define the
para-Dolbeault cohomology. 
This is a real analogue of the Dolbeault cohomology defined in a complex manifold.
For any $A \in \Gamma (L)$, $\alpha \in \Gamma (\tilde{L})$, the
interior products $\iop_A$, $\tilde{\iop}_\alpha$ are defined:
\begin{align}
	\iop_A : {\mathcal A}^{r,s}({\mathcal M}) \to {\mathcal A}^{r-1,s}({\mathcal M})
	\qquad \mbox{and} \qquad
	\tilde{\iop}_\alpha : {\mathcal A}^{r,s}({\mathcal M}) \to {\mathcal A}^{r,s-1}({\mathcal M}).
	\label{eq:interior-prod-def}
\end{align}
By these operations, we define the Lie derivatives on $L$ and $\tilde{L}$:
\begin{align}
	{\mathcal L}_A \xi
	&= (\dop \iop_A + \iop_A \dop) \xi, \qquad
	\tilde{\mathcal L}_\alpha \xi
	= (\tilde{\dop} \tilde{\iop}_\alpha
		+ \tilde{\iop}_\alpha \tilde{\dop}) \xi.
	\label{eq:Lie-deriv-Dolbeault}
\end{align}
Here $A \in \Gamma(L)$, $\alpha \in \Gamma(\tilde{L})$, $\xi \in
{\mathcal A}^{r,s}({\mathcal M})$ are arbitrary (multi-)vectors.
By a para-Hermitian metric $\eta$, there is a natural $C^\infty({\mathcal
M}, \mathbb{R})$-bilinear map on ${\mathcal A}^{1,0}({\mathcal M}) \times {\mathcal A}^{0,1}({\mathcal M})$.
We call this the (symmetric) pairing. 
The pairing is denoted as $(\alpha, A) \mapsto \pairing{\alpha}{A}$.
This is an analogue of the inner products between vectors and forms on
$T\mathcal{M}$ and $T^* \mathcal{M}$. Here we note that $L$ and $\tilde{L}$
are not necessarily dual with each other. 
When $A \in {\mathcal A}^{r,0}({\mathcal M})$, $\alpha \in {\mathcal
A}^{0,s}({\mathcal M})$, $r \neq s$, then the pairing is given by $\pairing{\alpha}{A} = 0$.
In particular, for $\alpha \in {\mathcal A}^{0,s}({\mathcal M})$ and $A_1,
\ldots, A_s \in {\mathcal A}^{1,0}({\mathcal M})$, we write
\begin{align}
\pairing{\alpha}{A_1 \wedge \cdots \wedge A_s} = \alpha (A_1, \ldots, A_s).
\end{align}
Similarly, for $A \in {\mathcal A}^{r,0}({\mathcal M})$ and 
$\alpha_1, \ldots, \alpha_r \in {\mathcal A}^{0,1}({\mathcal M})$ we
write 
\begin{align}
\pairing{\alpha_1 \wedge \cdots \wedge \alpha_r}{A} 
= A (\alpha_1, \ldots, \alpha_r).
\end{align}
Now we express the interior products \eqref{eq:interior-prod-def} by
these quantities.
For $\alpha \in {\mathcal A}^{0,s}({\mathcal M})$, $\iop_A \alpha$ is an
element of ${\mathcal A}^{0,s-1}({\mathcal M})$. Therefore, for 
$A_1, \ldots, A_{s-1} \in {\mathcal A}^{1,0}({\mathcal M})$, it is
written as 
\begin{align}
	\iop_A \alpha (A_1, \ldots, A_{s-1}) 
	&= \alpha (A, A_1, \ldots, A_{s-1}).
\end{align}
Similarly, for $A \in {\mathcal A}^{r,0}({\mathcal M})$, 
$\tilde{\iop}_\alpha A$ is an element of ${\mathcal
A}^{r-1,0}({\mathcal M})$.
Therefore by $\alpha_1, \ldots, \alpha_{r-1} \in {\mathcal
A}^{0,1}({\mathcal M})$, it is written as 
\begin{align}
	\tilde{\iop}_\alpha A (\alpha_1, \ldots, \alpha_{r-1})
	&= A (\alpha, \alpha_1, \ldots, \alpha_{r-1}).
\end{align}
The interior product $\iop_A$ (resp.\ $\tilde{\iop}_\alpha$) is a
degree $-1$ derivation on the exterior algebras of $\tilde{L}$ (resp.\ $L$):
\begin{align}
	\iop_A (\alpha \wedge \beta) 
	&= (\iop_A \alpha) \wedge \beta + (-1)^s \alpha \wedge \iop_A \beta, 
	\notag \\
	\tilde{\iop}_\alpha (A \wedge B)
	&= (\tilde{\iop}_\alpha A) \wedge B + (-1)^r A \wedge \tilde{\iop}_\alpha B.
\end{align}
Here $\alpha \in {\mathcal A}^{0,s}({\mathcal M})$, $\beta \in {\mathcal A}^{0,\bullet}({\mathcal M})$, 
$A \in {\mathcal A}^{r,0}({\mathcal M})$, $B \in {\mathcal A}^{\bullet,0}({\mathcal M})$.

\subsection{Doubled aspects of Vaisman algebroid in DFT}
Now we discuss the algebroid structure governed by the {\sf C}-bracket
\eqref{eq:C-bracket} in DFT.
The doubled space-time on which DFT is defined is given by a
flat para-Hermitian manifold $\mathcal{M}$ whose local coordinate is
$x^{M}$ \cite{Freidel:2017yuv}.
The tangent space $T{\mathcal M}$ is spanned by $\partial_M$ ($M =
1,\ldots,2D$). Vector fields on
$T{\mathcal M}$ are decomposed by the projection operators $P,\tilde{P}$
defined by the para-complex structure $K$. Namely, for $\Xi = \Xi^M \del_M \in T \mathcal{M}$, we have 
\begin{align}
	\Xi^M \partial_M &= A^\mu(x,\tilde{x}) \partial_\mu
		+ \alpha_\mu(x,\tilde{x}) \tilde{\partial}^\mu,
\end{align}
where $A \in \Gamma(L)$, $\alpha \in \Gamma (\tilde{L})$. 
Here $x^M = (x^{\mu}, \tilde{x}_{\mu})$ is the induced decomposition of
the local coordinate on the base space $\mathcal{M}$.
Therefore $L$ is spanned by $\partial_\mu$ ($\mu = 1,\ldots, D$) while
$\tilde{L}$ is spanned by $\tilde{\partial}^\mu$ in the DFT framework.
In a flat para-Hermitian manifold, there is always a local frame where 
the para-Hermitian metric $\eta$ is expressed as 
\begin{align}
\eta_{MN} = 
\begin{pmatrix}
0 & 1 \\
1 & 0
\end{pmatrix}.
\label{eq:O(d,d)_metric}
\end{align}
Since this metric induces a map $L \to \tilde{L}$:
\begin{align}
\eta_{MN} A^N = A_M
\end{align}
as the way obvious with its index position, there is a natural isomorphism
between $\tilde{L}$ and $L^*$.
With this isomorphism at hand, we can identify these spaces.
We note that the metric \eqref{eq:O(d,d)_metric} implies that 
the inner product among $X,Y \in \Gamma (L)$ is $\langle X, Y \rangle =
0$ and the same is true even for $\tilde{L}$.
This means that $L$ and $\tilde{L}$ are maximally isotropic subbundles and $T
\mathcal{M} = L \oplus \tilde{L}$.

Given these structures, one can define the space of multi-vectors $\hat{\mathcal{A}}^k
(\mathcal{M})$ and the canonical projectors $\pi^{r,s}$.
The projectors are defined, for example, as follows.
The projectors in a para-complex manifold with $K = \mathrm{diag}(-1,+1)$,
in their apparent representation, are given by 
\begin{align}
P &= 
\begin{pmatrix}
0 & 0 \\
0 & 1
\end{pmatrix}
, \qquad
\tilde{P} =
\begin{pmatrix}
1 & 0 \\
0 & 0
\end{pmatrix}.
\end{align}
We consider the case of $r=2$, $s=0$. 
The component expression of an element $T \in \hat{\mathcal A}^2 (\mathcal{M})$ is denoted by
\begin{align}
T^{MN} &= 
\begin{pmatrix}
t_{\mu\nu} & t_\mu{}^\nu \\
t^\mu{}_\nu & t^{\mu\nu}
\end{pmatrix}.
\end{align}
The canonical projector $\pi^{2,0}$ defined through $P$ is given by 
\begin{align}
P^M{}_K T^{KL} P^N{}_L &= 
\begin{pmatrix}
0 & 0 \\
0 & t^{\mu\nu}
\end{pmatrix}.
\end{align}
Here, $t^{\mu\nu}$ is the element of ${\mathcal A}^{2,0}({\mathcal M})$. 
This implies $\pi^{2,0}(T^{MN}) = t^{\mu\nu}$.
The other projectors $\pi^{1,1}, \pi^{0,2}$ are defined similarly.

Now we define the Lie bracket on $L$. For $A,B \in \Gamma (L)$, this is given by 
\begin{align}
	[A,B]_L &= [A,B]_L^\mu \partial_\mu
	= (A^\nu \partial_\nu B^\mu - B^\nu \partial_\nu A^\mu) \partial_\mu.
\end{align}
Since this is the ordinary Lie bracket in differential geometry, it
satisfies the Jacobi identity trivially. 
It is obvious that it also satisfies the Leibniz rule.
With this bracket and the trivial bundle map $\rho_L = \mathrm{id}_L$ as the anchor,
then $L$ is endowed with a Lie algebroid structure.
Note that since $L$ is involutive with respect to $[\cdot,\cdot]_L$, it is integrable and define a Dirac structure on $T \mathcal{M}$.
As discussed in Section \ref{sect:DD}, by introducing multi-vectors, we
generalize the Lie bracket to the Schouten-Nijenhuis bracket.
An explicit realization of the Schouten-Nijenhuis bracket in DFT is as follows. 
Given a $k$-vector $A \in \Gamma (\wedge^k L)$,
\begin{align}
	A = {1 \over k!} A^{\mu_1 \cdots \mu_k} \partial_{\mu_1} \wedge \cdots \wedge \partial_{\mu_k},
\end{align}
we introduce the ``odd coordinate'' $\zeta_\mu := \partial_\mu$.
Then the $k$-vector is expressed as 
\begin{align}
	A = {1 \over k!} A^{\mu_1 \cdots \mu_k} \zeta_{\mu_1} \cdots \zeta_{\mu_k}.
\end{align}
Note that $\zeta_\mu$ can be treated as a Grassmann number whose
differential $\partial / \partial \zeta_\mu$ is defined by the right
derivative.
Namely, 
\begin{align}
	{\partial \over \partial \zeta_{\mu_n}}
		(\zeta_{\mu_1} \cdots \zeta_{\mu_n} \cdots \zeta_{\mu_k})
	&= (\zeta_{\mu_1} \cdots \zeta_{\mu_n} \cdots \zeta_{\mu_k})
		{\overleftarrow{\partial} \over \partial \zeta_{\mu_n}}
	= (-1)^{k-n} \zeta_{\mu_1} \cdots \check{\zeta}_{\mu_n} \cdots \zeta_{\mu_k}.
\end{align}
Here the symbol $\check{\zeta}_{\mu_n}$ stands for that $\zeta_{\mu_n}$
is removed. By using this $\zeta_\mu$ derivative, the Schouten-Nijenhuis
bracket is explicitly given by 
\begin{align}
	[A,B]_{\rm S} = \left( {\partial \over \partial \zeta_\mu} A \right) \partial_\mu B
		- (-1)^{(p-1)(q-1)}
			\left( {\partial \over \partial \zeta_\mu} B \right) \partial_\mu A.
\end{align}
Here $A \in \Gamma (\wedge^p L), B \in \Gamma (\wedge^q L)$.
The discussion is totally parallel in $\tilde{L}$.
The same definition holds for $[\cdot,\cdot]_{\rm S}^*$ on $\tilde{L}$ where $\zeta_{\mu} =
\del_{\mu}$ is replaced by $\zeta^{* \mu} = \tilde{\del}^{\mu}$.
One can show that this expression satisfies the definition of the
Schouten-Nijenhuis bracket discussed in Section \ref{sect:DD}.
It is known that multi-vectors on a manifold define a Gerstenhaber
algebra by the Schouten-Nijenhuis bracket \cite{Tulczyjew}.
By the Vaintrob theorem~\cite{Vaintrob:1997}, a Lie algebroid structure
over a vector bundle $V \to M$ and a Gerstenhaber algebra over
multi-vectors $\Gamma(\wedge^\bullet V)$ are equivalent.

The symmetric pairing $\pairing{\alpha}{A}$ is defined, for example, 
\begin{align}
\alpha (A_1, \cdots, A_s) = \alpha_{\mu_1 \cdots \mu_s} A_1^{\mu_1}
 \cdots A_s^{\mu_s},
\end{align}
and so on. Since $\tilde{L}$ and $L^*$ are identified via the natural
isomorphism, the symmetric pairing $\pairing{\cdot}{\cdot}$ is the inner
product $\langle \cdot, \cdot \rangle$ in disguise.
A Lie algebroid coboundary operator that maps a $k$-vector to
a $(k+1)$-vector is given by the para-Dolbeault operator 
$\tilde{\dop} : \wedge^k L \to \wedge^{k+1} L$.
This is characterized by the following general relation:
\begin{align}
	\tilde{\dop}X(\alpha_1, \ldots, \alpha_{k+1})
	&= \sum_{i=1}^{k+1} (-1)^{i+1} \rhotilde (\alpha_i) \cdot
		(X(\alpha_1, \ldots, \check{\alpha}_i, \ldots, \alpha_{k+1})) \notag \\
	&\quad + \sum_{i<j} (-1)^{i+j} X(
		\dualSchouten{\alpha_i}{\alpha_j}, \alpha_1,
		\ldots, \check{\alpha}_i, \ldots, \check{\alpha}_j, \ldots, \alpha_{k+1}).
	\label{eq:def-dtilde}
\end{align}
Here $X \in \Gamma (\wedge^k L)$, $\alpha_i \in \Gamma (\tilde{L})$ and 
the symbol $\check{\alpha}_i$ stands for that the $i$-th $\alpha$ is
removed. 
The bracket $\dualSchouten{\cdot}{\cdot}$ and the anchor $\rhotilde$ is defined on $\tilde{L}$.
In particular, using the local coordinate, we find that the action of $\tilde{\dop}$ on a
$k$-vector $X$ is explicitly given by 
\begin{align}
	\tilde{\dop} X = {1 \over k!} \tilde{\partial}^\mu
		X^{\nu_1 \cdots \nu_k}(x,\tilde{x})
		\partial_\mu \wedge \partial_{\nu_1} \wedge \cdots \wedge \partial_{\nu_k}.
	\label{eq:winding-extderiv}
\end{align}
We confirm that this definition of $\tilde{\dop}$ is compatible with the bracket
$\dualSchouten{\cdot}{\cdot}$. By the definition of $\tilde{\dop}$
\eqref{eq:def-dtilde}, for $k=1$, $A \in \Gamma (L)$, $\alpha_1,
\alpha_2 \in \Gamma (\tilde{L})$ we have 
\begin{align}
	\tilde{\dop}A(\alpha_1, \alpha_2)
	&= (-1)^{2} \rhotilde (\alpha_1) \cdot (A(\alpha_2))
	 	+ (-1)^{3} \rhotilde (\alpha_2) \cdot (A(\alpha_1))
		+ (-1)^{3} A(\dualSchouten{\alpha_1}{\alpha_2}) \notag\\
	&= \rhotilde (\alpha_1) \cdot (A(\alpha_2))
	 	- \rhotilde (\alpha_2) \cdot (A(\alpha_1))
		- A(\dualSchouten{\alpha_1}{\alpha_2}).
\end{align}
Then in the DFT realization, since
$\rhotilde (\alpha_1) = \alpha_{1\mu} \tilde{\partial}^\mu$,
we have 
\begin{align}
	A(\dualSchouten{\alpha_1}{\alpha_2})
	&= \rhotilde (\alpha_1) \cdot (A(\alpha_2)) - \rhotilde (\alpha_2) \cdot (A(\alpha_1))
		- \tilde{\dop}A(\alpha_1, \alpha_2) \notag\\
	&= \alpha_{1\mu} \tilde{\partial}^\mu (A^\nu \alpha_{2\nu})
		- \alpha_{2\nu} \tilde{\partial}^\nu (A^\mu \alpha_{1\mu})
		- (\tilde{\partial}^\mu A^\nu - \tilde{\partial}^\nu A^\mu) \alpha_{1\mu} \alpha_{2\nu} \notag\\
	&= A^\mu (\alpha_{1\nu} \tilde{\partial}^\nu \alpha_{2\mu}
		- \alpha_{2\nu} \tilde{\partial}^\nu \alpha_{1\mu}).
\end{align}
Therefore we find that the exterior derivative $\tilde{\dop}$ on $L$ and the
bracket $\dualSchouten{\cdot}{\cdot}$ is compatible.
The same discussion holds also for the operator $\dop$ on the Lie algebroid $\tilde{L}$.

We next derive the Lie derivative in DFT.
For $A,B \in {\mathcal A}^{1,0}({\mathcal M})$ and $\alpha,\beta \in {\mathcal A}^{0,1}({\mathcal M})$, the
interior products (or the symmetric pairing) are realized as follows:
\begin{align}
	\iop_A \beta &= A^\mu \beta_\mu, \qquad 
	\iop_A \dop \beta 
	= (A^\nu \partial_\nu \beta_\mu - A^\nu \partial_\mu \beta_\nu) \tilde{\partial}^\mu, 
	\notag \\
	\tilde{\iop}_\alpha B &= \alpha_\mu B^\mu, \qquad
	\tilde{\iop}_\alpha \tilde{\dop} B
	= (\alpha_\nu \tilde{\partial}^\nu B^\mu - \alpha_\nu
 \tilde{\partial}^\mu B^\nu) \partial_\mu.
\label{eq:interior_product}
\end{align}
The Lie derivative defined in \eqref{eq:Lie-deriv-Dolbeault} is
therefore given by
\begin{align}
	{\mathcal L}_A \beta &= (\dop \iop_A + \iop_A \dop) \beta \notag\\
	&= \dop(\iop_A \beta) + \iop_A (\dop \beta)
	= \dop(A^\nu \beta_\nu)
		+ \iop_A (\partial_\mu \beta_\nu \tilde{\partial}^\mu \wedge \tilde{\partial}^\nu) \notag\\
	&= [(\partial_\mu A^\nu) \beta_\nu + A^\nu \partial_\mu \beta_\nu] \tilde{\partial}^\mu
		+ A^\mu \partial_\mu \beta_\nu \tilde{\partial}^\nu
		- A^\nu \partial_\mu \beta_\nu \tilde{\partial}^\mu \notag\\
	&= (A^\nu \partial_\nu \beta_\mu + \beta_\nu \partial_\mu A^\nu) \tilde{\partial}^\mu.
\end{align}
Similarly we have
\begin{align}
	\tilde{\mathcal L}_\alpha B
	&= (\tilde{\dop} \tilde{\iop}_\alpha
		+ \tilde{\iop}_\alpha \tilde{\dop}) B
	= \tilde{\dop} (\alpha_\nu B^\nu)
		+ \tilde{\iop}_\alpha (\tilde{\partial}^\mu B^\nu \partial_\mu \wedge \partial_\nu) \notag\\
	&= [(\tilde{\partial}^\mu \alpha_\nu) B^\nu
		+ \alpha_\nu \tilde{\partial}^\mu B^\nu] \partial_\mu
		+ \alpha_\mu \tilde{\partial}^\mu B^\nu \partial_\nu
		- \alpha_\nu \tilde{\partial}^\mu B^\nu \partial_\mu \notag\\
	&= (\alpha_\nu \tilde{\partial}^\nu B^\mu + B^\nu \tilde{\partial}^\mu \alpha_\nu)
	 	\partial_\mu.
\end{align}
We have consistently defined the Lie algebroid 
$(\wedge^\bullet L, [\cdot,\cdot]_{\rm S}, \dop)$ 
and its dual Lie algebroid 
($\wedge^\bullet \tilde{L}$, $[\cdot, \cdot]_{{\rm S}}^*$, $\tilde{\dop}$) in DFT. 

We are now in a position to discuss doubled structures of $(L,\tilde{L})$.
As we have discussed in Section \ref{sect:DD}, a 
Lie bialgebroid is defined by a Lie algebroid $(L, [\cdot,\cdot]_L,
\rho_L, \dop)$ and its dual Lie coalgebroid $ (L^*,
[\cdot,\cdot]_{L^*}, \rho_{L^*}, \dop_*)$ together with a
compatibility condition between them called the derivation condition
\eqref{eq:derivation}.
Again, this is given by 
\begin{align}
	\dop_* [X,Y]_{\mathrm{S}} &= [\dop_* X, Y]_{\mathrm{S}} + [X, \dop_* Y]_{\mathrm{S}},
	\qquad X,Y \in \Gamma (\wedge^{\bullet} L),
\label{eq:derivation_condition_DFT}
\end{align}
where $\dop: \wedge^k L^* \to \wedge^{k+1} L^*$ and $\dop_*
: \wedge^k L \to \wedge^{k+1} L$ are exterior derivatives defined above.
Now we examine the derivation condition in DFT by the explicit calculations. 
It is enough to show for $A, B \in
\Gamma (L)$. The left hand side of \eqref{eq:derivation_condition_DFT}
is given by 
\begin{align}
	\tilde{\dop} [A, B]_{\rm S}
	&= \tilde{\partial}^\mu [A, B]_{\rm S}^\nu
 \partial_\mu \wedge \partial_\nu 
\notag
\\
	&= \tilde{\partial}^\mu (A^\rho \partial_\rho B^\nu - B^\rho \partial_\rho A^\nu)
		\partial_\mu \wedge \partial_\nu 
\notag
\\
	&= (\tilde{\partial}^\mu A^\rho \partial_\rho B^\nu
		+ A^\rho \partial_\rho \tilde{\partial}^\mu B^\nu
		- \tilde{\partial}^\mu B^\rho \partial_\rho A^\nu
		- B^\rho \partial_\rho \tilde{\partial}^\mu A^\nu)
		\partial_\mu \wedge \partial_\nu,
\end{align}
while the right hand side is calculated by using the explicit form of
the Schouten-Nijenhuis bracket:
\begin{align}
	[\tilde{\dop} A, B]_{\rm S}
	&= \left( {\partial \over \partial \zeta_\rho} \tilde{\dop} A \right)
		\partial_\rho B
		- (-1)^0 \left( {\partial \over \partial \zeta_\rho} B \right)
		\partial_\rho \tilde{\dop} A 
\notag
\\
	&= (\tilde{\partial}^\mu A^\rho \zeta_\mu - \tilde{\partial}^\rho A^\mu \zeta_\mu)
		\partial_\rho B^\nu \zeta_\nu
		- B^\rho \partial_\rho \tilde{\partial}^\mu A^\nu
 \zeta_\mu \zeta_\nu 
\notag
\\
	&= (\tilde{\partial}^\mu A^\rho \partial_\rho B^\nu
		- \tilde{\partial}^\rho A^\mu \partial_\rho B^\nu
		- B^\rho \partial_\rho \tilde{\partial}^\mu A^\nu)
 \partial_\mu \wedge \partial_\nu, 
\notag \\
	[A, \tilde{\dop} B]_{\rm S}
	&= -[\tilde{\dop} B, A]_{\rm S} 
\notag
\\
	&= - (\tilde{\partial}^\mu B^\rho \partial_\rho A^\nu
		- \tilde{\partial}^\rho B^\mu \partial_\rho A^\nu
		- A^\rho \partial_\rho \tilde{\partial}^\mu B^\nu) \partial_\mu \wedge \partial_\nu.
\end{align}
From these expressions, we obtain
\begin{align}
	\tilde{\dop} [A, B]_{\rm S}
	&= [\tilde{\dop} A, B]_{\rm S}
		+ [A, \tilde{\dop} B]_{\rm S}
		+ (\tilde{\partial}^\rho A^\mu \partial_\rho B^\nu
			+ \tilde{\partial}^\rho B^\nu \partial_\rho
 A^\mu) \partial_\mu \wedge \partial_\nu.
\label{eq:DFT_derivation_violation}
\end{align}
The last contribution represents the violation of the derivation
condition \eqref{eq:derivation_condition_DFT}.
We have then explicitly shown that given the Lie algebroid structures
$L$ and $\tilde{L} \simeq L^*$ in DFT, they do not form a Lie bialgebroid in
general.
Although this is true, 
following the general discussion in Section \ref{sect:Vaisman},
the double $L \oplus L^*$ defines a Vaisman algebroid.
The anchor in the Vaisman algebroid is defined as $\rho_{\rm V} = \rho_L +
\rho_{L^*}$ while the bilinear form $\bbra{\Xi_1, \Xi_2}$ for $\Xi_i \in
\Gamma (T\mathcal{M})$ is given by 
\begin{align}
\bbra{\Xi_1, \Xi_2} = \bbra{A + \alpha, B + \beta} = 
\frac{1}{2} 
\Bigl\{
\pairing{\alpha}{B} + \pairing{\beta}{A}
\Bigr\}.
\end{align}
Here $\pairing{\cdot}{\cdot}$ is the symmetric pairing defined before.
The differential operator is defined as $\mathcal{D} = \dop + \tilde{\dop}$.
By using the Lie brackets $[\cdot, \cdot]_L$, $[\cdot, \cdot]_{L^*}$, Lie
derivatives $\mathcal{L}_A$, $\tilde{\mathcal{L}}_{\alpha}$ and
operators $\dop, \iop, \tilde{\dop}, \tilde{\iop}$, we define the
Vaisman bracket for vectors $\Xi_i \in \Gamma (T \mathcal{M})$:
\begin{align}
[\Xi_1,\Xi_2]_{\rm V}
= [A + \alpha, B + \beta]_{\rm V}
&=  [A,B]_{L}
	+ {\mathcal L}_A \beta - {\mathcal L}_B \alpha
	- {1 \over 2} \dop (\iop_A \beta - \iop_B \alpha)
\notag \\
& \quad
	+ [\alpha,\beta]_{\tilde{L}}
	+ \tilde{\mathcal L}_\alpha B - \tilde{\mathcal L}_\beta A
	- {1 \over 2} \tilde{\dop} (\tilde{\iop}_\alpha B - \tilde{\iop}_\beta A),
\label{eq:DFT_V_bracket}
\end{align}
This is nothing but the {\sf C}-bracket \eqref{eq:C-bracket}.
The quadruple $(L \oplus \tilde{L}, [\cdot, \cdot]_{\sf C}, \rho_{\rm V},
\bbra{\cdot,\cdot})$ then defines a Vaisman algebroid.

We note that the last term in \eqref{eq:DFT_derivation_violation} is rewritten as 
\begin{align}
	\tilde{\partial}^\rho A^\mu \partial_\rho B^\nu
		+ \tilde{\partial}^\rho B^\nu \partial_\rho A^\mu
	&= \eta^{KL} \partial_K A^\mu \partial_L B^\nu.
\end{align}
It is obvious that this vanishes when the strong constraint is imposed.
This means that the derivation condition between $L$ and $\tilde{L}$ is
satisfied and $(L,\tilde{L})$ becomes a Lie bialgebroid when the strong
constraint is imposed and the gauge transformation parameters are
restricted \cite{Hull:2009zb}.
In this case, the double $L \oplus \tilde{L}$ defines a Courant
algebroid following the general discussions \cite{LiWeXu, Deser:2014mxa}.
This completely agrees with the analysis in
\cite{Chatzistavrakidis:2018ztm} where the pre-DFT algebroid (Vaisman
algebroid) becomes a Courant algebroid after imposing the strong
constraint.
We again stress that an algebraic origin of the strong constraint is the
derivation condition that is a compatibility condition between $L$ and
$\tilde{L}$ which allows them to be a Lie bialgebroid.

\subsection{Gauge symmetries, foliations and generalized geometry}
In this subsection, we discuss the gauge symmetries associated with
$L,\tilde{L}$ and the relation to generalized geometry. 
As discussed in \cite{Vaisman:2012px, Svoboda:2018rci}, the structure of
the {\sf C}-bracket in DFT naturally arises as a Vaisman bracket on a para-Hermitian geometry.
The {\sf C}-bracket is recognized as a T-duality covariantized 
Lie bracket-like structure that
accommodates the diffeomorphism and the $B$-field gauge symmetry algebra in the NSNS
sector of supergravity.
The geometric realization of the {\sf C}-bracket does not necessarily require the
strong constraint.
In this sense, the {\sf C}-bracket governs the ``off-shell'' gauge
symmetry of DFT (a symmetry without the strong constraint).
Due to the para-complex structure underlying the doubled space-time
$\mathcal{M}$, there is a natural decomposition of the tangent bundle $T
\mathcal{M} = L \oplus \tilde{L}$ in which Lie algebroid structures are
found.
Since the distributions $L,\tilde{L}$ are Dirac structures and therefore
are integrable, they are given by tangent bundles of
foliations $\mathcal{F}, \tilde{\mathcal{F}}$ in $\mathcal{M}$.
A physical space-time is therefore
identified as a leaf defined by $\tilde{x}_\mu = \mathrm{const.}$ in a
para-Hermitian manifold.
With the natural isomorphism induced by an inner product defined by the
metric $\eta$, the vector components in $\tilde{L} = T \mathcal{F}
\simeq L^* = T^* \mathcal{F}$ is identified with 1-forms over a leaf in $\mathcal{M}$.
Therefore, one can understand that the Lie bracket $[\cdot,\cdot]_L$ 
over $L$ governs the diffeomorphism parametrized by vector gauge
parameters $\xi_i^{\mu}$ while the bracket $[\cdot,\cdot]_{\tilde{L}}$ over $\tilde{L}$ represents the $B$-field 
gauge symmetry parametrized by 1-forms $\tilde{\xi}_{i,\mu}$.
Since the Lie bracket for the 1-forms
$[\tilde{\xi}_1,\tilde{\xi}_2]_{\tilde{L}}$ is generically non-zero, 
the T-duality covariantized $B$-field gauge symmetry is effectively
enhanced to non-Abelian ``off-shell''.

\begin{figure}[tb]
\begin{center}
\begin{tikzpicture} 
%
\draw (0bp,100bp) node[above right] {{\sf C}-bracket};
\draw[->,thick] (15bp,95bp) -- (15bp,20bp);
\draw (15bp,85bp) node[right] {\footnotesize strong constraint $\tilde{\partial} * = 0$};
\draw (0bp,0bp) node[above right] {c-bracket};
%
%
\draw[dashed] (130bp,0bp) -- (130bp,120bp);
%
%
\draw (140bp,100bp) node[above right] {Vaisman bracket};
\draw[->,thick] (155bp,95bp) -- (155bp,70bp);
\draw (155bp,85bp) node[right] {\footnotesize derivation condition \eqref{eq:derivation}};
\draw (140bp,50bp) node[above right] {Courant bracket in 
\eqref{eq:Courant_bracket}
};
\draw[->,thick] (155bp,45bp) -- (155bp,20bp);
\draw (155bp,35bp) node[right] {\footnotesize $[\cdot,\cdot]_* = 0$};
\draw (140bp,0bp) node[above right] {c-bracket};
%
\draw[<->,thick,red] (120bp,105bp) -- (140bp,105bp);
\draw[<->,thick,red] (120bp,5bp) -- (140bp,5bp);
\end{tikzpicture}
\caption{Paths to the c-bracket in DFT and Vaisman algebroids.}
\label{fig:sc_dc}
\end{center}
\end{figure}
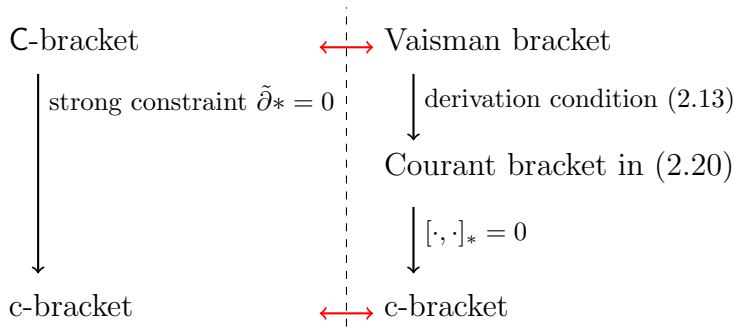

Upon the imposition of the strong constraint, the gauge algebra is closed by
the {\sf C}-bracket.
Therefore in order that the algebra 
given by the {\sf C}-bracket generates a symmetry, the strong constraint is necessarily
satisfied, ether implicit or explicitly.
A way to solve the strong constraint trivially is to make the winding
derivative be vanishing $\tilde{\partial} *= 0$.
In this case, the bracket including 1-forms vanish $[\tilde{\xi}_1,
\tilde{\xi}_2]_{\tilde{L}} = 0$ and the {\sf C}-bracket is reduced to
the c-bracket defined in \eqref{eq:c-bracket}.
This means that by imposing the section condition on any DFT fields and
gauge parameters and making the theory be ``on-shell'' 
({\it i.e.}\ defined on a physical subspace), the non-Abelian ``off-shell'' $B$-field
gauge symmetry becomes an Abelian symmetry ``on-shell''.
In this sense, the c-bracket is an ``on-shell'' counterpart of the {\sf C}-bracket.
From a mathematical point of view, the c-bracket is obtained by first
imposing the derivation condition \eqref{eq:derivation} on the
Vaisman bracket and then make the Lie bracket on $\tilde{L}$ be a zero-bracket
$[\cdot,\cdot]_{\tilde{L}} = 0$ (see Fig \ref{fig:sc_dc}).
As we have explicitly shown, with the adaptation of the derivation
condition, $(L,\tilde{L})$ forms a Lie bialgebroid.
Through the prescription by Liu-Weinstein-Xu~\cite{LiWeXu}, the
c-bracket defines a Courant algebroid.
This c-bracket is nothing but the original Courant bracket appeared in
generalized geometry \cite{Hitchin:2004ut}.

Given a para-Dolbeault cohomology, the ``on-shell'' fields and
 gauge parameters in DFT satisfying the strong constraint are
 characterized by para-holomorphic quantities defined by the para-Dolbeault
operators:
\begin{align}
\mbox{para-holomorphic}: \qquad & 
\tilde{\dop} \Phi = 0,
\end{align}
where $\Phi$ is any doubled fields and gauge parameters.
This is equivalent to say that the para-holomorphic quantities are restricted in leaves
in the foliation ${\mathcal F}$.
We note that this is not the unique solution to the strong constraint.
The other possibility
\begin{align}
\mbox{anti-para-holomorphic}: \qquad & \dop \Phi = 0,
\end{align}
also satisfies the strong constraint trivially.
The anti-para-holomorphic quantities are defined along the transverse
directions to leaves. 
Namely, they live in the winding space defined by $x^{\mu} =
\text{const}$. 
We note that this kind of winding dependent space-time actually appears in
solutions to DFT \cite{Kimura:2018hph}.

\section{Conclusion and discussions} \label{sect:conclusion}
In this paper, we studied the doubled aspects of the Vaisman
algebroid which governs the gauge symmetry in double field theory.
In the first half of the paper, we studied the Lie algebroid
and its doubled structures behind Vaisman algebroids from a mathematical viewpoint.
A Vaisman algebroid is obtained through an analogue of the Drinfel'd
double of a pair of Lie algebroids $(E,E^*)$. 
The exterior algebras based on the Schouten-Nijenhuis bracket are naturally defined in each $E,E^*$.
Due to the failure of the derivation condition of the exterior
derivative on the Schouten-Nijenhuis bracket, the pair $(E,E^*)$ does not form a
Lie bialgebroid in general.
Although $(E,E^*)$ fails to be a Lie bialgebroid, 
we show that their double $L \oplus L^*$, with appropriate definition of
a bracket, satisfies two axioms for Vaisman algebroids.
This is a weakened construction of the Drinfel'd double of Lie bialgebroids
for Courant algebroids \cite{LiWeXu}.
Indeed, when the derivation condition is imposed, the bracket
$[\cdot,\cdot]_{\rm V}$ satisfies three
additional axioms which are necessary for Courant algebroids.
We then consider Dirac structures, namely, maximally isotropic
integrable subbundles $L, \tilde{L} \simeq L^*$ of a Vaisman algebroid. 
Even though they are Lie algebroids, the derivation condition of the
exterior derivatives does not follow in general and $(L,L^*)$ does not
have Lie bialgebroid structure.

In the latter half of the paper, we investigated doubled structures
 in the gauge symmetry of DFT. 
The symmetry is generated by the {\sf C}-bracket on the doubled space-time.
We introduce the doubled space-time as a flat para-Hermitian manifold $\mathcal{M}$ with
dimension $2D$. By the para-complex structure $K$, the tangent bundle $T\mathcal{M}$ is
decomposed into the two eigenbundles $T\mathcal{M} = L \oplus \tilde{L}$
 associated with $K = \pm 1$.
Subsequently, the doubled vector $\Xi = \Xi^M \del_M \in \Gamma (T \mathcal{M})$ is separated into
 two parts $\Xi = A^{\mu} \del_{\mu} + \alpha_{\mu} \tilde{\del}^{\mu}$
 where $A \in \Gamma (L),\alpha \in \Gamma (\tilde{L})$. 
We discussed natural Lie algebroid structures on $L,\tilde{L}$.
With the Schouten-Nijenhuis bracket, they are extended into a Gerstenhaber algebra.
We then examined the exterior algebras and the para-Dolbeault cohomology on $L,\tilde{L}$.
We showed that the dual exterior derivative
$\tilde{\dop}$ does not satisfy the derivation condition on the Lie bracket $[\cdot,\cdot]_L$ in $L$. 
By the general discussion in Section \ref{sect:DD}, therefore, $(L,\tilde{L})$ never define a Lie
bialgebroid. 
Instead, the double $L \oplus \tilde{L}$ defines a Vaisman algebroid and
the {\sf C}-bracket is identified with the Vaisman bracket.
We showed that the failure of the derivation condition is resolved by
imposing the strong constraint.
With these results at hand, we found an algebraic origin of
the strong constraint.
Namely, it is an efficient condition for the derivation
condition that ensures that $(L,\tilde{L})$ becomes a Lie bialgebroid.

Based on these structures, we discussed physical aspects of the symmetry
generated by the {\sf C}-bracket.
The gauge transformation of the NSNS $B$-field should be effectively
non-Abelian due to the $O(D,D)$ covariantization of the symmetry.
The integrability condition of $L,\tilde{L}$
implies that they are given by foliations $\mathcal{F},
\tilde{\mathcal{F}}$ of $\mathcal{M}$, $L = T \mathcal{F}, \tilde{L} = T
\tilde{\mathcal{F}}$. 
Geometrically, this results in the fact that the base space of
$L$ is given by leaves determined by $\tilde{x}_{\mu} = \text{const}$. 
The physical space-time is a slice of doubled space-time
whose winding coordinates are fixed values.
With the foliated structure and the natural isomorphism $\tilde{L} \to
L^*$, the part of the doubled vectors $\alpha = \alpha_{\mu} \tilde{\del}^{\mu}$ is identified with 1-forms $\alpha =
\alpha_{\mu} {\dop}x^{\mu}$ over a leaf in $\mathcal{M}$.
A trivial way to solve the strong constraint is 
to consider all the DFT fields that take values in the (anti)holomorphic quantities.
Then the Vaisman algebroid is reduced to a Courant algebroid where the
$B$-field gauge field is realized as an Abelian symmetry.

In this paper, we have worked with a para-Hermitian geometry.
There are independent but maybe equivalent approaches based on
graded geometries \cite{Deser:2014mxa, Deser:2018oyg, Crow-Watamura:2018liw} which is traced back to an equivalence between Courant
algebroids and QP-manifolds \cite{Roytenberg:2002nu}.
It would be interesting to study the structure of the {\sf C}-bracket
with the derived brackets \cite{Kosmann-Schwarzbach2,Deser:2018oyg}.

Although the strong constraint is needed for the closure of the gauge
algebra in DFT and it makes a Vaisman algebroid be a Courant algebroid,
we stress that the strong constraint is not necessary in more general
setups \cite{Grana:2012rr}. This implies that Vaisman algebroids
would play important roles in applications of DFT.
For example, DFT in a group manifold is a key ingredient for studies on
Poisson-Lie T-duality \cite{Hassler:2017yza}.
Indeed, the Drinfel'd double and Courant algebroids play significant roles to
understand the Poisson-Lie T-duality in string theory
\cite{Lust:2018jsx, Marotta:2018swj, Severa:2018pag,Demulder:2018lmj}.

Finally, we comment on the gauge symmetry in DFT.
As we mentioned in the introduction, the gauge symmetry in DFT is
characterized by the {\sf C}-bracket.
This is in fact true for the infinitesimal gauge transformations.
The nature of the finite gauge transformations in DFT has been studied in various viewpoints \cite{Hohm:2012mf, Hohm:2012gk, Park:2013mpa, Berman:2014jba, Cederwall:2014kxa, Hull:2014mxa, Naseer:2015tia, Rey:2015mba}.
Mathematically, the finite gauge transformation in DFT is governed by an ``integrated'' version of Vaisman and Courant algebroids.
This is analogous to the fact that the infinitesimal object of a Lie group is
given by a Lie algebra ({\it cf.}\ the Lie's third theorem).
Indeed, it is known that a Lie bialgebra is the infinitesimal counterpart of
a Poisson-Lie group~\cite{Drinfeld:1983ky}. 
Similarly, a Lie algebroid and a Lie bialgebroid are infinitesimal objects of a
Lie and a Poisson groupoids \cite{Pradines, Mackenzie}.
It is also discussed that a kind of the Courant algebroid is an infinitesimal object of a groupoid \cite{Severa}.
It is therefore interesting to study the integration of Vaisman and Courant 
algebroids \cite{Severa2} from mathematical and physical viewpoints. 
These kind of issues are known as the ``coquecigrue problem'' -- finding
an imaginary creature that appears in the famous texts Gargantua and Pantagruel.
This was first proposed by J.~L.~Loday as an analogue of the Lie's third
fundamental theorem for Lie groups.
The doubled structure discussed in this paper would provide a
mathematical foundation towards the coquecigrue problem for gauge
symmetry in DFT.
We believe that revealing a geometric origin of DFT gauge symmetry is important to understand the stringy
 winding effects to space-times \cite{Kimura:2013fda, Berkeley:2014nza,
 Berman:2014jsa, Bakhmatov:2016kfn, Kimura:2018hph}.
We will come back to these issues in future works.

\subsection*{Acknowledgments}
The work of S.S.\ is supported by the Japan Society for the Promotion of
Science (JSPS) KAKENHI Grant Number {\tt JP17K14294}.
The work of K.S.\ is supported by the Sasakawa Scientific Research Grant (No.\ 2019-2027) from the Japan Science Society.

\begin{appendix}

\section{A quick introduction to DFT} \label{sect:DFT_review}
Double Field Theory (DFT) \cite{Hull:2009mi} is an effective theory of string theory that realizes T-duality manifestly.
The T-duality transformation is expressed as an $O(D,D)$ rotation in $2D$-dimensional doubled space-time while 
the physical space-time is $D$-dimensional.
The doubled space-time is characterized by the coordinate:
\begin{align}
x^M &= 
\begin{pmatrix}
\tilde{x}_\mu \\ x^\mu
\end{pmatrix}
\qquad (M = 1, \ldots, 2D; \; \mu = 1, \ldots, D).
\end{align}
Here $x^\mu$ is the coordinate of a conventional space-time probed by particles.
The coordinate $\tilde{x}_\mu$, called the winding coordinate, is the Fourier dual of the winding modes of strings.
The doubled space-time inherits the $O(D,D)$ invariant metric and its inverse:
\begin{align}
	\eta_{MN} &= 
		\begin{pmatrix}
		0 & \delta^\mu{}_\nu \\
		\delta_\mu{}^\nu & 0
		\end{pmatrix}
	, \qquad
	\eta^{MN} = 
		\begin{pmatrix}
		0 & \delta_\mu{}^\nu \\
		\delta^\mu{}_\nu & 0
		\end{pmatrix}.
\end{align}
All the indices of $O(D,D)$ tensors are raised and lowered by
$\eta_{MN}$ and $\eta^{MN}$.

The dynamical objects in DFT is the so-called generalized metric ${\mathcal H}_{MN}$ and the generalized dilaton $d$ defined by 
\begin{align}
{\mathcal H}_{MN} &= 
\begin{pmatrix}
g^{\mu\nu} & -g^{\mu\rho} B_{\rho\nu} \\
B_{\mu\rho} g^{\rho\nu} & g_{\mu\nu} - B_{\mu\rho} g^{\rho\sigma} B_{\sigma\nu}
\end{pmatrix}
, \qquad
e^{-2d} = \sqrt{-g} e^{-2\phi}.
\end{align}
All the component fields $(g, B, \phi)$ depend on the doubled coordinate $x^M$.
Here $g_{\mu \nu}, g^{\mu\nu}$ correspond to the metric of the physical space-time when it is restricted to the space spanned by $x^\mu$.
Similarly, $B_{\mu\nu}$, $\phi$ correspond to the Kalb-Ramond $B$-field
and the dilaton in type II supergravities, respectively.
The inverse of $\mathcal{H}_{MN}$ satisfies the following relation, 
\begin{align}
{\mathcal H}^{MN} &= \eta^{MK} \eta^{NL} {\mathcal H}_{KL}.
\end{align}
The generalized metric ${\mathcal H}_{MN}$ parametrized the coset space $O(D,D)/(O(D) \times O(D))$.
The $O(D,D)$ invariant DFT action is given by~\cite{Hohm:2010pp}
\begin{align}
S_{\rm DFT} &= \int \dop^{2D} x  \; e^{-2d} {\mathcal R}({\mathcal H}, d), 
\label{eq:DFTaction} \\
{\mathcal R} &= 
	{1 \over 8} {\mathcal H}^{MN} \partial_M {\mathcal H}^{KL} \partial_N {\mathcal H}_{KL}
	- {1 \over 2} {\mathcal H}^{MN} \partial_M {\mathcal H}^{KL} \partial_K {\mathcal H}_{NL}
\notag \\
& \quad
	+ 4 {\mathcal H}^{MN} \partial_M \partial_N d 
	- \partial_M \partial_N {\mathcal H}^{MN} 
	- 4 {\mathcal H}^{MN} \partial_M d \, \partial_N d
	+ 4 \partial_M {\mathcal H}^{MN} \partial_N d.
\end{align}
The $O(D,D)$ invariance is manifest in this action.
There is an additional $\mathbb{Z}_2$ symmetry that corresponds to reversing the orientation of closed strings.
In addition, the action \eqref{eq:DFTaction} is invariant under the {\it DFT gauge transformation}.

The gauge symmetry in DFT originates from the T-duality
convariantized diffeomorphism and the $U(1)$ gauge symmetry of the NSNS $B$-field.
The infinitesimal gauge transformation $\delta_\Xi$ of $O(D,D)$ tensors
is given by the generalized Lie derivative $\widehat{\mathcal L}_\Xi$. 
The generalized Lie derivative on a doubled vector $V^M$ with weight $w(V)$ is defined by~\cite{Hohm:2010pp},
\begin{align}
\widehat{\mathcal L}_\Xi V^M
&= \Xi^K \partial_K V^M + (\partial^M \Xi_K - \partial_K \Xi^M) V^K
	+ w(V) V^M \partial_K \Xi^K.
\label{eq:def-genLiederiv}
\end{align}
Here $\Xi$ is a gauge parameter which takes value in a doubled vector.
The weight of the generalized metric is $w({\mathcal H}) = 0$ while that of the generalized dilaton is $w(e^{-2d}) = 1$.
Sometimes the generalized Lie derivative is called the {\sf D}-bracket
(a generalization of the Dorfman bracket in generalized geometry).
This defines the algebra of the DFT gauge symmetry, namely, the commutator of the generalized Lie derivative
$\widehat{\mathcal L}_\Xi$ is calculated as 
\begin{align}
[\widehat{\mathcal L}_{\Xi_1}, \widehat{\mathcal L}_{\Xi_2}]
&= \widehat{\mathcal L}_{[\Xi_1,\Xi_2]_{\sf C}} + F(\Xi_1, \Xi_2, \cdot).
	\label{eq:com-genLiederiv}
\end{align}
Here $\Xi_i^M$ ($i = 1,2$) are gauge parameters.
As its notation stands for, the algebra \eqref{eq:com-genLiederiv} is governed by the {\sf C}-bracket:
\begin{align}
[\Xi_1, \Xi_2]_{\sf C}^M
&= \Xi_1^K \partial_K \Xi_2^M - \Xi_2^K \partial_K \Xi_1^M
- {1 \over 2} \eta_{KL}
	(\Xi_1^K \partial^M \Xi_2^L - \Xi_2^K \partial^M \Xi_1^L).
	\label{eq:C-bracket_ap}
\end{align}
The extra term $F$ in \eqref{eq:com-genLiederiv} is given by 
\begin{align}
	F(\Xi_1, \Xi_2, V)^M
	&= {1 \over 2} \eta_{KL}
		(\Xi_1^K \partial^P \Xi_2^L - \Xi_2^K \partial^P \Xi_1^L) \partial_P V^M
		- (\partial^P \Xi_1^M \partial_P \Xi_2^K - \partial^P \Xi_2^M \partial_P \Xi_1^K) V_K.
	\label{eq:Cbr-SCterm}
\end{align}
Due to the non-zero contributions of $F$, the commutator of the generalized Lie derivative does not close by the {\sf C}-bracket in general.
In order that the algebra closes, one should impose the following {\it strong constraint} on the DFT fields and the gauge parameters:
\begin{align}
\eta^{MN} \partial_M * \partial_N * = 0.
\label{eq:strong-constraint}
\end{align}
Here $*$ stands for arbitrary DFT fields and gauge parameters.
Solutions to the constraint \eqref{eq:strong-constraint} determine a
physical space-time.
A trivial way to solve the strong constraint is to impose the condition $\tilde{\partial}^\mu * = 0$.
This means that any DFT fields and gauge parameters depend only on 
the coordinate $x^{\mu}$ that is the Fourier dual of the KK-modes.
Therefore the strong constraint is needed to make DFT be a physical theory.
Strictly speaking, the physical condition necessary for the DFT fields is given by the section condition $\eta^{MN} \partial_M \partial_N * = 0$
which is nothing but the level matching condition for closed strings. 
Contrast to the strong constraint, this is called the weak constraint. 
We stress that $\tilde{\partial}^\mu * = 0$ provides a solution to
\eqref{eq:strong-constraint} but it is not the unique way to find solutions.
We also note that the strong constraint is a sufficient condition for the closure of the DFT gauge algebra but it is not a necessary condition 
of the theory.

The gauge transformation of the generalized Ricci scalar ${\mathcal R}$ is~\cite{Grana:2012rr}
\begin{align}
\delta_\Xi {\mathcal R} = \widehat{\mathcal L}_\Xi {\mathcal R}
&= \Xi^K \partial_K {\mathcal R} + G(\Xi, {\mathcal H}, d),
\end{align}
where $G$ is a contribution which vanishes under the strong constraint.
Although the DFT action ~\eqref{eq:DFTaction} is an $O(D,D)$ scalar, it
is not invariant under the DFT gauge transformation in general.
In order to make it be gauge invariant, one needs the strong constraint~\eqref{eq:strong-constraint}.
When all the fields do not depend on the winding coordinate $\tilde{x}_\mu$, namely, 
when we impose the trivial constraint $\tilde{\partial}^\mu * =0$, then the DFT action 
~\eqref{eq:DFTaction} reduces to the one for the NSNS sector of type II supergravity:
\begin{align}
S_{\rm DFT} \xrightarrow{\tilde{\partial}^\mu * = 0}
S_{\rm NSNS} &= 
	\int \dop^D x \sqrt{-g} e^{-2\phi} 
	\left[ R + 4 (\partial \phi)^2 - {1 \over 12} (H_{\it 3})^2 \right].
\end{align}
Here $R$ is the Ricci scalar constructed from the metric $g_{\mu\nu}$.
$H_{\it 3} = \dop B$ is the field strength of the Kalb-Ramond
$B$-field.
In the same way, the {\sf D}-bracket is reduced to the Dorfman bracket
in generalized geometry by the condition $\tilde{\partial}^\mu * = 0$:
\begin{align}
\widehat{\mathcal L}_{\Xi_1} \Xi_2^M = [\Xi_1, \Xi_2]_{\sf D}^M 
\xrightarrow{\tilde{\partial}^\mu * = 0}
[A + \alpha, B + \beta]_{\rm d} &= [A, B]_{L} + {\mathcal L}_A \beta - \iop_B \dop\alpha.
\end{align}
Here the doubled vector is decomposed as $\Xi_1 = A + \alpha$, $\Xi_2 = B + \beta$ and 
$[A,B]_{L}$ is the ordinary Lie bracket for vectors on the
$D$-dimensional space-time.
The ``lower components'' of the doubled vectors $\alpha, \beta$ are
recognized as 1-forms on the physical space-time.
Finally, the {\sf C}-bracket is reduced to the Courant bracket
\eqref{eq:c-bracket} in
generalized geometry by the condition $\tilde{\partial}^\mu * = 0$:
\begin{align}
[\Xi_1, \Xi_2]_{\sf C} \xrightarrow{\tilde{\partial}^\mu * = 0}
[A + \alpha, B + \beta]_{\rm c} &=
	[A, B]_{L} + {\mathcal L}_A \beta - {\mathcal L}_B \alpha
	- {1 \over 2} \dop (\iop_A \beta - \iop_B \alpha).
\end{align}

\section{Calculus on {\sf C}-bracket} \label{sect:C-bracket}
In this section, we introduce relevant properties of the {\sf C}-bracket
in DFT.
The relation between the {\sf C}-bracket and the {\sf D}-bracket is as follows:
\begin{align}
[{\p}, {\Q}]_{\sf D}^M &= [{\p}, {\Q}]_{\sf C}^M + {1 \over 2} \eta^{MN} \partial_N (\eta_{KL} {\p}^K {\Q}^L).
\label{eq:differCD}
\end{align}
We note that the {\sf D}-bracket is not skew symmetric with respect to
its arguments.
Then the {\sf C}-bracket is defined as the anti-symmetric combination of the {\sf D}-bracket:
\begin{align}
[{\p}, {\Q}]_{\sf C}^M 
	&= {1 \over 2} (\widehat{\mathcal L}_{{\p}} {\Q}^M - \widehat{\mathcal L}_{{\Q}} {\p}^M)
	= {1 \over 2} ([{\p}, {\Q}]_{\sf D}^M - [{\Q}, {\p}]_{\sf D}^M).
\label{eq:Cis-antiD}
\end{align}
In order to see the role of the strong constraint in the Jacobi identity, 
we evaluate the Jacobiator of the {\sf C}-bracket:
\begin{align}
J_{\sf C}({\p}, {\Q}, {\R})^M 
	&= [[{\p},{\Q}]_{\sf C}, {\R}]_{\sf C}^M 
		+  [[{\Q},{\R}]_{\sf C}, {\p}]_{\sf C}^M +  [[{\R},{\p}]_{\sf C}, {\Q}]_{\sf C}^M.
\label{eq:C-Jacobi}
\end{align}
In the following calculus, we never impose the constraint \eqref{eq:strong-constraint}.
Before the discussion, we first show the (left-)Leibniz identity (Jacobi-like identity) of the {\sf D}-bracket.
The {\sf D}-bracket can be written as 
$[{\p},{\Q}]_{\sf D}^M = [{\p},{\Q}]^M + {\Q}^K \partial^M {\p}_K$ by using the ordinary Lie bracket $[\cdot,\cdot]$ on $T\mathcal{M}$.
The ordinary Leibniz identity for the Lie bracket $[\cdot,\cdot]$ is given by
\begin{align}
[{\p}, [{\Q}, {\R}]] &= [[{\p},{\Q}], {\R}] + [{\Q}, [{\p}, {\R}]].
\end{align}
Here ${\p},{\Q},{\R}$ are vectors in the doubled space-time.
We now calculate the corresponding terms of the {\sf D}-bracket.
The first term $[{\p}, [{\Q},{\R}]_{\sf D}]_{\sf D}$ is evaluated as 
\begin{align}
& [{\p}, [{\Q}, {\R}]_{\sf D}]_{\sf D}^M 
\notag \\
&= [{\p}, [{\Q}, {\R}]_{\sf D}]^M + \eta_{KL} [{\Q}, {\R}]_{\sf D}^K \partial^M {\p}^L 
\notag \\
&= [{\p}, [{\Q}, {\R}]]^M 
	+ ({\p}^N \partial_N (\eta_{KL} {\R}^K \partial^M {\Q}^L) 
		- (\eta_{KL} {\R}^K \partial^N {\Q}^L) \partial_N {\p}^M) 
\notag \\
& \quad
	+ \eta_{KL} [{\Q}, {\R}]^K \partial^M {\p}^L
	+ \eta_{KL} (\eta_{IJ} {\R}^I \partial^K {\Q}^J) \partial^M {\p}^L.
\label{eq:D-Leibniz-LHS}
\end{align}
On the other hand, the term $[[{\p}, {\Q}]_{\sf D}, {\R}]_{\sf D}$ is calculated to be 
\begin{align}
& [[{\p}, {\Q}]_{\sf D}, {\R}]_{\sf D}^M \notag\\
&= [[{\p}, {\Q}]_{\sf D}, {\R}]^M + \eta_{KL} {\R}^K \partial^M [{\p}, {\Q}]_{\sf D}^L \notag\\
&= [[{\p}, {\Q}], {\R}]^M + ((\eta_{KL} {\Q}^K \partial^N {\p}^L) \partial_N {\R}^M 
	- {\R}^N \partial_N (\eta_{KL} {\Q}^K \partial^M {\p}^L)) \notag\\
& \quad
	+ \eta_{KL} {\R}^K \partial^M [{\p}, {\Q}]^L 
	+ \eta_{KL} {\R}^K \partial^M (\eta_{IJ} {\Q}^I \partial^L {\p}^J).
\end{align}
We then obtain 
\begin{align}
& [[{\p}, {\Q}]_{\sf D}, {\R}]_{\sf D}^M + [{\Q}, [{\p}, {\R}]_{\sf D}]_{\sf D}^M 
\notag \\
& = [{\p}, [{\Q}, {\R}]]^M 
	+ \eta_{KL} ([{\Q},{\R}]^K \partial^M {\p}^L
	+ (\eta_{IJ} {\R}^I \partial^K {\Q}^J) \partial^M {\p}^L
	+ {\p}^N \partial_N ({\R}^K \partial^M {\Q}^L)) 
\notag \\
& \quad 
	+ \eta_{KL} ({\Q}^K \partial^N {\p}^L \partial_N {\R}^M 
		- {\R}^K \partial^N {\p}^L \partial_N {\Q}^M).
\label{eq:D-Leibniz-RHS}
\end{align}
Again by using \eqref{eq:D-Leibniz-LHS}, we obtain the following relation,
\begin{align}
[{\p}, [{\Q}, {\R}]_{\sf D}]_{\sf D}^M 
&= [[{\p}, {\Q}]_{\sf D}, {\R}]_{\sf D}^M + [{\Q}, [{\p}, {\R}]_{\sf D}]_{\sf D}^M 
	+ \mathrm{SC}_{\rm D}({\p},{\Q},{\R})^M,
\label{eq:D-Leibniz-mod} \\
\mathrm{SC}_{\sf D} ({\p},{\Q},{\R})^M
&= \eta_{KL} ({\R}^K \partial^N {\p}^L \partial_N {\Q}^M 
		- {\Q}^K \partial^N {\p}^L \partial_N {\R}^M
		- {\R}^K \partial^N {\Q}^L \partial_N {\p}^M).
\label{eq:D-Leibniz-SC}
\end{align}
This is the (left-)Leibniz identity for the {\sf D}-bracket.
Note that the term $\mathrm{SC}_{\sf D}$ vanishes under the imposition of the strong constraint.

By using the (left-)Leibniz identity of the {\sf D}-bracket~\eqref{eq:D-Leibniz-mod},
we calculate the Jacobiator of the {\sf C}-bracket~\eqref{eq:C-Jacobi}.
The analysis is based on the proposition 3.16 in \cite{Gualtieri}. 
We first evaluate the term $[[{\p}, {\Q}]_{\sf C}, {\R}]_{\sf C}$.
Since the {\sf C}-bracket is the anti-symmetric combination of the {\sf D}-bracket
~\eqref{eq:Cis-antiD}, we have
\begin{align}
[[{\p},{\Q}]_{\sf C}, {\R}]_{\sf C} 
&= {1 \over 2} ([[{\p},{\Q}]_{\sf C}, {\R}]_{\sf D} - [{\R}, [{\p},{\Q}]_{\sf C}]_{\sf D}) 
\notag \\
&= {1 \over 4} ([[{\p},{\Q}]_{\sf D}, {\R}]_{\sf D} - [[{\Q},{\p}]_{\sf D}, {\R}]_{\sf D} 
	- [{\R}, [{\p},{\Q}]_{\sf D}]_{\sf D} + [{\R}, [{\Q},{\p}]_{\sf D}]_{\sf D}).
\label{eq:part-of-JacC1}
\end{align}
Then, by the (left-)Leibniz identity~\eqref{eq:D-Leibniz-mod}, 
the relation 
\eqref{eq:part-of-JacC1} becomes 
\begin{align}
[[{\p},{\Q}]_{\sf C}, {\R}]_{\sf C} 
&= {1 \over 4} (
	 [{\p}, [{\Q}, {\R}]_{\sf D}]_{\sf D} 
	 - [{\Q}, [{\p}, {\R}]_{\sf D}]_{\sf D}
	- \mathrm{SC}_{\sf D}({\p},{\Q},{\R}) 
\notag \\
& \qquad
	- [{\Q}, [{\p}, {\R}]_{\sf D}]_{\sf D}
	+ [{\p}, [{\Q}, {\R}]_{\sf D}]_{\sf D}
	+ \mathrm{SC}_{\sf D}({\Q},{\p},{\R}) 
\notag \\
& \qquad
	- [{\R}, [{\p},{\Q}]_{\sf D}]_{\sf D} 
	+ [{\R}, [{\Q},{\p}]_{\sf D}]_{\sf D}
	).
\label{eq:part-of-JacC2}
\end{align}
Therefore the summation over the cyclic permutations of the above expression gives 
\begin{align}
& [[{\p},{\Q}]_{\sf C}, {\R}]_{\sf C} + \text{c.p.} \notag \\
& = {1 \over 4} ([{\p}, [{\Q},{\R}]_{\sf D}]_{\sf D} - [{\Q}, [{\p},{\R}]_{\sf D}]_{\sf D} 
	- \mathrm{SC}_{\sf D}({\p},{\Q},{\R}) + \mathrm{SC}_{\sf D}({\Q},{\p},{\R}) + \text{c.p.}).
\end{align}
By using the Leibniz identity again, we find
\begin{align}
[[{\p},{\Q}]_{\sf C}, {\R}]_{\sf C} + \text{c.p.}
&= {1 \over 4} ([[{\p},{\Q}]_{\sf D},{\R}]_{\sf D} - \mathrm{SC}_{\sf D}({\Q},{\p},{\R}) + \text{c.p.}).
\label{eq:C-Jacobi-2}
\end{align}
Using the relation between {\sf C}- and {\sf D}-brackets~\eqref{eq:differCD}, we obtain 
\begin{align}
[[{\p},{\Q}]_{\sf C}, {\R}]_{\sf C}
&= [[{\p},{\Q}]_{\sf C}, {\R}]_{\sf D} 
- 
\partial^\bullet \etaprod{ [{\p},{\Q}]_{\sf C}, {\R} }
\notag \\
&= [[{\p},{\Q}]_{\sf D}, {\R}]_{\sf D} 
- 
[\partial^\bullet \etaprod{ {\p},{\Q} }, {\R}]_{\sf D}
- 
\partial^\bullet \etaprod{ [{\p},{\Q}]_{\sf C}, {\R} }.
\label{eq:part-of-JacC3}
\end{align}
Here $\partial^\bullet$ is a differential operator whose index is raised by $\eta^{MN}$ and 
$\etaprod{ {\p}, {\Q} } = {1 \over 2} \eta_{MN} {\p}^M {\Q}^N$.
The second term in \eqref{eq:part-of-JacC3} is calculated as 
\begin{align}
[\partial^\bullet \etaprod{ {\p}, {\Q} }, {\R}]_{\sf D}^M
&= {1 \over 2} 
\Big(
\partial^N ({\p}^K {\Q}_K) \partial_N {\R}^M
	+ (\partial^M \partial_N ({\p}^K {\Q}_K) - \partial_N \partial^M ({\p}^K {\Q}_K)) {\R}^N 
\Big)
\notag \\
&= {1 \over 2} 
\partial^N ({\p}^K {\Q}_K) \partial_N {\R}^M.
\end{align}
We note that this part vanishes under the strong constraint.
Finally, by using the result \eqref{eq:part-of-JacC3}, the 
Jacobiator of the {\sf C}-bracket~\eqref{eq:C-Jacobi-2} is 
\begin{align}
& [[{\p},{\Q}]_{\sf C}, {\R}]_{\sf C} + \text{c.p.} \notag \\
& = {1 \over 4} ([[{\p},{\Q}]_{\sf C},{\R}]_{\sf C} 
	+ 
	\partial^\bullet \etaprod{ [{\p},{\Q}]_{\sf C}, {\R} }
	+ 
	[\partial^\bullet \etaprod{ {\p},{\Q} }, {\R}]_{\sf D}
	- \mathrm{SC}_{\sf D}({\Q},{\p},{\R}) + \text{c.p.})
\notag \\
&= {1 \over 4} ( J_{\sf C}({\p},{\Q},{\R}) + 3 \partial^\bullet N_{\sf C}({\p},{\Q},{\R}) + 3\mathrm{SC}_{\sf C}({\p},{\Q},{\R}) ),
\end{align}
where
\begin{align}
N_{\sf C}({\p},{\Q},{\R}) 
	&= {1 \over 3} 
	\big( \etaprod{ [{\p},{\Q}]_{\sf C}, {\R} } + \text{c.p.} \big), \\
\mathrm{SC}_{\sf C} ({\p},{\Q},{\R}) 
	&= {1 \over 3} 
	\big( [\partial^\bullet \etaprod{ {\p},{\Q} }, {\R}]_{\sf D} 
	- 
	\mathrm{SC}_{\sf D}({\Q},{\p},{\R}) + \text{c.p.} \big).
\end{align}
Therefore we obtain
\begin{align}
J_{\sf C}({\p},{\Q},{\R}) &= \partial^\bullet N_{\sf C} ({\p},{\Q},{\R}) + \mathrm{SC}_{\sf C} ({\p},{\Q},{\R}).
\end{align}
Here $N_{\sf C}$ is the Nijenhuis operator and the contribution 
$\mathrm{SC}_{\sf C}$ vanishes under the strong constraint.
It is clear that the {\sf C}-bracket never satisfies the axiom
\ref{axiom:C1} of the Courant algebroid unless the strong constraint is imposed.

In order to see the doubled aspects of the gauge symmetry, 
we decompose the {\sf C}-bracket into the components.
The gauge parameters are decomposed into their KK and winding parts:
\begin{align}
{\p}^M &= 
\begin{pmatrix}
\alpha_\mu \\ A^\mu
\end{pmatrix}
,  \qquad
{\Q}^M = 
\begin{pmatrix}
\beta_\mu \\ B^\mu
\end{pmatrix}.
\end{align}
Then the {\sf C}-bracket is rewritten as 
\begin{align}
[{\p}, {\Q}]_{\sf C}^M
&= {\p}^K \partial_K {\Q}^M - {\Q}^K \partial_K {\p}^M
	- {1 \over 2} \eta_{KL} ({\p}^K \partial^M {\Q}^L - {\Q}^K \partial^M {\p}^L) \notag\\
&= \alpha_\nu \tilde{\partial}^\nu {\Q}^M + A^\nu \partial_\nu {\Q}^M
	- \beta_\nu \tilde{\partial}^\nu {\p}^M - B^\nu \partial_\nu {\p}^M \notag\\
& \quad
	- {1 \over 2} (\alpha_\nu \partial^M B^\nu + A^\nu \partial^M \beta_\nu
		- \beta_\nu \partial^M A^\nu - B^\nu \partial^M \alpha_\nu).
\end{align}
Now we consider the doubled basis 
$\e^M = (\e_\mu, \te^\mu)$.
Then, by contracting this with the {\sf C}-bracket, we obtain
\begin{align}
[ {\p}, {\Q} ]_{\sf C} =
[{\p},{\Q}]_{\sf C}^M \eta_{MN} \e^N
&= \alpha_\nu \tilde{\partial}^\nu \beta_\mu \te^\mu
	+ A^\nu \partial_\nu \beta_\mu \te^\mu
	- \beta_\nu \tilde{\partial}^\nu \alpha_\mu \te^\mu
	- B^\nu \partial_\nu \alpha_\mu \te^\mu \notag\\
& \quad
	+ \alpha_\nu \tilde{\partial}^\nu B^\mu \e_\mu
	+ A^\nu \partial_\nu B^\mu \e_\mu
	- \beta_\nu \tilde{\partial}^\nu A^\mu \e_\mu
	- B^\nu \partial_\nu A^\mu \e_\mu \notag\\
& \quad
	- {1 \over 2} (\alpha_\nu \tilde{\partial}^\mu B^\nu
		+ A^\nu \tilde{\partial}^\mu \beta_\nu
		- \beta_\nu \tilde{\partial}^\mu A^\nu
		- B^\nu \tilde{\partial}^\mu \alpha_\nu) \e_\mu \notag\\
& \quad
	- {1 \over 2} (\alpha_\nu \partial_\mu B^\nu
		+ A^\nu \partial_\mu \beta_\nu
		- \beta_\nu \partial_\mu A^\nu
		- B^\nu \partial_\mu \alpha_\nu) \te^\mu.
\end{align}
Each part is given by 
\begin{align}
[A,B]_{L}
	&= [A,B]_{L}^\mu \e_\mu
	= (A^\nu \partial_\nu B^\mu - B^\nu \partial_\nu A^\mu) \e_\mu, 
	\notag
	\\
[\alpha,\beta]_{\tilde{L}}
	&= \big( [\alpha, \beta]_{\tilde{L}} \big)_\mu \te^\mu
	= (\alpha_\nu \tilde{\partial}^\nu \beta_\mu
		- \beta_\nu \tilde{\partial}^\nu \alpha_\mu) \te^\mu,
	\notag
	 \\
\dop \iop_A \beta
	&= \dop (A^\nu \beta_\nu)
	= \partial_\mu (A^\nu \beta_\nu) \te^\mu
	= (\beta_\nu \partial_\mu A^\nu + A^\nu \partial_\mu \beta_\nu) \te^\mu, 
	\notag\\
\tilde{\dop} \iop_A \beta
	&= \tilde{\dop}(A^\nu \beta_\nu)
	= \tilde{\partial}^\mu (A^\nu \beta_\nu) \e_\mu
	= (\beta_\nu \tilde{\partial}^\mu A^\nu
		+ A^\nu \tilde{\partial}^\mu \beta_\nu) \e_\mu,
	\notag \\
\tilde{\mathcal L}_\alpha B
	&= (\alpha_\nu \tilde{\partial}^\nu B^\mu
		+ B^\nu \tilde{\partial}^\mu \alpha_\nu) \e_\mu, 
	\notag
	\\
{\mathcal L}_A \beta
	&= (A^\nu \partial_\nu \beta_\mu + \beta_\nu \partial_\mu A^\nu) \te^\mu.
\end{align}
Then, ${\mathcal L}$ becomes a Lie derivative by a vector field, while 
$\tilde{\mathcal L}$ becomes a (T-dualized) Lie derivative by a ``winding vector field''.
$[\cdot, \cdot]_{L}$ is the ordinary Lie bracket while $[\cdot,
\cdot]_{\tilde{L}}$ is the Lie bracket for the ``winding vector field''.
$\dop$ is the exterior derivative and $\tilde{\dop}$ is the
winding exterior derivative. Then the {\sf C}-bracket is decomposed as 
\begin{align}
[{\p},{\Q}]_{\sf C}^M \e_M
&= ([\alpha,\beta]_{\tilde{L}})_\mu \te^\mu
	+ A^\nu \partial_\nu \beta_\mu \te^\mu
	- B^\nu \partial_\nu \alpha_\mu \te^\mu
	+ [A, B]_{L}^\mu \e_\mu
	+ \alpha_\nu \tilde{\partial}^\nu B^\mu \e_\mu
	- \beta_\nu \tilde{\partial}^\nu A^\mu \e_\mu \notag\\
& \quad
	- {1 \over 2} ( 2 A^\nu \tilde{\partial}^\mu \beta_\nu - (\tilde{\dop} \iop_A \beta)^\mu
		+ (\tilde{\dop} \iop_B \alpha)^\mu - 2 B^\nu \tilde{\partial}^\mu \alpha_\nu) \e_\mu \notag\\
& \quad
	- {1 \over 2} ( (\dop \iop_A \beta)_\mu - 2 \beta_\nu \partial_\mu A^\nu
		- (\dop \iop_B \alpha)_\mu + 2 \alpha_\nu \partial_\mu B^\nu) \te^\mu \notag\\
&= \left( [A,B]_{L}^\mu + \tilde{\mathcal L}_\alpha B^\mu - \tilde{\mathcal L}_\beta A^\mu
	+ {1 \over 2} (\tilde{\dop} (\iop_A \beta - \iop_B \alpha))^\mu \right) \e_\mu \notag\\
& \quad
	+ \left( ([\alpha,\beta]_{\tilde{L}})_\mu + {\mathcal L}_A \beta_\mu - {\mathcal L}_B \alpha_\mu
	- {1 \over 2} (\dop (\iop_A \beta - \iop_B \alpha))_\mu \right) \te^\mu.
\end{align}
This is explicitly written as a sum of 
the Courant bracket-like structures~\cite{Hull:2009zb}:
\begin{align}
[{\p},{\Q}]_{\sf C}
= [A + \alpha, B + \beta]_{\sf C}
&=  [A,B]_{L}
	+ {\mathcal L}_A \beta - {\mathcal L}_B \alpha
	- {1 \over 2} \dop (\iop_A \beta - \iop_B \alpha)
\notag \\
& \quad
	+ [\alpha,\beta]_{\tilde{L}}
	+ \tilde{\mathcal L}_\alpha B - \tilde{\mathcal L}_\beta A
	+ {1 \over 2} \tilde{\dop} (\iop_A \beta - \iop_B \alpha).
\label{eq:C-bracket-comp}
\end{align}
From this expression, when we consider the supergravity frame
$\tilde{\partial} * = 0$ the second part vanishes and the {\sf C}-bracket
reduces to the original Courant bracket (c-bracket) \eqref{eq:c-bracket}.

\section{Detailed calculations on algebroids} \label{sect:calculations}
In this section, we exhibit detailed calculations on \eqref{C1_4} and \eqref{eq:C4_V}.

\subsection{Calculations on \eqref{C1_4}}
We here derive eq.~\eqref{C1_4}.
By its definition, the Lie derivative on $[\xi_1,\xi_2]_{\mathrm{V}}$ is calculated as 
\begin{align}
  \mathcal{L}_{X_3}[\xi_1,\xi_2]_{\rm V}
  &= ({\dop} \iop_{X_3} + \iop_{X_3}{\dop})[\xi_1,\xi_2]_{\rm V} \notag\\
  &= {\dop} \langle X_3,[\xi_1,\xi_2]_{\rm V} \rangle + 
\iop_{X_3}\mathcal{L}_{\xi_1}{\dop}\xi_2 - \iop_{X_3}\mathcal{L}_{\xi_2}{\dop}\xi_1 \notag\\
  &\quad + \iop_{X_3}({\dop}[\xi_1,\xi_2]_{\rm V} -
 \mathcal{L}_{\xi_1}{\dop}\xi_2 + \mathcal{L}_{\xi_2}{\dop}\xi_1),
\label{eq:C1_0}
\end{align}
where we have extracted out the derivation condition part.

In order to evaluate the cyclic permutations of the above expression, it
is convenient to rewrite the interior product parts 
$\iop_{X_3}\mathcal{L}_{\xi_1}{\dop}\xi_2 - \iop_{X_3}\mathcal{L}_{\xi_2}{\dop}\xi_1$.
To this end, we first show that the following relation holds:
\begin{equation}
  \iop_X\mathcal{L}_{\xi}{\dop}\eta = [\xi,\mathcal{L}_X\eta]_{\rm V} -
  \mathcal{L}_{\mathcal{L}_\xi X}\eta + [{\dop} \langle \eta,X
  \rangle,\xi] + {\dop}(\rho_*(\xi) \cdot \langle \eta,X \rangle) - {\dop}
  \langle [\xi,\eta]_{\rm V},X \rangle.
  \label{C1_5}
\end{equation}
Deriving this expression needs a little effort.
First, the Leibniz rule of the Lie derivative in
\eqref{eq:Lie_derivative_relations} results in 
\begin{equation}
 \braket{\iop_X \mathcal{L}_\xi \dop \eta ,Y} =
  \mathcal{L}_\xi(\dop\eta(X,Y)) - \dop\eta(\mathcal{L}_\xi X,Y) -
  \dop\eta(X,\mathcal{L}_\xi Y).
\label{eq:C1_6}
\end{equation}
The first term in the above gives $\mathcal{L}_{\xi} \dop \eta (X,Y) =
\rho_* (\xi) \cdot ( \dop \eta (X,Y) )$ while 
the remaining terms are evaluated by the definition of the exterior derivative:
\begin{equation}
\dop\xi(X,Y) = \rho (X) \cdot \braket{\xi,Y} - \rho (Y) \cdot \braket{\xi,X} - \braket{\xi,[X,Y]_E}.
\end{equation}
Therefore, \eqref{eq:C1_6} is written as
\begin{align}
 \braket{\iop_X \mathcal{L}_\xi \dop \eta ,Y}
 &= \rho_*(\xi) \rho(X) \cdot \braket{\xi,Y} - \rho_* (\xi) \rho
 (Y) \cdot \braket{\eta,X} 
- \rho_* (\xi)  \cdot \braket{\eta,[X,Y]_E} \notag\\
 &\quad  
- \rho (\mathcal{L}_\xi X) \cdot  \braket{\eta,Y} 
+ \rho (Y) \cdot \braket{\eta,\mathcal{L}_\xi X} 
+ \braket{\eta , [\mathcal{L}_\xi
 X,Y]_E}
\notag\\
 &\quad  
- \rho (X) \cdot  \braket{\eta,\mathcal{L}_\xi Y} 
+ \rho (\mathcal{L}_\xi Y) \cdot \braket{\eta, X} 
+ \braket{\eta , [X,\mathcal{L}_\xi Y]_E}.
\label{eq:C1_a}
\end{align}
Using \eqref{Liederiv_f1},\eqref{Liederiv_f2}, the first terms in each
line in the right hand side of \eqref{eq:C1_a} is evaluated as 
 \begin{align}
  &\rho_*(\xi) \rho(X) \cdot \braket{\xi,Y} =
  \rho_*(\xi) \cdot \braket{\mathcal{L}_X\eta,Y} +
  \rho_*(\xi) \cdot \braket{\eta,[X,Y]_E}, 
\notag \\
  &\rho (\mathcal{L}_\xi X) \cdot  \braket{\eta,Y} =
  \braket{\mathcal{L}_{\mathcal{L}_\xi X}\eta,Y} +
  \braket{\eta,[\mathcal{L}_\xi X,Y]_E},
\notag  \\
  &\rho (\mathcal{L}_\xi Y) \cdot  \braket{\eta,X} 
  = \braket{\mathcal{L}_{\mathcal{L}_\xi Y}\eta,X} 
  + \braket{\eta,[\mathcal{L}_\xi Y,X]_E}.
 \end{align}
Therefore we find
 \begin{align}
 \braket{\iop_X \mathcal{L}_\xi \dop \eta ,Y}
 &= \rho_*(\xi) \cdot \braket{\mathcal{L}_X\eta,Y} 
 - \rho_*(\xi)\rho(Y) \cdot \braket{\eta,X} 
 - \braket{\mathcal{L}_{\mathcal{L}_\xi X}\eta, Y } 
 \notag\\
 & \quad
 + \rho(Y) \cdot \braket{\eta,\mathcal{L}_\xi X} 
 - \rho(X) \cdot \braket{\eta ,\mathcal{L}_\xi Y} 
 + \braket{\mathcal{L}_{\mathcal{L}_\xi Y}\eta,X}.
\end{align}
The right hand side is rewritten as the inner products with $Y$ as
follows. Again, by the defining properties of the Lie derivative
\eqref{eq:Lie_derivative_relations}, we have
\begin{align}
  \rho(Y) \cdot \braket{\eta,\mathcal{L}_\xi X}
  &= \rho(Y)\rho_*(\xi) \cdot \braket{\eta,X} 
  - \rho(Y) \cdot \braket{[\xi,\eta]_{\mathrm{V}},X} 
  \notag\\
  &= \braket{\dop(\rho_*(\xi) \cdot \braket{\eta,X}),Y} 
  - \braket{\dop\braket{[\xi,\eta]_{\mathrm{V}},X},Y}.
\end{align}
Using this we obtain
\begin{align}
  \braket{\iop_X \mathcal{L}_\xi \dop \eta ,Y}
  &= \braket{\dop(\rho_*(\xi) \cdot \braket{\eta,X}),Y} 
  + \braket{\dop\braket{[\xi,\eta]_{\mathrm{V}},X},Y} 
  - \braket{\mathcal{L}_{\mathcal{L}_\xi X}\eta, Y } 
  + \braket{[\xi,\mathcal{L}_X\eta]_{\mathrm{V}},Y}
  \notag\\
  &\quad 
  - \mathcal{L}_\xi\braket{\mathcal{L}_Y\eta,X} 
  - \mathcal{L}_\xi\braket{\eta,\mathcal{L}_YX} 
  - \braket{\eta,\mathcal{L}_X \mathcal{L}_\xi Y} 
  + \braket{\mathcal{L}_{\mathcal{L}_\xi Y}\eta,X}.
\end{align}
Since we have the following relation,
\begin{align}
  \braket{\mathcal{L}_{\mathcal{L}_\xi Y}\eta,X} - \braket{\eta,\mathcal{L}_X \mathcal{L}_\xi Y}
  &= \braket{\mathcal{L}_{\mathcal{L}_\xi Y}\eta,X} + \braket{\eta,\mathcal{L}_{\mathcal{L}_\xi Y}X} \notag\\
  &= \mathcal{L}_{\mathcal{L}_\xi Y} \braket{\eta,X} \notag\\
  &= \braket{\dop\braket{\eta,X},\mathcal{L}_\xi Y} \notag\\
  &= \mathcal{L}_\xi\braket{\dop\braket{\eta,X},Y} - \braket{[\xi,\dop\braket{\eta,X}]_{\mathrm{V}},Y}.
\end{align}
We find
\begin{align}
  \braket{\iop_X \mathcal{L}_\xi \dop \eta ,Y}
  &= \braket{\dop(\rho_*(\xi) \cdot \braket{\eta,X}),Y} 
  + \braket{\dop\braket{[\xi,\eta]_{\mathrm{V}},X},Y} 
  - \braket{\mathcal{L}_{\mathcal{L}_\xi X}\eta, Y } 
  \notag\\
  &\quad 
  + \braket{[\xi,\mathcal{L}_X\eta]_{\mathrm{V}},Y}  
  - \braket{[\xi ,\dop\braket{\eta,X}]_{\mathrm{V}},Y} 
  \notag\\
  &\quad 
  - \mathcal{L}_\xi\braket{\mathcal{L}_Y\eta,X} 
  - \mathcal{L}_\xi\braket{\eta,\mathcal{L}_YX} 
  + \mathcal{L}_\xi\braket{Y,\dop\braket{\eta,X}}.
\end{align}
We note that the terms in the third line vanish due to the following relation,
\begin{align}
   \mathcal{L}_\xi\braket{Y, \dop \braket{\eta,X}}
   &= \mathcal{L}_\xi\mathcal{L}_Y \braket{\eta,X} \notag\\
   &= \mathcal{L}_\xi\braket{\mathcal{L}_Y\eta,X} + \mathcal{L}_\xi\braket{\eta,\mathcal{L}_YX}.
\end{align}
Then, the expression \eqref{C1_5} follows.

Now we go back to the evaluation of \eqref{eq:C1_0}.
Using the relation \eqref{C1_5}, the interior product parts 
$\iop_{X_3}\mathcal{L}_{\xi_1}{\dop}\xi_2 -
\iop_{X_3}\mathcal{L}_{\xi_2}{\dop}\xi_1$ in \eqref{eq:C1_0} are
calculated. Then we find
\begin{align}
 \mathcal{L}_{X_3}[\xi_1,\xi_2]_{\mathrm{V}}
 &= - {\dop} \langle [\xi_1,\xi_2]_{\mathrm{V}},X_3 \rangle \notag\\
 &\quad 
 	+ [\xi_1,\mathcal{L}_{X_3}\xi_2]_{\mathrm{V}}
        - \mathcal{L}_{\mathcal{L}_{\xi_1} X_3}\xi_2
        + [{\dop} \langle \xi_2,X_3\rangle,\xi_1]_{\mathrm{V}}
        + {\dop}(\rho_*(\xi_1)  \cdot  \langle \xi_2,X_3 \rangle)\notag\\
 &\quad 
 	- [\xi_2,\mathcal{L}_{X_3}\xi_1]_{\mathrm{V}}
        + \mathcal{L}_{\mathcal{L}_{\xi_2} X_3}\xi_1
        - [{\dop} \langle \xi_1,X_3 \rangle,\xi_2]_{\mathrm{V}}
        - {\dop}(\rho_*(\xi_2)  \cdot  \langle \xi_1,X_3 \rangle) \notag\\
 &\quad 
 	+ \iop_{X_3}({\dop}[\xi_1,\xi_2]_{\mathrm{V}} 
		- \mathcal{L}_{\xi_1}{\dop}\xi_2 
		+ \mathcal{L}_{\xi_2}{\dop}\xi_1).
 \label{eq:Lie-deriv-braket}
\end{align}
The cyclic permutations of the second and the third terms are found to
be
\begin{align}
  &[\xi_1,\mathcal{L}_{X_3}\xi_2]_{\mathrm{V}} - [\xi_2,\mathcal{L}_{X_3}\xi_1]_{\mathrm{V}} + \text{c.p.}
   = [\mathcal{L}_{X_1}\xi_2 - \mathcal{L}_{X_2}\xi_1,\xi_3]_{\mathrm{V}} + \text{c.p.}, \notag\\
  &- \mathcal{L}_{\mathcal{L}_{\xi_1} X_3}\xi_2 + \mathcal{L}_{\mathcal{L}_{\xi_2} X_3}\xi_1 + \text{c.p.}
   = \mathcal{L}_{\mathcal{L}_{\xi_1}X_2 - \mathcal{L}_{\xi_2}X_1}\xi_3 + \text{c.p.}, \notag\\
  &[{\dop} \langle \xi_2,X_3\rangle,\xi_1]_{\mathrm{V}} - [{\dop} \langle \xi_1,X_3 \rangle,\xi_2]_{\mathrm{V}} + \text{c.p.}
   = + 2[{\dop} \mbra{e_1,e_2},\xi_3]_{\mathrm{V}} + \text{c.p.},\notag\\
  &{\dop}(\rho_*(\xi_1)  \cdot  \langle \xi_2,X_3 \rangle) 
  - {\dop}(\rho_*(\xi_2)  \cdot  \langle \xi_1,X_3 \rangle) + \text{c.p.}
   = 2{\rm d}(\rho_* (\xi_3) \cdot \mbra{e_1,e_2}) + \text{c.p.}
\end{align}
We then finally obtain 
\begin{align}
  \mathcal{L}_{X_3}[\xi_1,\xi_2]_{\mathrm{V}} + \text{c.p.}
  &= - {\dop} \langle [\xi_1,\xi_2]_{\mathrm{V}}, X_3 \rangle 
\notag\\
  &\quad + [\mathcal{L}_{X_1}\xi_2 - \mathcal{L}_{X_2}\xi_1,\xi_3]_{\mathrm{V}}
     +\mathcal{L}_{\mathcal{L}_{\xi_1}X_2 - \mathcal{L}_{\xi_2}X_1}\xi_3 
\notag\\
  &\quad + 2[{\dop} \mbra{e_1,e_2},\xi_3]_{\mathrm{V}}
     + 2{\rm d}(\rho_* (\xi_3) \cdot \mbra{e_1,e_2}) 
\notag\\
  &\quad + \iop_{X_3}({\dop}[\xi_1,\xi_2]_{\mathrm{V}} - \mathcal{L}_{\xi_1}{\dop}\xi_2
     + \mathcal{L}_{\xi_2}{\dop}\xi_1) + \text{c.p.}
\end{align}
This is the equation \eqref{C1_4}.

\subsection{Calculations on \eqref{eq:C4_V}}
We here derive the relation \eqref{eq:C4_V}.
For any $X,Y \in \Gamma (E)$, $f \in C^{\infty} (M)$, we have 
\begin{align}
[X,f Y]_E = f [X,Y]_E + (\rho (X) \cdot f) Y.
\end{align}
Then, we obtain
\begin{align}
\dop_* [X,f Y]_E =& \ 
\dop_* f \wedge [X,Y] + f \dop_* [X,Y]_E 
+ \dop_* \left( \rho (X) \cdot f \right) \wedge Y + (\rho (X) \cdot f )
 \dop_* Y
\notag \\
=& \ \dop_* f \wedge [X,Y] 
+ f 
\Bigl(
- \mathcal{L}_Y \dop_* X + \mathcal{L}_X \dop_* Y
\Bigr)
+
 f 
\Bigl(
\dop_* [X,Y]_E 
+ \mathcal{L}_Y \dop_* X - \mathcal{L}_X \dop_* Y
\Bigr)
\notag \\
&  
+ \dop_* \left( \rho (X) \cdot f \right) \wedge Y + (\rho (X) \cdot f )
 \dop_* Y.
\label{eq:AppC1}
\end{align}
Here we have extracted out the derivation condition part.

On the other hand, since we have
\begin{align}
& - \mathcal{L}_{f Y} \dop_* X + \mathcal{L}_X \dop_* (f Y) 
\notag \\
&=  
- \mathcal{L}_{fY} \dop_* X + \mathcal{L}_X (\dop_* f \wedge Y + f
 \dop_* Y) 
\notag \\
& =  
- \mathcal{L}_{f Y} \dop_* X + (\mathcal{L}_X \dop_* f) \wedge Y +
 \dop_* f \wedge \mathcal{L}_X Y + (\rho (X) \cdot f) \dop_* Y + f
 \mathcal{L}_X \dop_* Y,
\end{align}
then we find
\begin{align}
\dop_* [X,fY]_E =& \
\Bigl(
\dop_* [X,fY]_E - \mathcal{L}_X \dop_* (fY) + \mathcal{L}_{fY} \dop_* X
\Bigr)
+
\Bigl(
-\dop_* [X,fY]_E + \mathcal{L}_X \dop_* (fY) + \mathcal{L}_{fY} \dop_* X
\Bigr)
\notag \\
=& \ 
\Bigl(
\dop_* [X,fY]_E - \mathcal{L}_X \dop_* (fY) + \mathcal{L}_{fY} \dop_* X
\Bigr)
\notag \\
& 
- \mathcal{L}_{f Y} \dop_* X + (\mathcal{L}_X \dop_* f) \wedge Y +
 \dop_* f \wedge \mathcal{L}_X Y + (\rho (X) \cdot f) \dop_* Y + f
 \mathcal{L}_X \dop_* Y.
\label{eq:AppC2}
\end{align}
By comparing the two expressions \eqref{eq:AppC1} and \eqref{eq:AppC2}, 
we find
\begin{align}
\Bigl\{
\mathcal{L}_{X} \dop_* f - \iop_{\dop f} \dop_* X - \dop_* ( \rho (X)
 \cdot f)
\Bigr\} \wedge Y
=& \ 
f 
\Bigl(
\dop_* [X,Y]_E + \mathcal{L}_Y \dop_* X - \mathcal{L}_{X} \dop_* Y
\Bigr)
\notag \\
& 
-
\Bigl(
\dop_* [X,f Y]_E - \mathcal{L}_X \dop_* (f Y) + \mathcal{L}_{f Y} \dop_* X
\Bigr),
\end{align}
where we have used the relation $\mathcal{L}_{fY} \dop_* X = f
\mathcal{L}_Y \dop_* X - Y \wedge \iop_{\dop f} \dop_* X$ which follows
from a property in \eqref{eq:Lie_derivative_relations}.
Again, using the properties in \eqref{eq:Lie_derivative_relations}, we
have
\begin{align}
\mathcal{L}_X \dop_* f - \iop_{\dop f} \dop_* X - \dop_* (\rho (X) \cdot
 f) 
=& \ \mathcal{L}_X \dop_* f - \mathcal{L}_{\dop f} X + \dop_*
 (\iop_{\dop f} X) - \dop_* (\rho (X) \cdot f )
\notag \\
=& \ - \mathcal{L}_{\dop f} X + \mathcal{L}_X \dop_* f + \dop_* 
\Bigl(
\iop_{\dop f} X - \rho (X) \cdot f
\Bigr)
\notag \\
=& \ - \mathcal{L}_{\dop f} X + [X,\dop_* f]_E.
\end{align}
Here we have used $\iop_{\dop f} X = \langle \dop f, X \rangle = \rho
(X) \cdot f$. Then, the equation \eqref{eq:C4_V} follows.

\end{appendix}


\end{document}